\documentclass[10pt]{article}

\usepackage{amsmath}
\usepackage{amssymb}
\usepackage{tikz}
\usepackage{graphicx}
\usepackage[footnotesize,hang]{caption}
\usepackage{subfig}
\usepackage{color}
\usepackage{rotating}
\usepackage{mathtools}
\usepackage{xcolor}

\usetikzlibrary{decorations.pathmorphing}

\usepackage{setspace}
\title{Three-dimensional capillary waves due to a submerged source with small surface tension}
\author{C. J. Lustri$^1$\footnote{Electronic address: christopher.lustri@mq.edu.au; corresponding author}, R. Pethiyagoda$^2$, and S. J. Chapman$^3$.}
\date{%
    $^1$Department of Mathematics and Statistics, 12 Wally's Walk, Macquarie University, New South Wales 2109, Australia\\
    $^2$School of Mathematical Sciences, Queensland University of Technology, Brisbane QLD 4001, Australia\\
    $^3$Oxford Center for Industrial and Applied Mathematics, Mathematical Institute, University of Oxford, Oxford OX1 3LB, UK\\[2ex]%
}                                     % Activate to display a given date or no date

\begin{document}
\maketitle

\begin{abstract}
Steady and unsteady linearised flow past a submerged source are
studied in the small-surface-tension limit,  in the absence of
gravitational effects. The free-surface capillary waves generated are
exponentially small in the surface tension, and are determined using
the theory of  exponential asymptotics.
In the steady problem, capillary waves are found to extend upstream
from the source, switching on across curves on the free surface known
as Stokes lines. Asymptotic predictions and
compared with computational solutions for the position of the free surface.

In the unsteady problem, transient effects cause  the solution to
display more complicated asymptotic behaviour, such as higher-order
Stokes lines. The theory of  exponential asymptotics
is applied to show how the capillary waves evolve over time, and
eventually tend to the steady solution. 
\end{abstract}

%\begin{keywords}
%exponential asymptotics; water waves; free surface flows; low-Bond number limit
%\end{keywords}

\section{Introduction}

\subsection{Background}

Free surface waves induced by flow over a
submerged obstacle can broadly be divided into waves in which the
dominant effect is gravity, and waves 
in which the dominant effect is surface tension, known as capillary
waves. The behaviour of 
waves on the length scale of large obstacles, such as ships or
submarines, tend to be dominated by gravitational effects; however,
surface tension plays an important role at smaller scales. In this
study, we perform an asymptotic study of capillary waves caused by
flow over a submerged source in the small-surface-tension limit.  

Much of the early theoretical work on the behaviour of capillary waves on a steady stream is summarised in \cite{Whitham1}, which contains a demonstration that capillary waves caused by flow past an obstacle have a group velocity faster than the flow speed in two dimensions, and therefore any steady wavetrain must be found upstream from the obstacle. Early studies on the behaviour of capillary waves in the absence of gravity include \cite{Crapper1}, who derived an exact closed-form solution for two-dimensional capillary waves on infinitely deep water, and \cite{Kinnersley1}, who obtained similar solutions for capillary waves on water of finite depth. 

Further studies on related two-dimensional capillary wave systems
include \cite{VandenBroeck6}, who discovered a new family of nonlinear
capillary wave solutions on deep water. \cite{Hogan1,Hogan2, Hogan3}
studied particle trajectories in capillary wave systems, finding a
range of symmetric and anti-symmetric wave patterns. More recently,
\cite{Vandenbroeck7} extended the results of \cite{Crapper1} and
\cite{Kinnersley1} to axisymmetric domains, while \cite{Crowdy1}
extended the solutions by \cite{Kinnersley1} in order to consider
capillary waves on the surface of multiply-connected fluid
domains. \cite{Blyth1}, and \cite{Crowdy2} used complex variable
methods to determine the behaviour of capillary waves on a flat and
curved fluid sheet respectively, and \cite{Vandenbroeck8} studied the
behaviour of capillary waves in systems with varying surface
tension. \cite{Vandenbroeck9} provides a broad summary of known
results on two-dimensional nonlinear capillary wave systems, including
flows past geometry and around obstacles. Importantly, we note that
\cite{Chapman6} studied the behaviour of exponentially small capillary
waves in two dimensions, using asymptotic methods similar to those
applied in the present study. 

Each of these previous investigations concentrated on capillary waves
in two dimensions. Capillary waves in three-dimensional systems are
less well-studied. There do exist a large number of numerical studies
into the behaviour of gravity-capillary waves induced by flow past an
obstacle or pressure distribution in three dimensions. Many of these
are found in \cite{Dias1} and \cite{Vandenbroeck10}, and the
references therein. Other three-dimensional studies have concentrated
on exploring capillary wave behaviour caused by rotating obstacles,
such as whirligig beetles \cite{Tucker1, Chepelianskii1}. In
particular, we will note that our results are qualitatively similar to
Figure 2 of \cite{Tucker1}, which depicts an experimental image of a
whirligig beetle travelling through water in a straight line. This
experimental fluid regime is consistent with the parameter regime in
the present study, though, of course, the beetle is not submerged. 

%Finally, we note that capillary waves have been studied in the contexts of other fluids, such as liquid gallium \cite{Regan1}, and silicon \cite{Fork1}. While different fluids have different physical properties, the methodology described in the present study can be applied to any fluid flow past an obstacle.

Here we are concerned with three-dimensional capillary waves in the
small-surface-tension limit. This is a challenging limit to explore,
as the behaviour of capillary waves past a submerged obstacle in this
limit cannot be captured by an asymptotic power series: the capillary wave
amplitude is typically exponentially small in the
surface tension.  

We can motivate this claim by considering a steady train of
two-dimensional surface waves on deep water. From \cite{Whitham1}, we
find that after neglecting gravitational effects, the wavelength of
two-dimensional linearised capillary waves, denoted by $\lambda$, on a
steady flow with velocity $U$, representative length-scale $L$,
surface tension $\sigma$ and density $\rho$ is given by 
\begin{equation}
\frac{\lambda}{L} = \frac{2\pi\sigma}{\rho LU^2 } = 2\pi \epsilon,
\end{equation}
where $\epsilon = \sigma/\rho LU^2$ is the inverse Weber
number.
Since the velocity potential $\phi$ satisfies Laplace's equation,
solutions which oscillate with the required wavelength are given (up
to a phase shift) by 
\begin{equation}
\phi = \left(C_1 \mathrm{e}^{y/\epsilon} + C_2
  \mathrm{e}^{-y/\epsilon}\right)\sin(x/\epsilon), 
\end{equation}
where all distances have been scaled by $L$, and 
$C_1$ and $C_2$ are arbitrary constants.
Thus we might expect the impact of a submerged object on waves on the
free surface to decay exponentially with the distance of the object
from the free surface relative to $\epsilon$.
Consequently, asymptotic studies of capillary wave behaviour in the
  limit $\epsilon \rightarrow 0$ require the use of exponential asymptotic
techniques. 

These techniques,  described in \cite{Berry5}, \cite{Berry4},
\cite{Daalhuis1}, \cite{Chapman1}, \cite{Bennett1}, and elsewhere,
have been developed for the purpose of calculating such exponentially
small behaviour. More general introductions to exponential asymptotic
methods are found in \cite{Paris1}, \cite{Boyd3} and \cite{Jones1},
while comprehensive discussions are found in \cite{Segur1} and
\cite{Boyd2}. In section \ref{S1.methodology}, we will describe the
exponential asymptotic methodology used in the present study, which is
based on the work of \cite{Daalhuis1}, \cite{Chapman1},
\cite{Chapman4}, and the theory in \cite{Howls1}. 
As already noted, \cite{Chapman6} used
exponential asymptotic techniques to determine the behaviour of
nonlinear capillary waves in two dimensions. 

While the neglect of gravity in the present study limits the range of
its applicability, it provides a step towards  an
 analysis of a system containing both gravity and capillary
waves. In two dimensions an exponential asymptotic analysis has been
performed on 
gravity-capillary systems  with small surface tension
and small Froude number by \cite{Trinh3, Trinh4}. These studies found
an intricate interplay between gravity and capillary waves with a
structure considerably more complicated than that of each effect
considered in isolation.
An equivalent three-dimensional analysis would be challenging, and the
present analysis at least provides an answer in the limit that
capillarity dominates gravity.

Exponential asymptotic methods have been applied in order to study a
range of other wave problems arising in fluid dynamics. \cite{Chapman3}
applied exponential asymptotic methods  to resolve the low speed
 paradox, finding the behaviour of exponentially small
two-dimensional gravity waves over a step in the small-Froude-number
limit. This was extended in \cite{Lustri1} to describe the waves
caused by flow past submerged slopes, ridges, or trenches, by
\cite{Lustri3} to describe gravity waves caused by flow past a
submerged line source, and by \cite{Trinh1,Trinh2} in order to
describe gravity waves caused by ship hulls in two dimensions. These
ideas were extended to three dimensions by \cite{Lustri2} for the case
of steady linearised flow past a source, and \cite{Lustri4} for the
corresponding unsteady problem. The present analysis follows a similar
methodology to these studies of three-dimensional flow.  

Shallow gravity-capillary waves with small surface tension, modelled
by the singularly-perturbed fifth-order Korteweg de Vries (KdV)
equation, have also been the subject of a number of studies using
exponential asymptotics, including early work in \cite{Pomeau1}, and
subsequent studies by \cite{Boyd3, Boyd4, Grimshaw1, Grimshaw2,
  Trinh5,Yang1}. These studies showed that any solitary wave solution
to the singularly-perturbed fifth-order KdV equation must have a
wavetrain with exponentially small, non-decaying amplitude, proven
rigorously in \cite{Beale1, Iooss2, Sun1}.

Other applications of exponential asymptotics in fluid dynamics include 
\cite{Keller2} and
\cite{Ward1}, who studied slow flow past a cylinder with small Reynolds
number, using exponential asymptotics to determine the flow asymmetry
and drag, and  \cite{Chapman11}, who  obtained self-similar
solutions for thin film rupture.  

Aside from \cite{Lustri2, Lustri4}, each of the previous exponential
asymptotic studies on fluid flow behaviour have been performed in two
dimensions. We will therefore follow the methodology of \cite{Lustri2,
  Lustri4}, in which we linearise the problem around the source
strength in order to fix the position of the boundary in the
linearised regime. We will then apply exponential asymptotic
techniques directly to the flow equations in order to determine the
fluid potential, and the free surface position. 

\subsection{Fluid Regime}

The combination of inviscid flow and capillary-dominated surface waves
is an unusual one. This implies that we are considering systems in
which both viscous effects and gravitational effects are negligible compared to 
surface tension. For the inviscid fluid flow model to be valid, we require
that the inverse of the Reynolds number, given by $1/Re = \mu/\rho U
L$ where $\mu$ is the dynamic viscosity of the fluid, to be  small. In
this case, the Bernoulli 
condition for gravity-capillary waves, expressed in terms of the
surface tension parameter $\epsilon$ and the Froude number $F$, is
given by 
\begin{equation}
\frac{1}{2}(|\nabla\phi|^2 - 1) + \frac{\xi}{F^2} + \epsilon \kappa = 0\qquad \mathrm{on} \qquad z = \xi,
\end{equation}
where $\phi$ is the complex potential, $\xi$ is the free-surface
position, $\kappa$ is the surface curvature, and  the Froude
number given by $F = U/\sqrt{gL}$. We see that for the surface
behaviour to be dominated by surface tension effects, we require
$1/F^2 \ll \epsilon$. Consequently, the regime under consideration is
given by 
\begin{equation}\label{1.Cond}
 Re^{-1},\  F^{-2} \ll \epsilon \ll 1.
\end{equation}
In contrast to the associated analysis on gravity waves performed by
\cite{Lustri2,Lustri4}, this set of scalings is not relevant to the
study of submerged obstacles such as submarines. These scalings are
instead associated with ripples caused by small submerged objects,
such as fish or insects, which move rapidly, or by thin sheets of fast
moving fluid.

For example, using values of the density, viscosity, and surface
tension of water/air at $20^{\circ}$C from \cite{Batchelor1}
we require that
\[ U \ll 72 \,\mathrm{m}\, \mathrm{s}^{-1}, \qquad
  L \ll 2.7\, \mathrm{mm}, \qquad
  L U^2 \gg 0.000073 \,\mathrm{m}^3 \mathrm{s}^{-2}
  .\]
Using values $L= 1\,$mm and $U = 0.5\,$m\,s$^{-1}$, gives $Re^{-1}
\approx 0.002008$, $F^{-2} 
\approx 0.03922$, and $\epsilon \approx 0.2917$.
Studies of capillary waves in liquids other than water (such
as liquid silicon or liquid gallium, whose capillary waves were
investigated in \cite{Fork1} and \cite{Regan1} respectively), will
necessarily produce different parameter regimes in which the current
analysis is valid. 

%
%\begin{table}
%\centering
%\includegraphics{Table_Values.png};
%\caption{Physically realistic values which can be approximated by the regime \eqref{1.Cond}.}\label{1.table1}
%\end{table}

%  \begin{table}
% \centering
% \subfloat[Physical quantities]{
%  \begin{tabular}{ | c || l | }
%    \hline
%    $\rho$ & 998.2 m/s$^2$ \\ \hline
%    $\mu$  & 1.002$\times$10$^{-3}$  kg/m s \\ \hline
%    $\sigma$  & 72.8$\times$10$^{-3}$ N/m\\ \hline
%    $g$  & 9.806 m/s$^2$ \\ \hline
%    $L$ & 1$\times$10$^{-3}$ m \\ \hline
%    $U$ & 1\,m \\ \hline 
%    \end{tabular}
%    }
%    \hspace{20mm}
%    \subfloat[Dimensionless constants]{
%    \begin{tabular}{| c || l |}
%    \hline
%    $Re^{-1}$ & 1.003$\times$10$^{-3}$ \\ \hline
%    $F^{-2}$ & 9.806$\times$10$^{-3}$ \\ \hline
%    $\epsilon$ & 0.07293 \\ 
%    \hline
%  \end{tabular}
%  }
% \end{table}

\subsection{Methodology}\label{S1.methodology}

In order to study the behaviour of capillary waves due to flow past
submerged obstacles, we will adapt the methodology of \cite{Lustri2}
for the steady flow case, and \cite{Lustri4} for the unsteady flow
case. These studies considered flow past submerged obstacles in the
small-Froude-number limit. The surface waves were found to be
exponentially small in this limit, and therefore could not
be studied using classical asymptotic power series
techniques. Instead, exponential asymptotic techniques were applied to
determine the solution, and it was found that the gravity waves were
switched on as certain curves on the free surface, known as Stokes
curves, were crossed. We will see that similar behaviour is present in
the solution to 
the capillary wave problem in the small-surface-tension limit.

\cite{Stokes1} first observed that a function containing multiple exponential terms in the complex plane can contain curves along which the behaviour of the subdominant exponential changes rapidly. These curves are known as \textit{Stokes lines}. This investigation will apply the exponential asymptotic technique developed by \cite{Daalhuis1} and extended by \cite{Chapman1} for investigating the smooth, rapid switching of exponentially small asymptotic contributions across Stokes lines. 

The first step in this technique is to express the solution as an asymptotic power series, such as
\begin{equation*}
f(x; \epsilon) \sim \sum_{n=0}^{\infty} \epsilon^n f_n(x)  \qquad \mathrm{as} \quad \epsilon \rightarrow 0.
\end{equation*}
As the capillary wave problem is singularity perturbed in the small-surface-tension limit, the series will be divergent. However, the error of the divergent series approximation can be minimised by truncating the series after some finite number of terms, known as the optimal truncation point.  To find the optimal truncation point, we follow the commonly-used heuristic described by \cite{Boyd1}, in which the series is truncated at its smallest term. \cite{Chapman1} observed that the optimal truncation point tends to become large in the asymptotic limit, and hence knowledge of the behaviour of the \textit{late-order terms} of the series (that is, the form of $a_n$ in the limit that $n \rightarrow \infty$) is sufficient to truncate the asymptotic series optimally.

In singular perturbation problems, \cite{Dingle1} noted that successive terms in the asymptotic series expansion are typically obtained by repeated differentiation of an earlier term in the series. Consequently, singularities present in the early terms of the series expression will persist into later terms. Furthermore, as these singularities are repeatedly differentiated, the series terms will diverge as the ratio between a factorial and the increasing power of a function $\chi$ which is zero at the singularity. \cite{Chapman1} therefore propose that the asymptotic behaviour of the series terms may be expressed as a sum of factorial-over-power ansatz expressions, each associated with a different early-order singularity, such as
\begin{equation}\label{ch1:ansatz}
f_n \sim \frac{F\, \Gamma(n+\gamma)}{\chi^{n+\gamma}} \qquad \mathrm{as} \quad n \rightarrow \infty,
\end{equation}
where $\Gamma$ is the gamma function defined in \cite{Abramowitz1},  $F$, $\gamma$ and $\chi$ are functions that do not depend on $n$, and $\chi = 0$ at the singularity in the early series terms.  They conclude that the correct late-term behaviour may be represented as the sum of these ansatz expressions, each associated with a different singularity of the leading order solution. The global behaviour of the functions $F$, $\gamma$ and $\chi$ may be found by substituting this ansatz directly into the equations governing the terms of the asymptotic series, and then matching to the local behaviour in the neighbourhood of the singularity under consideration.

The late-order term behaviour given in (\ref{ch1:ansatz}) is related
to applying a WKB ansatz of the form $F\mathrm{e}^{-\chi/\epsilon}$ to the
equation for $f$ linearised about the truncated expansion. The
behaviour of $\chi$, or the \textit{singulant}, therefore plays an
important role in understanding the Stokes line behaviour. In fact,
\cite{Dingle1} notes that Stokes switching takes place on curves where
the switching exponential is maximally subdominant to the
leading-order behaviour; this occurs where the singulant is purely
real and positive. Hence, the singulant provides a useful condition to
determine the possible location of Stokes lines: 
\begin{equation}\label{1:StokesCond}
\mathrm{Re}(\chi) > 0,\qquad \mathrm{Im}(\chi) = 0.
\end{equation}
We also note another interesting class of curves, known as anti-Stokes
lines. These are curves across which an exponentially
small solution contribution switches to instead be exponentially large
in the asymptotic limit. From the WKB ansatz of the exponential
contribution, it is apparent that anti-Stokes lines correspond to
curves satisfying  
\begin{equation}\label{1:AntiStokesCond}
\mathrm{Re}(\chi) = 0.
\end{equation}

Once the form of the late-order terms is established, we may find the smallest term in the series, and hence truncate the series optimally. This gives
\begin{equation*}
f(x; \epsilon) = \sum_{n=0}^{N-1} \epsilon^n a_n(x) + R_N,
\end{equation*}
where $N(x;\epsilon)$ is the optimal truncation point, and $R_N$ is the now exponentially small remainder term.

The method of \cite{Daalhuis1} now involves substituting the truncated
series expression back into the original problem, obtaining an
equation for the remainder term. The switching behaviour of this
remainder is found by solving the remainder equation in the
neighbourhood of Stokes lines; in fact, the condition
(\ref{1:StokesCond}) for the position of the Stokes lines can be found
directly through this process. This methodology is sufficient to
determine the capillary wave behaviour in steady three-dimensional
flow over a submerged source.  

However, this methodology alone is not enough to explain the behaviour
seen in the unsteady capillary wave problem. The exponential
asymptotic methodology of \cite{Daalhuis1} and \cite{Chapman1} was
developed for investigating ordinary differential equations. Because
the flow surface is two-dimensional, we require the extension of these
techniques to partial differential equations which was developed by
\cite{Chapman4}. 

Initially, the method is identical, however in some partial
differential equations (and indeed, in higher-order differential
equations), further variants of Stokes switching may occur. If the
remainder itself is expanded as
\begin{equation*}
R_N \sim \mathrm{e}^{-\chi/\epsilon} \sum_{n=0}^{\infty} \epsilon^n  R_N^{(n)} \qquad \mathrm{as} \quad \epsilon \rightarrow 0,
\end{equation*}
then again applying the method of \cite{Daalhuis1} and \cite{Chapman1}, we truncate optimally, giving 
\begin{equation*}
R_N = \mathrm{e}^{-\chi/\epsilon} \sum_{n=0}^{M-1} \epsilon^n R_N^{(n)} + S_M,
\end{equation*}
where $S_M$ is the new (doubly) exponentially subdominant remainder term. It is obviously possible to formulate problems in which the remainder $S_M$ may be expanded as another exponentially subdominant divergent asymptotic series, and so on. Hence, we find that a hierarchy of increasingly exponentially subdominant late-order contributions may be present in the asymptotic expression. 

The unsteady capillary wave problem contains two distinct pairs of exponentials in the asymptotic solution, related to steady and transient rippling behaviour, and the switching interaction between these pairs plays an important role in describing the solution. In this case, switching also occurs when one exponential component is maximally subdominant to another exponential component. To find curves along which an exponential component with singulant $\chi_1$ can switch a subdominant exponential component with singulant $\chi_2$, the switching condition instead becomes
\begin{equation*}
\mathrm{Re}(\chi_2) > \mathrm{Re}(\chi_1),\qquad \mathrm{Im}(\chi_2) = \mathrm{Im}(\chi_1).
\end{equation*}
Note that setting $\chi_1 = 0$, corresponding to the leading-order algebraic
contribution to the solution, reproduces the condition
\eqref{1:StokesCond}. 

Finally, in order to fully describe the solution behaviour for the
unsteady capillary wave problem, we must also consider a further
variant of Stokes switching behaviour, known as higher-order Stokes
phenomenon. Higher-order Stokes switching was first observed by
\cite{Aoki1}, and explained in detail by \cite{Howls1} and
\cite{Chapman4}. These studies found that higher-order Stokes
switching behaviour typically plays a role when there are three or
more singulants contributing to the solution (including the
leading-order singulant $\chi=0$). 

When an ordinary Stokes line is crossed, an exponentially small
contribution is switched on, the size of which is governed by a Stokes
switching parameter. When a higher-order Stokes line is crossed, this
switching parameter itself is switched on or off. The effect of this
higher-order switching is that ordinary Stokes lines themselves are
switched on or off as higher-order Stokes lines are
crossed. Higher-order Stokes lines originate at the intersection of
multiple Stokes lines in the complex plane, known as Stokes crossing
points. The practical effect of this switching is that higher-order
Stokes lines can terminate at Stokes crossing points, rather than
continuing indefinitely. 

The study \cite{Howls1} showed that when a problem contains three or more singulant contributions, associated with $\chi_1$, $\chi_2$ and $\chi_3$, higher-order Stokes lines can follow curves satisfying the criterion 
\begin{equation*}
\mathrm{Im}\left[\frac{\chi_3 - \chi_2}{\chi_3 - \chi_1}\right] = 0.
\end{equation*}
Unsteady free-surface flow, such as the unsteady gravity wave problem considered in \cite{Lustri4}, does contain three interacting  contributions (exponentially small steady and transient ripples, and algebraic effects); hence, higher-order Stokes lines must play a role in the solution. It is therefore not sufficient to find the ordinary Stokes lines in this problem, as the Stokes structure would be incorrect. Instead we must also determine the higher-order Stokes line behaviour, and therefore the location at which the ordinary Stokes lines are switched on and off. This will permit us to determine the full asymptotic free-surface wave behaviour.

\section{Steady Flow}\label{STEADY}
\subsection{Formulation}\label{S_Formulation}

We consider the steady-state problem of uniform flow past a submerged point source in three dimensions. We suppose that the strength of the source is small so that the problem may be linearised.

\subsubsection{Full problem}
\begin{figure}
\centering
\begin{tikzpicture}
[xscale=0.75,>=stealth,yscale=0.65]
\draw[->] (-4,0) -- (4,0) node[right] {\scriptsize{$x$}};
\draw[->] (0,-3) -- (0,3) node[above] {\scriptsize{$z$}};
\draw[->] (-2.5,-2.5) -- (2.5,2.5) node[above right] {\scriptsize{$y$}};;
\draw[->] (-3+0.5,2-1) node[left] {\scriptsize{$z = \xi(x,y)$}} .. controls (-2.5+0.5,2-1) and (-2+0.5,2-1) .. (-2+0.5,1.5-1);
\draw[thick,-] (0.1,-2.38) -- (-0.1,-2.62);
\draw[thick,-] (0.1,-2.62) -- (-0.1,-2.38);
\draw[->] (-7,-2.75) -- (-6.25,-2.75) node[above] {\scriptsize{Flow Direction}} -- (-5.5,-2.75);
\draw[->,white] (7,-2.75)  -- (5.5,-2.75);
\node at (-0.1,-2.5) [left] {\scriptsize{$(0,0,-h)$}};
\node at (0.1,-2.5) [below right] {\scriptsize{Source}};
\fill[opacity=0.2,rounded corners=20] (3,0) -- (5,2) -- (-1,2) -- (-3,0) -- (-5,-2) -- (1,-2) -- cycle;

\begin{scope}
\clip[rounded corners=20] (3,0) -- (5,2) -- (-1,2) -- (-3,0) -- (-5,-2) -- (1,-2) -- cycle;
%\fill[opacity=0.2] (4,-1.75) -- (0,0) -- (4,1) -- cycle;
\end{scope}

%\draw[->] (3+0.5,-2+1) node[right] {\scriptsize{Downstream}} .. controls (2.5+0.5,-2+1) and (2+0.5,-2+1) .. (2+0.5,-1.5+1);
%\node at (3+0.78,-2+0.5) [right] {\scriptsize{wavetrain}};

\end{tikzpicture}\caption{Prescribed fluid configuration for three-dimensional flow over a source. The shaded region represents the position of the free surface $\xi(x,y)$, and the cross represents the position of the source. The flow region lies below the free surface, and the mean flow is moving from left to right, with flow velocity $U$ in the unscaled problem. The waves form upstream from the obstacle, which is consistent with the theory of \cite{Whitham1} for two-dimensional capillary waves.
}
\label{1:3dsteadyconfig}
\end{figure}
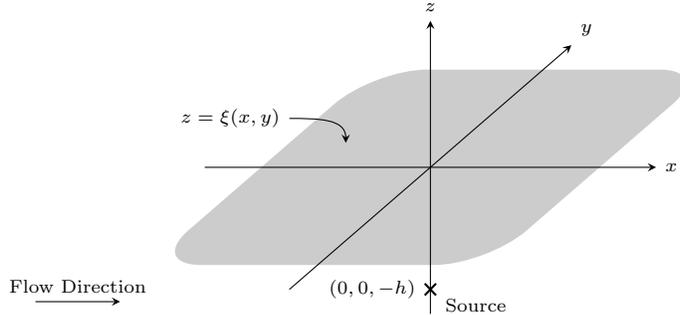

We consider a three-dimensional incompressible, irrotational, inviscid
free-surface flow of infinite depth with a submerged point source at
depth $H$ and upstream flow velocity $U$. We normalise the fluid
velocity with $U$ and distance with an typical length $L$, giving
nondimensionalised source depth $h = H/L$, shown schematically in
figure \ref{1:3dsteadyconfig}. %The natural choice for $L$ is $H$, but
                               %we prefer to separate $H$ and $L$ so
                               %that we may later more easily take the
                               %limit $h \rightarrow 0$. [NB. I will
                               %put in some plots of this later] 

Denoting the (nondimensional) position of the free surface by $z = \xi(x,y)$, the (nondimensional) velocity potential satisfies

\begin{equation}
\label{1:nlgeq} \nabla^2\phi = 0,\qquad -\infty < z < \xi(x,y),
\end{equation}
with kinematic and dynamic boundary conditions  
\begin{alignat}{2}
\label{1:nlbc1}\xi_x\phi_x + \xi_y\phi_y &= \phi_z,\qquad & z &= \xi(x,y),\\
\label{1:nlbc2}\frac{1}{2}\left(|\nabla\phi|^2-1\right) +\epsilon\kappa  &= 0,\qquad & z &= \xi(x,y),
\end{alignat}
where $\kappa$ represents the curvature of the free surface, positive
if the centre of curvature lies in the fluid region, and the
inverse Weber number $\epsilon = \sigma
/\rho L U^2 $, where $\sigma$ represents the surface tension parameter
and $\rho$ represents the fluid density. The curvature is given by 
\begin{equation}
\kappa = -\nabla_s \cdot\left[\frac{\nabla_s \xi}{\sqrt{1+|\nabla_s\xi|^2}}\right],
\end{equation}
where $\nabla_s$ represents the surface gradient of the flow. We are concerned with the free-surface behaviour in the limit $0 < \epsilon \ll 1$, in which the surface tension effects become small. Since the flow is uniform in the far field, $\phi_x \rightarrow 1$ there, while at the source
\begin{equation}
\label{1:nlsc1} \phi \sim \frac{\delta}{4\pi\sqrt{x^2 + y^2 + (z+h)^2}} \qquad \mathrm{as} \quad (x,y,z) \rightarrow (0,0,-h).
\end{equation}
We will be concerned with the limit $0 < \delta \ll \epsilon$, so that the disturbance to the free stream is weak and the equations may be linearised in $\delta$.

Finally, we incorporate a radiation condition which states that the
steady surface capillary waves must propagate upstream.

\subsubsection{Linearisation}

\begin{figure}
\centering

\includegraphics[width=0.99\textwidth]{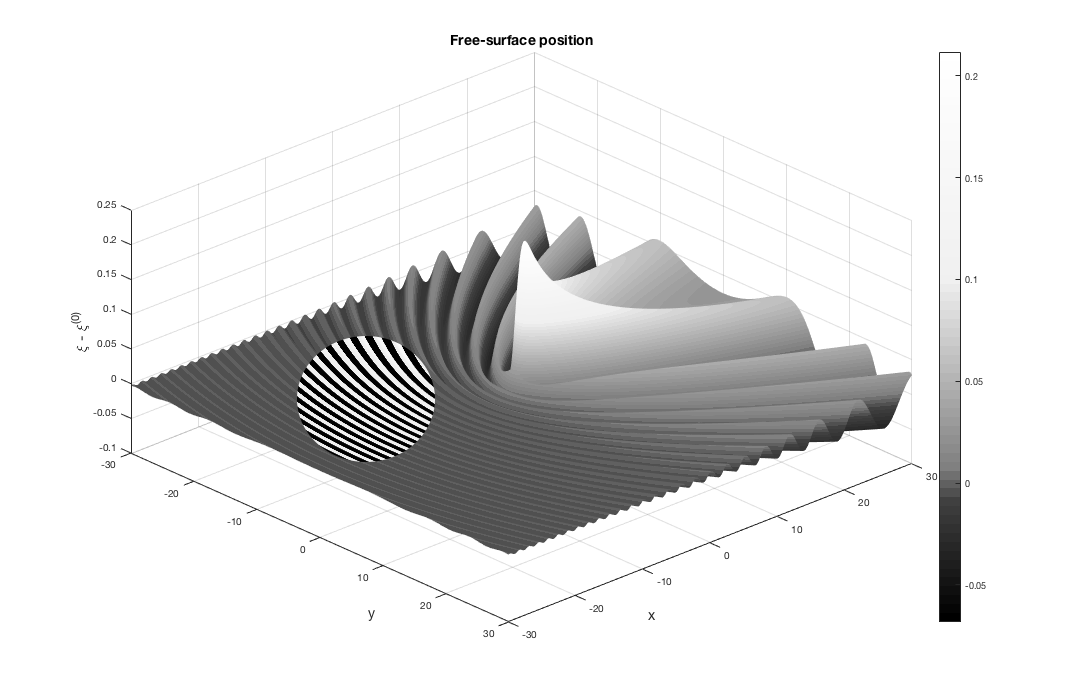};

\caption{Surface plot of the modified free-surface position $\xi - \xi^{(0)}$, where $\xi^{(0)}$ is the leading-order free-surface profile given in \eqref{1:xi0}, with $h = 1$ and $\epsilon = 0.2$, with flow in the positive $x$-direction. The figure depicts capillary waves extending in the direction of the flow origin, in addition to lower order effects above and ahead of the disturbance; these visible lower order effects are not exponentially small ripples, which are located behind the disturbance. The contrast is increased in a circular region in order to make the position of the capillary waves visually distinguishable.}
\label{F:3DNum}
\end{figure}

We linearise about uniform flow by setting
\begin{equation*}
 \phi = x + \delta\tilde{\phi},\qquad \xi = \delta\tilde{\xi},
\end{equation*}
to give, at leading order in $\delta$
\begin{alignat}{2}
\label{1:lgeq} \nabla^2\tilde{\phi} &= 0 ,\qquad &-\infty < z& < 0,\\
\label{1:lbc1}\tilde{\phi}_z - \tilde{\xi}_x &= 0,\qquad & z&= 0,\\
\label{1:lbc2}\tilde{\phi}_x -\epsilon \left(\tilde{\xi}_{xx} + \tilde{\xi}_{yy}\right)&=0  ,\qquad & z &= 0,
\end{alignat}
where the boundary conditions are now applied on the fixed surface $z = 0$. The far-field conditions imply that $\tilde{\phi} \rightarrow 0$ as $x^2 + y^2 + z^2 \rightarrow \infty$, while near the source
\begin{equation}
\label{1:lsc1}\tilde{\phi} \sim \frac{1}{4\pi\sqrt{x^2 + y^2 + (z+h)^2}} \qquad \mathrm{as} \quad (x,y,z) \rightarrow (0,0,-h).
\end{equation}
With the addition of a radiation condition, specifying that capillary
waves must propagate upstream from the source, the system described in
(\ref{1:lgeq})--(\ref{1:lsc1}) completely specifies the linearied
version of the three-dimensional problem shown in figure
\ref{1:3dsteadyconfig}. We can solve the linearised problem
numerically using a modified version of the algorithm from
\cite{Lustri2, Lustri4} to obtain free-surface profiles such as that
illustrated in figure \ref{F:3DNum}.  

We analytically continue the free surface such that $x, y \in \mathbb{C}$, with the free surface still satisfying $z = 0$. This does not change the form of (\ref{1:lgeq})--(\ref{1:lsc1}), but it does mean that the two-dimensional physical free surface is now a subset of a four-dimensional complexified free surface.

\subsection{Series expression}\label{SERIES}

Following the approach of \cite{Lustri2}, we first expand the fluid potential and free-surface position as a power series in $\epsilon$,
\begin{equation}
 \label{1:series}\tilde{\phi} \sim \sum_{n=0}^{\infty}\epsilon^n\phi^{(n)},\qquad \tilde{\xi} \sim \sum_{n=0}^{\infty}\epsilon^n\xi^{(n)},
\end{equation}
to give for $n \geq 0$
\begin{alignat}{2}
\label{1:sergeq} \nabla^2 \phi^{(n)} &= 0,\qquad &-\infty < z& < 0,\\
\label{1:serbc1}\phi^{(n)}_z - \xi^{(n)}_x &= 0,\qquad & z&= 0,\\
\label{1:serbc2}{\phi}_x^{(n)} -{\xi}_{xx}^{(n-1)} - {\xi}_{yy}^{(n-1)} &=0 ,\qquad & z &= 0,
\end{alignat}
with the convention that $\phi^{(-1)} = 0$. The far-field behaviour tends to zero at all orders of $n$, and the singularity condition (\ref{1:lsc1}) is applied to the leading-order expression, giving
\begin{equation}
\label{1:sersc1}\phi^{(0)} \sim \frac{1}{4\pi\sqrt{x^2 + y^2 + (z+h)^2}} \qquad \mathrm{as} \quad (x,y,z) \rightarrow (0,0,-h).
\end{equation}

The leading-order solution is given by
\begin{align}
 \label{1:phi0} \phi^{(0)} &= \frac{1}{4\pi\sqrt{x^2 + y^2 + (z+h)^2}} - \frac{1}{4\pi\sqrt{x^2 + y^2 + (z-h)^2}},\\
 \label{1:xi0} \xi^{(0)} &= -\frac{x h}{2 \pi (y^2 + h^2)\sqrt{x^2+y^2+h^2}} - \frac{1}{2\pi(y^2+h^2)},
\end{align}
where the leading-order free surface behaviour is set to be
undisturbed far ahead of the source. 

Through repeated iteration of (\ref{1:serbc1})--(\ref{1:serbc2}), we find that the position of singularities in subsequent terms of the series (\ref{1:series}) remains constant, while the power of the singularity increases at each order, as we expect for such singular perturbation problems (see \cite{Chapman3}).

\subsection{Late-Order Terms}\label{LOT_steady}

In order to optimally truncate the asymptotic series prescribed in (\ref{1:series}), we must determine the form of the late-order terms. To accomplish this, we make a factorial-over-power ansatz (see \cite{Chapman1}), having the form
\begin{equation}
 \label{1:ansatz1} \phi^{(n)} \sim \frac{\Phi(x,y,z)\Gamma(n+\gamma)}{\chi(x,y,z)^{n+\gamma}},\qquad \xi^{(n)} \sim \frac{\Xi(x,y)\Gamma(n+\gamma)}{\chi(x,y,0)^{n+\gamma}}, \qquad \mathrm{as} \quad n \rightarrow \infty,
\end{equation}
where $\gamma$ is a constant. In order that (\ref{1:ansatz1}) is the power series developed in section \ref{SERIES}, we require that the singulant, $\chi$, satisfies
\begin{equation}
 \label{1:singularitycondition}  \chi = 0 \qquad \mathrm{on} \qquad x^2 + y^2 + (z\pm h)^2 = 0,
\end{equation}
where the sign chosen depends upon which of the two singularities is being considered. For complex values of  $x$, $y$ and $z$, this defines a four-dimensional hypersurface. Irrespective of which singularity is under consideration, this hypersurface intersects the four-dimensional complexified free surface on the two-dimensional hypersurface satisfying $x^2 + y^2 + h^2 = 0$.  

It is important to note that the expression for $\xi^{(n)}$ is restricted to $z = 0$, as it describes the free-surface position. This does not pose a problem for the subsequent analysis, but does ensure that care must be taken at each stage to determine whether we are considering the full flow region, or just the free surface.

\subsubsection{Calculating the singulant}\label{CH5_SINGULANT1}

Applying the ansatz expressions in (\ref{1:ansatz1}) to the governing equation (\ref{1:sergeq}) and taking the first two orders as $n \rightarrow \infty$ gives
\begin{align}
\label{1:ansgov1} \chi_x^2 + \chi_y^2 + \chi_z^2 &= 0,\\
\label{1:ansgov2} 2\Phi_x\chi_x + 2\Phi_y\chi_y + 2\Phi_z\chi_z &= -(\chi_{xx}+\chi_{yy}+\chi_{zz}),
\end{align}
while the boundary conditions on $z = 0$ become, to leading order,
\begin{align}
 \label{1:ansbc1} -\chi_z\Phi + \chi_x \Xi &= 0,\\
\label{1:ansbc2} \chi_x\Phi + (\chi_x^2 + \chi_y^2)\Xi  &= 0.
\end{align}
The system in (\ref{1:ansbc1})--(\ref{1:ansbc2}) has nonzero solutions when 
\begin{equation}
\label{1:eik0} \chi^2_x = -\chi_z\left(\chi_x^2 + \chi_y^2\right).
\end{equation}
which gives the result
\begin{equation}
\label{1:chiz} \chi_z = -\frac{\chi_x^2}{\chi_x^2+\chi_y^2}.
\end{equation}
Using (\ref{1:chiz}), we find a relationship between $\Phi$ and $\Xi$ by rearranging (\ref{1:ansbc1}) to obtain
\begin{equation}\label{1:phixi}
\Xi = -\frac{\chi_x}{\chi_x^2+\chi_y^2} \Phi.
\end{equation}

Applying (\ref{1:chiz}) to (\ref{1:ansgov1}) evaluated on $z=0$ gives a singulant equation for $\chi$ on the free surface,
\begin{equation}
\label{1:eik}  \chi_x^4 + \left(\chi_x^2 + \chi_y^2\right)^3 = 0.
\end{equation}

Here though, because the singularity lies below the fluid surface, we must solve (\ref{1:eik}) for complex $x$ and $y$ with the boundary condition
\begin{equation}
\label{1:eik2}  \chi = 0 \quad \mathrm{on} \quad x^2 + y^2 + h^2 = 0.
\end{equation}
Parametrising (\ref{1:eik2}) as
\begin{equation}
 \label{1:sscauchy}x_0 = s, \qquad y_0 = \pm\mathrm{i}\sqrt{s^2+h^2},\qquad \chi_0 = 0.
\end{equation}
and solving (\ref{1:eik}) using Charpit's method (see \cite{Ockendon1}) gives
\begin{equation}
 \label{1:sschiray2} \chi = \pm\frac{9x(x^2+y^2)s^3 + h(2h^2+9x^2-6y^2)s^2 + 6x(y^2-h^2)s - 4h(h^2+y^2)}{3(2h^2+3x^2)},
\end{equation}
where $s$ is a solution to
\begin{equation}
 \label{1:sspoly} 9(x^2+y^2)s^4 + 12xhs^3 + (4h^2+9x^2+12y^2)s^2 + 12xhs + 4(h^2+y^2) = 0.
\end{equation}

\begin{figure}
\centering

\subfloat[Contour plots for $\chi_{S1}$]{
\centering
\begin{tikzpicture}
[xscale=0.35,>=stealth,yscale=0.35]

\fill[opacity = 0.07]  (5,5) -- (5,-5) -- (-5,-5) -- (-5,5) -- cycle;
\fill[white] plot[smooth] file {capsing_rem0p9.txt} -- (-5,5) -- (-5,-5) -- cycle;
\fill[opacity = 0.03]  (5,5) -- (5,-5) -- (-5,-5) -- (-5,5) -- cycle;

\fill[opacity = 0.07] plot[smooth] file {capsing_rem0p8.txt} -- (5,5) -- (5,-5) -- cycle;
\fill[opacity = 0.07] plot[smooth] file {capsing_rem0p7.txt} -- (5,5) -- (5,-5) -- cycle;
\fill[opacity = 0.07] plot[smooth] file {capsing_rem0p6.txt} -- (5,-5) -- (5,5) -- cycle;
\fill[opacity = 0.07] plot[smooth] file {capsing_rem0p5.txt} -- (5,5) -- (5,-5) -- cycle;
\fill[opacity = 0.07] plot[smooth] file {capsing_rem0p4.txt} -- (5,-5) -- (5,5) -- cycle;
\fill[opacity = 0.07] plot[smooth] file {capsing_rem0p3.txt} -- (5,-5) -- (5,5) -- cycle;
\fill[opacity = 0.07] plot[smooth] file {capsing_rem0p2.txt} -- (5,5) -- (5,-5) -- cycle;
\fill[opacity = 0.07] plot[smooth] file {capsing_rem0p1.txt} -- (5,5) -- (5,-5) -- cycle;
\fill[opacity = 0.15] plot[smooth] file {capsing_re0.txt} -- (5,-5) -- (5,5) -- cycle;
\fill[opacity = 0.2] plot[smooth] file {capsing_re0p1.txt}  -- cycle;
%\fill[opacity = 0.15] plot[smooth] file {capsing_rem0p2.txt} -- (5,5) --  (5,-5) -- cycle;

%\fill[white]  plot[smooth] file {capsing_re0p1.txt} -- cycle;
%
%\fill[opacity = 0.03]  (5,5) -- (5,-5) -- (-5,-5) -- (-5,5) -- cycle;
\draw[line width=0.5mm] plot[smooth] file {capsing_re0.txt};

\draw (-5,5) -- (-5,-5) -- (5,-5) -- (5,5) -- cycle;
\draw (-5,4) -- (-5.2,4) node[left] {\scriptsize{$4$}};
\draw (-5,2) -- (-5.2,2) node[left] {\scriptsize{$2$}};
\draw (-5,0) -- (-5.2,0) node[left] {\scriptsize{$0$}};
\draw (-5,-2) -- (-5.2,-2) node[left] {\scriptsize{$-2$}};
\draw (-5,-4) -- (-5.2,-4) node[left] {\scriptsize{$-4$}};
\draw (4,-5) -- (4,-5.2) node[below] {\scriptsize{$4$}};
\draw (2,-5) -- (2,-5.2) node[below] {\scriptsize{$2$}};
\draw (0,-5) -- (0,-5.2) node[below] {\scriptsize{$0$}};
\draw (-2,-5) -- (-2,-5.2) node[below] {\scriptsize{$-2$}};
\draw (-4,-5) -- (-4,-5.2) node[below] {\scriptsize{$-4$}};

\node at (0,7) [above] {\small{$\mathrm{Re}(\chi)$}};

\draw[->] (-5,0) -- (5.8,0) node [right] {\scriptsize{$x$}};
\draw[->] (0,-5) -- (0,5.8) node [above] {\scriptsize{$y$}};

\node at (-2.5,3) {\scriptsize{Anti-Stokes}};
\node at (-2.5,2.25) {\scriptsize{line}};
\draw[->] (-1.5,2.25) -- (0.5,2.25) .. controls (1.5,2.25) and (2,1.75) .. (2,1.25) -- (2,0.65);

%\draw (-5+15,5) -- (-5+15,-5) -- (5+15,-5) -- (5+15,5) -- cycle;
%\begin{scope}
%\clip (-5+15,5) -- (-5+15,-5) -- (5+15,-5) -- (5+15,5) -- cycle;
%\draw[line width=0.5mm] (0+15,5) -- (0+15,-5);
%\end{scope}

\fill[opacity = 0.03]  (5+15,5) -- (5+15,-5) -- (-5+15,-5) -- (-5+15,5) -- cycle;
\draw[line width=0.5mm] plot[smooth] file {capsing_im0.txt};
\fill[opacity = 0.15]  plot[smooth] file {capsing_im0.txt} -- cycle;
\fill[opacity = 0.07] plot[smooth] file {capsing_im0p5.txt} -- (5+15,-5) -- (5+15,5) -- cycle;
\fill[opacity = 0.07] plot[smooth] file {capsing_im1.txt} -- (5+15,-5) -- (5+15,5) -- cycle;
\fill[opacity = 0.07] plot[smooth] file {capsing_im1p5.txt} -- (5+15,5) -- (5+15,-5) -- cycle;
\fill[opacity = 0.07] plot[smooth] file {capsing_im2.txt} -- (5+15,-5) -- (5+15,5) -- cycle;
\fill[opacity = 0.07] plot[smooth] file {capsing_im2p5.txt} -- (5+15,-5) -- (5+15,5) -- cycle;
\fill[opacity = 0.07] plot[smooth] file {capsing_im3.txt} -- (5+15,-5) -- (5+15,5) -- cycle;
\fill[opacity = 0.07] plot[smooth] file {capsing_im3p5.txt} -- (5+15,-5) -- (5+15,5) -- cycle;
\fill[opacity = 0.07] plot[smooth] file {capsing_im4.txt} -- (5+15,-5) -- (5+15,5) -- cycle;
\fill[opacity = 0.07] plot[smooth] file {capsing_im4p5.txt} -- (5+15,5) -- (5+15,-5) -- cycle;
%\draw[-] plot[smooth] file {capsing_im5.txt};
\draw[line width = 0.5mm] (0+15,0) -- (5+15,0);

\fill[opacity = 0.20] plot[smooth] file {capsing_imm0p5.txt} -- cycle;
\fill[opacity = 0.20] plot[smooth] file {capsing_imm0p5b.txt} -- cycle;

\draw (-5+15,5) -- (-5+15,-5) -- (5+15,-5) -- (5+15,5) -- cycle;
\draw (-5+15,4) -- (-5.2+15,4) node[left] {\scriptsize{$4$}};
\draw (-5+15,2) -- (-5.2+15,2) node[left] {\scriptsize{$2$}};
\draw (-5+15,0) -- (-5.2+15,0) node[left] {\scriptsize{$0$}};
\draw (-5+15,-2) -- (-5.2+15,-2) node[left] {\scriptsize{$-2$}};
\draw (-5+15,-4) -- (-5.2+15,-4) node[left] {\scriptsize{$-4$}};
\draw (4+15,-5) -- (4+15,-5.2) node[below] {\scriptsize{$4$}};
\draw (2+15,-5) -- (2+15,-5.2) node[below] {\scriptsize{$2$}};
\draw (0+15,-5) -- (0+15,-5.2) node[below] {\scriptsize{$0$}};
\draw (-2+15,-5) -- (-2+15,-5.2) node[below] {\scriptsize{$-2$}};
\draw (-4+15,-5) -- (-4+15,-5.2) node[below] {\scriptsize{$-4$}};

\node at (0+15,7) [above] {\small{$\mathrm{Im}(\chi)$}};

\draw[->] (-5+15,0) -- (5.8+15,0) node [right] {\scriptsize{$x$}};
\draw[->] (0+15,-5) -- (0+15,5.8) node [above] {\scriptsize{$y$}};

%%%%%

\draw (7+17,5) -- (8+17,5) -- (8+17,-5) -- (7+17,-5) -- cycle;

\fill[opacity = 0.20]  (7+17,-5) -- (8+17,-5) -- (8+17,-5+1*0.8333) -- (7+17,-5+1*0.8333) -- cycle;
\fill[opacity = 0.15]  (7+17,-5) -- (8+17,-5) -- (8+17,-5+2*0.8333) -- (7+17,-5+2*0.8333) -- cycle;
\fill[opacity = 0.07]  (7+17,-5) -- (8+17,-5) -- (8+17,-5+3*0.8333) -- (7+17,-5+3*0.8333) -- cycle;
\fill[opacity = 0.07]  (7+17,-5) -- (8+17,-5) -- (8+17,-5+4*0.8333) -- (7+17,-5+4*0.8333) -- cycle;
\fill[opacity = 0.07]  (7+17,-5) -- (8+17,-5) -- (8+17,-5+5*0.8333) -- (7+17,-5+5*0.8333) -- cycle;
\fill[opacity = 0.07]  (7+17,-5) -- (8+17,-5) -- (8+17,-5+6*0.8333) -- (7+17,-5+6*0.8333) -- cycle;
\fill[opacity = 0.07]  (7+17,-5) -- (8+17,-5) -- (8+17,-5+7*0.8333) -- (7+17,-5+7*0.8333) -- cycle;
\fill[opacity = 0.07]  (7+17,-5) -- (8+17,-5) -- (8+17,-5+8*0.8333) -- (7+17,-5+8*0.8333) -- cycle;
\fill[opacity = 0.07]  (7+17,-5) -- (8+17,-5) -- (8+17,-5+9*0.8333) -- (7+17,-5+9*0.8333) -- cycle;
\fill[opacity = 0.07]  (7+17,-5) -- (8+17,-5) -- (8+17,-5+10*0.8333) -- (7+17,-5+10*0.8333) -- cycle;
\fill[opacity = 0.07]  (7+17,-5) -- (8+17,-5) -- (8+17,-5+11*0.8333) -- (7+17,-5+11*0.8333) -- cycle;

\fill[opacity = 0.03]  (7+17,5) -- (8+17,5) -- (8+17,0) -- (7+17,0) -- cycle;

\draw[line width=0.5mm] (6.8+17,-5+2*0.8333) -- (8+17,-5+2*0.8333);

\draw (7+17,5) -- (6.8+17,5) node[left] {\scriptsize{$5$}};
\draw (7+17,-5+2*0.8333) -- (6.8+17,-5+2*0.8333) node[left] {\scriptsize{$0$}};
\draw (7+17,-5) -- (6.8+17,-5) node[left] {\scriptsize{$-1$}};

\draw (-9,5) -- (-10,5) -- (-10,-5) -- (-9,-5) -- cycle;
\fill[opacity = 0.20]  (-9,-5) -- (-10,-5) -- (-10,-5+1*0.8333) -- (-9,-5+1*0.8333) -- cycle;
\fill[opacity = 0.15]  (-9,-5) -- (-10,-5) -- (-10,-5+2*0.8333) -- (-9,-5+2*0.8333) -- cycle;
\fill[opacity = 0.07]  (-9,-5) -- (-10,-5) -- (-10,-5+3*0.8333) -- (-9,-5+3*0.8333) -- cycle;
\fill[opacity = 0.07]  (-9,-5) -- (-10,-5) -- (-10,-5+4*0.8333) -- (-9,-5+4*0.8333) -- cycle;
\fill[opacity = 0.07]  (-9,-5) -- (-10,-5) -- (-10,-5+5*0.8333) -- (-9,-5+5*0.8333) -- cycle;
\fill[opacity = 0.07]  (-9,-5) -- (-10,-5) -- (-10,-5+6*0.8333) -- (-9,-5+6*0.8333) -- cycle;
\fill[opacity = 0.07]  (-9,-5) -- (-10,-5) -- (-10,-5+7*0.8333) -- (-9,-5+7*0.8333) -- cycle;
\fill[opacity = 0.07]  (-9,-5) -- (-10,-5) -- (-10,-5+8*0.8333) -- (-9,-5+8*0.8333) -- cycle;
\fill[opacity = 0.07]  (-9,-5) -- (-10,-5) -- (-10,-5+9*0.8333) -- (-9,-5+9*0.8333) -- cycle;
\fill[opacity = 0.15]  (-9,-5) -- (-10,-5) -- (-10,-5+10*0.8333) -- (-9,-5+10*0.8333) -- cycle;
\fill[opacity = 0.07]  (-9,-5) -- (-10,-5) -- (-10,-5+11*0.8333) -- (-9,-5+11*0.8333) -- cycle;

\fill[opacity = 0.03]  (-9,5) -- (-10,5) -- (-10,0) -- (-9,0) -- cycle;

\draw[line width=0.5mm] (-8.8,-5+2*0.8333) -- (-10,-5+2*0.8333);
\draw (-9,5) -- (-8.8,5) node[right] {\scriptsize{$1$}};
\draw (-9,-5+2*0.8333) -- (-8.8,-5+2*0.8333) node[right] {\scriptsize{$0$}};
\draw (-9,-5) -- (-8.8,-5) node[right] {\scriptsize{$-0.2$}};

\node at (-2.75+15,3) {\scriptsize{Stokes line}};
\draw[->] (-0.65+15,3) -- (0.5+15,3) .. controls (1.5+15,3) and (2+15,2.5) .. (2+15,2) -- (2+15,1.65);

\end{tikzpicture}
}

\subfloat[Contour plots for $\chi_{S2}$]{
\centering
\begin{tikzpicture}
[xscale=0.35,>=stealth,yscale=0.35]

\fill[opacity = 0.07]  (5,5) -- (5,-5) -- (-5,-5) -- (-5,5) -- cycle;
\fill[white] plot[smooth] file {capsing2_rem0p9.txt} -- (5,5) -- (5,-5) -- cycle;
\fill[opacity = 0.03]  (5,5) -- (5,-5) -- (-5,-5) -- (-5,5) -- cycle;

\fill[opacity = 0.07] plot[smooth] file {capsing2_rem0p8.txt} -- (-5,5) -- (-5,-5) -- cycle;
\fill[opacity = 0.07] plot[smooth] file {capsing2_rem0p7.txt} -- (-5,5) -- (-5,-5) -- cycle;
\fill[opacity = 0.07] plot[smooth] file {capsing2_rem0p6.txt} -- (-5,-5) -- (-5,5) -- cycle;
\fill[opacity = 0.07] plot[smooth] file {capsing2_rem0p5.txt} -- (-5,5) -- (-5,-5) -- cycle;
\fill[opacity = 0.07] plot[smooth] file {capsing2_rem0p4.txt} -- (-5,-5) -- (-5,5) -- cycle;
\fill[opacity = 0.07] plot[smooth] file {capsing2_rem0p3.txt} -- (-5,-5) -- (-5,5) -- cycle;
\fill[opacity = 0.07] plot[smooth] file {capsing2_rem0p2.txt} -- (-5,5) -- (-5,-5) -- cycle;
\fill[opacity = 0.07] plot[smooth] file {capsing2_rem0p1.txt} -- (-5,5) -- (-5,-5) -- cycle;
\fill[opacity = 0.15] plot[smooth] file {capsing2_re0.txt} -- (-5,-5) -- (-5,5) -- cycle;
\fill[opacity = 0.2] plot[smooth] file {capsing2_re0p1.txt}  -- cycle;
%\fill[opacity = 0.15] plot[smooth] file {capsing_rem0p2.txt} -- (5,5) --  (5,-5) -- cycle;

%\fill[white]  plot[smooth] file {capsing_re0p1.txt} -- cycle;
%
%\fill[opacity = 0.03]  (5,5) -- (5,-5) -- (-5,-5) -- (-5,5) -- cycle;
\draw[line width=0.5mm] plot[smooth] file {capsing2_re0.txt};

%\fill[opacity = 0.07]  (5,5) -- (5,-5) -- (-5,-5) -- (-5,5) -- cycle;
%\fill[white]  plot[smooth] file {capsing2_re0p1.txt} -- cycle;
%
%\fill[opacity = 0.03]  (5,5) -- (5,-5) -- (-5,-5) -- (-5,5) -- cycle;
%\draw[line width=0.5mm] plot[smooth] file {capsing2_re0.txt};
%
%\fill[opacity = 0.15] plot[smooth] file {capsing2_re0.txt} -- (-5,-5) -- (5,-5) -- (5,5) -- (-5,5) -- cycle;
%\fill[opacity = 0.07] plot[smooth] file {capsing2_rem0p1.txt} -- (-5,5) -- (5,5) -- (5,-5) -- (-5,-5) -- cycle;
%\fill[opacity = 0.07] plot[smooth] file {capsing2_rem0p2.txt} -- (-5,5) -- (5,5) -- (5,-5) -- (-5,-5) -- cycle;
%
%\fill[opacity = 0.07] plot[smooth] file {capsing2_rem0p3.txt} -- (5,-5) -- (5,5) -- cycle;
%\fill[opacity = 0.07] plot[smooth] file {capsing2_rem0p4.txt} -- (5,-5) -- (5,5) -- cycle;
%\fill[opacity = 0.07] plot[smooth] file {capsing2_rem0p5.txt} -- (5,5) -- (5,-5) -- cycle;
%\fill[opacity = 0.07] plot[smooth] file {capsing2_rem0p6.txt} -- (5,-5) -- (5,5) -- cycle;
%\fill[opacity = 0.07] plot[smooth] file {capsing2_rem0p7.txt} -- (5,5) -- (5,-5) -- cycle;
%\fill[opacity = 0.07] plot[smooth] file {capsing2_rem0p8.txt} -- (5,5) -- (5,-5) -- cycle;
%\fill[opacity = 0.07] plot[smooth] file {capsing2_rem0p9.txt} -- (5,5) -- (5,-5) -- cycle;

\draw (-5,5) -- (-5,-5) -- (5,-5) -- (5,5) -- cycle;
\draw (-5,4) -- (-5.2,4) node[left] {\scriptsize{$4$}};
\draw (-5,2) -- (-5.2,2) node[left] {\scriptsize{$2$}};
\draw (-5,0) -- (-5.2,0) node[left] {\scriptsize{$0$}};
\draw (-5,-2) -- (-5.2,-2) node[left] {\scriptsize{$-2$}};
\draw (-5,-4) -- (-5.2,-4) node[left] {\scriptsize{$-4$}};
\draw (4,-5) -- (4,-5.2) node[below] {\scriptsize{$4$}};
\draw (2,-5) -- (2,-5.2) node[below] {\scriptsize{$2$}};
\draw (0,-5) -- (0,-5.2) node[below] {\scriptsize{$0$}};
\draw (-2,-5) -- (-2,-5.2) node[below] {\scriptsize{$-2$}};
\draw (-4,-5) -- (-4,-5.2) node[below] {\scriptsize{$-4$}};

\node at (0,7) [above] {\small{$\mathrm{Re}(\chi)$}};

\draw[->] (-5,0) -- (5.8,0) node [right] {\scriptsize{$x$}};
\draw[->] (0,-5) -- (0,5.8) node [above] {\scriptsize{$y$}};

\fill[opacity = 0.03]  (5+15,5) -- (5+15,-5) -- (-5+15,-5) -- (-5+15,5) -- cycle;
\draw[line width=0.5mm] plot[smooth] file {capsing2_im0.txt};
\fill[opacity = 0.15]  plot[smooth] file {capsing2_im0.txt} -- cycle;
\fill[opacity = 0.07] plot[smooth] file {capsing2_im0p5.txt} -- (-5+15,-5) -- (-5+15,5) -- cycle;
\fill[opacity = 0.07] plot[smooth] file {capsing2_im1.txt} -- (-5+15,-5) -- (-5+15,5) -- cycle;
\fill[opacity = 0.07] plot[smooth] file {capsing2_im1p5.txt} -- (-5+15,5) -- (-5+15,-5) -- cycle;
\fill[opacity = 0.07] plot[smooth] file {capsing2_im2.txt} -- (-5+15,-5) -- (-5+15,5) -- cycle;
\fill[opacity = 0.07] plot[smooth] file {capsing2_im2p5.txt} -- (-5+15,-5) -- (-5+15,5) -- cycle;
\fill[opacity = 0.07] plot[smooth] file {capsing2_im3.txt} -- (-5+15,-5) -- (-5+15,5) -- cycle;
\fill[opacity = 0.07] plot[smooth] file {capsing2_im3p5.txt} -- (-5+15,-5) -- (-5+15,5) -- cycle;
\fill[opacity = 0.07] plot[smooth] file {capsing2_im4.txt} -- (-5+15,-5) -- (-5+15,5) -- cycle;
\fill[opacity = 0.07] plot[smooth] file {capsing2_im4p5.txt} -- (-5+15,5) -- (-5+15,-5) -- cycle;
%\draw[-] plot[smooth] file {capsing2_im5.txt};

\fill[opacity = 0.20] plot[smooth] file {capsing2_imm0p5.txt} -- cycle;
\fill[opacity = 0.20] plot[smooth] file {capsing2_imm0p5b.txt} -- cycle;
%\draw[-] plot[smooth] file {capsing2_imm1.txt};
%\draw[-] plot[smooth] file {capsing2_imm1b.txt};

\draw (-5+15,5) -- (-5+15,-5) -- (5+15,-5) -- (5+15,5) -- cycle;
\draw (-5+15,4) -- (-5.2+15,4) node[left] {\scriptsize{$4$}};
\draw (-5+15,2) -- (-5.2+15,2) node[left] {\scriptsize{$2$}};
\draw (-5+15,0) -- (-5.2+15,0) node[left] {\scriptsize{$0$}};
\draw (-5+15,-2) -- (-5.2+15,-2) node[left] {\scriptsize{$-2$}};
\draw (-5+15,-4) -- (-5.2+15,-4) node[left] {\scriptsize{$-4$}};
\draw (4+15,-5) -- (4+15,-5.2) node[below] {\scriptsize{$4$}};
\draw (2+15,-5) -- (2+15,-5.2) node[below] {\scriptsize{$2$}};
\draw (0+15,-5) -- (0+15,-5.2) node[below] {\scriptsize{$0$}};
\draw (-2+15,-5) -- (-2+15,-5.2) node[below] {\scriptsize{$-2$}};
\draw (-4+15,-5) -- (-4+15,-5.2) node[below] {\scriptsize{$-4$}};

\node at (0+15,7) [above] {\small{$\mathrm{Im}(\chi)$}};

\draw[->] (-5+15,0) -- (5.8+15,0) node [right] {\scriptsize{$x$}};
\draw[->] (0+15,-5) -- (0+15,5.8) node [above] {\scriptsize{$y$}};

\node at (2.5,3) {\scriptsize{Anti-Stokes}};
\node at (2.5,2.25) {\scriptsize{line}};
\draw[->] (1.5,2.25) -- (-0.5,2.25) .. controls (-1.5,2.25) and (-2,1.75) .. (-2,1.25) -- (-2,0.65);

%%%%%

\draw (7+17,5) -- (8+17,5) -- (8+17,-5) -- (7+17,-5) -- cycle;

\fill[opacity = 0.20]  (7+17,-5) -- (8+17,-5) -- (8+17,-5+1*0.8333) -- (7+17,-5+1*0.8333) -- cycle;
\fill[opacity = 0.15]  (7+17,-5) -- (8+17,-5) -- (8+17,-5+2*0.8333) -- (7+17,-5+2*0.8333) -- cycle;
\fill[opacity = 0.07]  (7+17,-5) -- (8+17,-5) -- (8+17,-5+3*0.8333) -- (7+17,-5+3*0.8333) -- cycle;
\fill[opacity = 0.07]  (7+17,-5) -- (8+17,-5) -- (8+17,-5+4*0.8333) -- (7+17,-5+4*0.8333) -- cycle;
\fill[opacity = 0.07]  (7+17,-5) -- (8+17,-5) -- (8+17,-5+5*0.8333) -- (7+17,-5+5*0.8333) -- cycle;
\fill[opacity = 0.07]  (7+17,-5) -- (8+17,-5) -- (8+17,-5+6*0.8333) -- (7+17,-5+6*0.8333) -- cycle;
\fill[opacity = 0.07]  (7+17,-5) -- (8+17,-5) -- (8+17,-5+7*0.8333) -- (7+17,-5+7*0.8333) -- cycle;
\fill[opacity = 0.07]  (7+17,-5) -- (8+17,-5) -- (8+17,-5+8*0.8333) -- (7+17,-5+8*0.8333) -- cycle;
\fill[opacity = 0.07]  (7+17,-5) -- (8+17,-5) -- (8+17,-5+9*0.8333) -- (7+17,-5+9*0.8333) -- cycle;
\fill[opacity = 0.07]  (7+17,-5) -- (8+17,-5) -- (8+17,-5+10*0.8333) -- (7+17,-5+10*0.8333) -- cycle;
\fill[opacity = 0.07]  (7+17,-5) -- (8+17,-5) -- (8+17,-5+11*0.8333) -- (7+17,-5+11*0.8333) -- cycle;

\fill[opacity = 0.03]  (7+17,5) -- (8+17,5) -- (8+17,0) -- (7+17,0) -- cycle;

\draw[line width=0.5mm] (6.8+17,-5+2*0.8333) -- (8+17,-5+2*0.8333);
\draw (7+17,5) -- (6.8+17,5) node[left] {\scriptsize{$5$}};
\draw (7+17,-5+2*0.8333) -- (6.8+17,-5+2*0.8333) node[left] {\scriptsize{$0$}};
\draw (7+17,-5) -- (6.8+17,-5) node[left] {\scriptsize{$-1$}};

\draw (-9,5) -- (-10,5) -- (-10,-5) -- (-9,-5) -- cycle;
\fill[opacity = 0.20]  (-9,-5) -- (-10,-5) -- (-10,-5+1*0.8333) -- (-9,-5+1*0.8333) -- cycle;
\fill[opacity = 0.15]  (-9,-5) -- (-10,-5) -- (-10,-5+2*0.8333) -- (-9,-5+2*0.8333) -- cycle;
\fill[opacity = 0.07]  (-9,-5) -- (-10,-5) -- (-10,-5+3*0.8333) -- (-9,-5+3*0.8333) -- cycle;
\fill[opacity = 0.07]  (-9,-5) -- (-10,-5) -- (-10,-5+4*0.8333) -- (-9,-5+4*0.8333) -- cycle;
\fill[opacity = 0.07]  (-9,-5) -- (-10,-5) -- (-10,-5+5*0.8333) -- (-9,-5+5*0.8333) -- cycle;
\fill[opacity = 0.07]  (-9,-5) -- (-10,-5) -- (-10,-5+6*0.8333) -- (-9,-5+6*0.8333) -- cycle;
\fill[opacity = 0.07]  (-9,-5) -- (-10,-5) -- (-10,-5+7*0.8333) -- (-9,-5+7*0.8333) -- cycle;
\fill[opacity = 0.07]  (-9,-5) -- (-10,-5) -- (-10,-5+8*0.8333) -- (-9,-5+8*0.8333) -- cycle;
\fill[opacity = 0.07]  (-9,-5) -- (-10,-5) -- (-10,-5+9*0.8333) -- (-9,-5+9*0.8333) -- cycle;
\fill[opacity = 0.15]  (-9,-5) -- (-10,-5) -- (-10,-5+10*0.8333) -- (-9,-5+10*0.8333) -- cycle;
\fill[opacity = 0.07]  (-9,-5) -- (-10,-5) -- (-10,-5+11*0.8333) -- (-9,-5+11*0.8333) -- cycle;

\fill[opacity = 0.03]  (-9,5) -- (-10,5) -- (-10,0) -- (-9,0) -- cycle;

\draw[line width=0.5mm] (-8.8,-5+2*0.8333) -- (-10,-5+2*0.8333);
\draw (-9,5) -- (-8.8,5) node[right] {\scriptsize{$1$}};
\draw (-9,-5+2*0.8333) -- (-8.8,-5+2*0.8333) node[right] {\scriptsize{$0$}};
\draw (-9,-5) -- (-8.8,-5) node[right] {\scriptsize{$-0.2$}};

\draw[line width = 0.5mm] (0+15,0) -- (-5+15,0);

\node at (2.75+15,3) {\scriptsize{Stokes line}};
\draw[->] (0.65+15,3) -- (-0.5+15,3) .. controls (-1.5+15,3) and (-2+15,2.5) .. (-2+15,2) -- (-2+15,1.65);

\end{tikzpicture}
}

\caption{Singulants obtained by solving (\ref{1:eik}) with boundary
  data (\ref{1:eik2}). The remaining singulant expressions are given
  by $-\chi_{S1,2}$, as well as $\pm\overline{\chi}_{S1,2}$, where the
  bar denotes complex conjugation. However, the contributions
  associated with $\pm\chi_{S2}$, $\pm\overline{\chi}_{S2}$, as well
  as $-\chi_{S1}$ and $-\overline{\chi}_{S1}$, will either produce
  waves directly downstream from the source, or contain no Stokes
  switching behaviour at all; hence, these singulants cannot generate
  exponentially small wave behaviour. Conversely, surface behaviour
  associated with $\chi_{S1}$ and $\overline{\chi}_{S1}$ will produce
  exponentially small waves in the upstream far field which are
  switched on across the Stokes line satisfying $\mathrm{Im}(\chi) =
  0$.
Note that there is no switching across the curve $y=0$, $x>0$ even
though $\mathrm{Im}(\chi) =  0$ there because $\mathrm{Re}(\chi)<0$ in
this region.  
}
\label{1:realrays}
\end{figure}
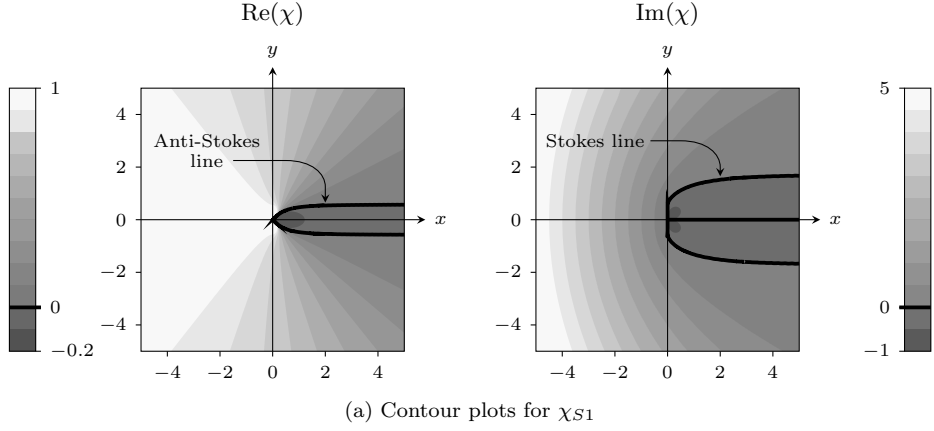
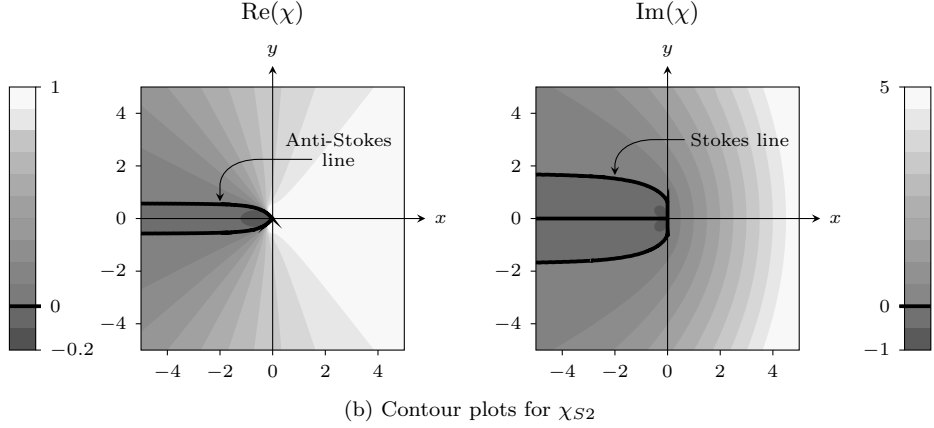

Equations (\ref{1:sschiray2})--(\ref{1:sspoly}) give eight possible expressions for the singulant (corresponding to the choice of sign in (\ref{1:sschiray2}) and the four solutions to (\ref{1:sspoly})). These therefore give eight possible sets of late-order behaviour in the problem.

Contour plots illustrating the behaviour of the singulant terms are
presented in figure \ref{1:realrays} for $h = 1$. To indicate that we
are considering the steady behaviour, we have labelled the singulants
as $\chi_{S1}$ and $\chi_{S2}$, where the number indicates two
different solutions from which all eight singulant expressions may be
easily obtained. Specifically, the eight expressions are given by
$\pm\chi_{S1,2}$ and $\pm{\overline{\chi}}_{S1,2}$, where the bar
denotes complex conjugation. We also note that $\chi_{S2}$ may be
obtained by reflecting $\chi_{S1}$ in the $y$-axis.

We recall from the methodology description that Stokes switching of
exponentially small contributions to the solution occurs across curves
known as Stokes lines, which must satisfy the condition on the
singulant given in \eqref{1:StokesCond}. As we have obtained explicit
expressions for the singulant, we are able to determine the location
of the Stokes lines in the solution, illustrated in figure
\ref{1:realrays} for $\chi_{S1}$ and $\chi_{S2}$. The Stokes lines are
illustrated by bold curves in the plots of $\mathrm{Im}(\chi)$ on the
right-hand side of figure \ref{1:realrays}. While we used the
condition \eqref{1:StokesCond} to identify the Stokes line locations,
it  appears also as a consequence of the matched asymptotic analysis,
which may be seen in Appendix \ref{CH5_STEADYSMOOTHING}.  

We may also determine the location of anti-Stokes lines using
\eqref{1:AntiStokesCond}, which are depicted as bold curves on the
plots of $\mathrm{Re}(\chi)$ on the left-hand side of this
figure. These curves are important, as we know that the corresponding
exponential contribution must be inactive in any region containing
anti-Stokes lines, as otherwise it would become exponentially large (and
therefore dominant) as the anti-Stokes lines is crossed.  

From figure \ref{1:realrays}, we therefore see that any free surface
behaviour associated with $\chi_{S2}$ or $\overline{\chi}_{S2}$ must
be switched on in the downstream region, and hence produce capillary
waves in the downstream far field, which violates the radiation
condition. Furthermore, both $-\chi_{S2}$ or $-\overline{\chi}_{S2}$
have $\mathrm{Re}(\chi) < 0$ across the Stokes line (satisfying
$\mathrm{Im}(\chi) = 0$), and hence no Stokes switching can
occur. This is also true of $-\chi_{S1}$ and
$-\overline{\chi}_{S1}$. Consequently, none of these singulant
contributions will produce exponentially small free surface capillary
waves.  

However, $\chi_{S1}$ and $\overline{\chi}_{S1}$ have
$\mathrm{Re}(\chi) > 0$ across the Stokes line, as well as in the
entire region in which the associated exponentially small wave
behaviour is switched on. Additionally, the wave behaviour is upstream
from the obstacle. From this, we conclude that the capillary wave
behaviour on the free surface is caused by the late-order terms associated
with $\chi_{S1}$ and $\overline{\chi_{S1}}$. The full Stokes structure
of the solution is therefore depicted in figure \ref{1:realrays} (a). 

%Finally, we note that no switching occurs across the Stokes line following $y = 0$, as $\mathrm{Re}(\chi) < 0$ in this region for $\chi_{S1}$ and $\overline{\chi}_{S1}$. 

Comparing the Stokes structure in figure \ref{1:realrays} (a) with the numerical free surface plot in  figure \ref{F:3DNum}, we see that the region upstream of the Stokes line in which the exponentially small ripples are present in the solution corresponds to the numerically calculated ripples in the surface plot. There are other features in the numerical plot which do not correspond to exponentially small ripples, and are present on both sides of the Stokes line; these features are not waves, but rather non-wavelike disturbances to the undisturbed flow found at algebraic orders of $\epsilon$ in the small-surface-tension limit.

As $\chi_{S1}$ and $\overline{\chi}_{S1}$ are the only contributions that contribute to the steady capillary wave behaviour, we will subsequently denote these as $\chi_S$ and $\overline{\chi}_S$ respectively. 

\subsubsection{Calculating the prefactor}\label{CH5_PREFACTOR}

In order to obtain a complete expression for the late-order terms (\ref{1:ansatz1}), we require an expression for the prefactors, $\Phi$ and $\Xi$. To find the prefactor equation, we consider the next order in (\ref{1:serbc1})--(\ref{1:serbc2}) as $n\rightarrow\infty$. Expanding the prefactors in the form of a power series in $n$ as $n\rightarrow \infty$,
\begin{equation}
  \label{1:phiseries}\Phi = \Phi_0 + \frac{1}{n}\Phi_1 + \ldots,\qquad \Xi = \Xi_0 + \frac{1}{n}\Xi_1 + \ldots
\end{equation}
and applying the late-order ansatz to (\ref{1:ansgov1})--(\ref{1:ansbc2}) now gives
\begin{align*}
-\chi_z\Phi_1 +\chi_x\Xi_1 &= - \Phi_{0,z} + \Xi_{0,x} ,\\
\chi_x\Phi_1 + (\chi_x^2 + \chi_y^2)\Xi_1  &=  \Phi_{0,x}+2\chi_x\Xi_{0,x} + 2\chi_y\Xi_{0,y} + (\chi_{xx}+\chi_{yy})\Xi_0  .
\end{align*}
This system only has nontrivial solutions for $\Phi_1$ and $\Xi_1$ when 
\begin{equation}
\chi_x(\Phi_{0,z} - \Xi_{0,x}) = \chi_z\left( \Phi_{0,x} +2\chi_x\Xi_{0,x} + 2\chi_y\Xi_{0,y}+ (\chi_{xx}+\chi_{yy})\Xi_0\right).
\end{equation}
Since we are presently interested in the leading-order behaviour of the prefactor, for ease of notation we now omit the subscripts and denote $\Xi_0$ by $\Xi$ and $\Phi_0$ by $\Phi$. This therefore gives
\begin{equation}
\label{1:phiz} \Phi_z = \Xi_x + \frac{\chi_z}{\chi_x}(\Phi_x+ 2\chi_x\Xi_x + 2\chi_y\Xi_y + (\chi_{xx}+\chi_{yy})\Xi).
\end{equation}

Now, to solve the prefactor equation (\ref{1:ansgov2}), we use this result, as well as (\ref{1:chiz}), to express the original equation entirely in terms of $x$ and $y$ derivatives. The resultant expression has the same ray equations as the singulant. Hence, we can express the prefactor equation using the characteristic variable of the singulant, $s$, which is given in terms of physical variables in (\ref{1:sspoly}). To fully determine the prefactor, we must subsequently match the solution of the prefactor equation to the behaviour of the flow in the neighbourhood of the singularity, as described in \cite{Chapman1}. This analysis is performed in Appendix \ref{PREFACTOR_APP}, and gives
%\begin{equation}
%\label{1:Phiss} \Phi= \frac{\sqrt{2}s}{8\pi^{3/2} h^{3/2}}\left[1+\frac{3(s^2+2h^2)(s-x)}{s h^{11}  (3s^2+2)}\right]^{\tfrac{(9s^2+2h^2)%(s^2+h^2) - 3\mathrm{i} s(2h^2+3s^2)\sqrt{s^2+h^2}}{2h^2(s^2+2h^2)}}, 
%\end{equation}
\begin{equation}
 \label{1:Phiss} \Phi= \frac{s\sqrt{2}}{4\pi^{3/2}h^{3/2}}\left[1-\frac{3h^4(4s^4+6s^2h^2-3h^4)(s-x)}{s^3(3s^2+2h^2)(2s^4+3h^4)}\right]^{\tfrac{s^2(24s^8+34s^6h^2+36s^4h^4+35h^6s^2+14h^8)}{2h^6(3h^4-6s^2h^2-4s^4)}},\end{equation}
where $s$ is the solution of (\ref{1:sspoly}) corresponding to the singulant illustrated in figure \ref{1:realrays}.

Finally, to find $\gamma$, we ensure that the strength of the singularity in the late-order behaviour $\phi^{(n)}$, given in (\ref{1:ansatz1}) is consistent with the leading-order behaviour $\phi^{(0)}$, which has strength $1/2$. It is clear from the recurrence relation \eqref{2:serbc2} that the strength of the singularity will increase by one between $\phi^{(n-1)}$ and $\phi^{(n)}$. This implies that near the singularity at $x^2 + y^2 + h^2 = 0$,
\begin{equation}
\label{1:gammamatch}\frac{\Phi\Gamma(\gamma)}{\chi^{\gamma}} \rightarrow \frac{\alpha(x,y)}{(x^2+y^2+h^2)^{1/2}},
\end{equation}
where $\alpha$ is of order one in the limit. From (\ref{1:Phiss}), we see that the prefactor is also order one in this limit, while the local analysis near the singularity (\ref{1:innerchi1}) showed that  $1/\chi$ will be a singularity with strength one at $x^2 + y^2 + h^2 = 0$. Consequently, matching the order of the expressions in (\ref{1:gammamatch}) gives $\gamma = 1/2$.

We have therefore completely described the late-order terms given in (\ref{1:ansatz1}), where (\ref{1:phixi}) is used to determine the value of $\Xi$.

In Appendix \ref{CH5_STEADYSMOOTHING}, we use the form of the
late-order terms ansatz in \eqref{1:ansatz1}) in order to apply the
matched asymptotic expansion methodology of \cite{Daalhuis1}. We
optimally truncate the asymptotic series, and then find an equation
for the exponentially small truncation remainder. Using this
expression, we determine where the exponentially small remainder
varies rapidly, which corresponds to the location of Stokes lines. If
we had not applied the condition in \eqref{1:StokesCond}, this would
have been necessary to determine the Stokes structure of the
solution. Finally, we use matched asymptotic expansions in the
neighbourhood of the Stokes lines in order to determine the quantity
that is switched on as the Stokes lines are crossed. 

Using this method, we find that the exponentially small contributions to the fluid potential (denoted $\phi_{\mathrm{exp}}$) and free surface position (denoted $\xi_{\mathrm{exp}}$) are switched on across the Stokes line, and in regions in which they are active, they are given by
\begin{equation}
\phi_{\mathrm{exp}} \sim \frac{2\pi\mathrm{i} \Phi}{\sqrt{\epsilon}}\mathrm{e}^{-\chi_S/\epsilon} + \mathrm{c.c.},\qquad \xi_{\mathrm{exp}} \sim \frac{2\pi\mathrm{i} \Xi}{\sqrt{\epsilon}}\mathrm{e}^{-\chi_S/\epsilon} + \mathrm{c.c.},\label{1:RnSn}
\end{equation}
where c.c. denotes the complex conjugate contribution. In particular, the expression for $\xi_{\mathrm{exp}}$ contains exponentially small oscillations representing the capillary ripples on the free surface.

\subsection{Results and Comparison}\label{CH5_STEADYRESULTS}

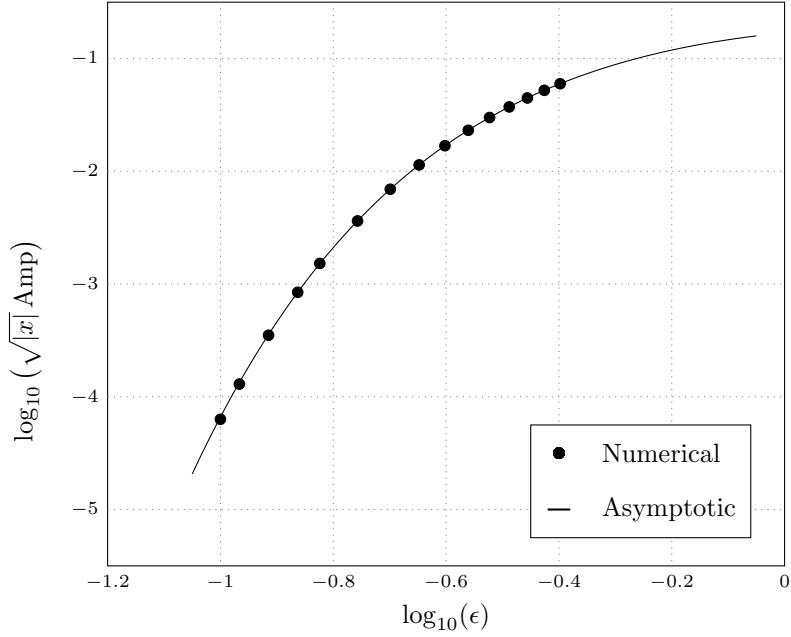
\begin{figure}
\centering

\begin{tikzpicture}
[x=0.75*10.00cm,y=0.75*2.00cm]

\draw (0,-5.5) -- (-1.2,-5.5) -- (-1.2,-0.5) -- (0,-0.5) -- cycle;

\draw[gray,dotted] (-1.2,-4) node[black,left] {\scriptsize{$-4$}} -- (0,-4);
\draw[gray,dotted] (-1.2,-3) node[black,left] {\scriptsize{$-3$}} -- (0,-3);
\draw[gray,dotted] (-1.2,-2) node[black,left] {\scriptsize{$-2$}} -- (0,-2);
\draw[gray,dotted] (-1.2,-1) node[black,left] {\scriptsize{$-1$}} -- (0,-1);
\draw[gray,dotted] (-1.2,-5) node[black,left] {\scriptsize{$-5$}} -- (0,-5);

\draw[gray,dotted] (-1,-0.5) -- (-1,-5.5) node[black,below] {\scriptsize{$-1$}};
\draw[gray,dotted] (-0.8,-0.5) -- (-0.8,-5.5) node[black,below] {\scriptsize{$-0.8$}};
\draw[gray,dotted] (-0.6,-0.5) -- (-0.6,-5.5) node[black,below] {\scriptsize{$-0.6$}};
\draw[gray,dotted] (-0.4,-0.5) -- (-0.4,-5.5) node[black,below] {\scriptsize{$-0.4$}};
\draw[gray,dotted] (-0.2,-0.5) -- (-0.2,-5.5) node[black,below] {\scriptsize{$-0.2$}};
\node at (0,-5.5) [below] {\scriptsize{$0$}};
\node at (-1.2,-5.5) [below] {\scriptsize{$-1.2$}};

\node at (-1.3,-3.5) [above,rotate=90]  {{$\log_{10}\big(\sqrt{|x|}\,\mathrm{Amp}\big)$}};
\node at (0.1,-3.5) [above,rotate=-90,white] {{$\log_{10}\big(\sqrt{|x|}\,\mathrm{Amp}\big)$}};
\node at (-0.6,-5.75) [below] {{$\log_{10}({\epsilon})$}};

\draw plot[smooth] file {Cap_Steady_Asymp.txt};
 \draw plot[mark=*, only marks] file {Cap_Steady_Num_2.txt};
 
   \fill[white] (-0.45,-4.25) -- (-0.06,-4.25) -- (-0.06,-5.25) -- (-0.45,-5.25) -- cycle;
 \draw (-0.45,-4.25) -- (-0.06,-4.25) -- (-0.06,-5.25) -- (-0.45,-5.25) -- cycle;
  
  \draw plot[mark=*, only marks] (-0.4,-4.5);
  \draw[thick] (-0.42,-5) -- (-0.38,-5);
  \node at (-0.34,-4.5) [right] {{Numerical}};
    \node at (-0.34,-5) [right] {{Asymptotic}};

\end{tikzpicture}

\caption{Numerical (dots) verses scaled asymptotic (line) amplitude of
  capillary waves in the far field ($x \rightarrow -\infty$) along $y =
  0$ for $h=1$.} 
\label{F:SteadyAsympNum}
\end{figure}

Evaluating the amplitude of the waves along $y = 0$ gives
\begin{equation}
\xi_{\mathrm{exp}}\sim \frac{\mathrm{i} }{\sqrt{2\pi  \epsilon (4h-3\mathrm{i} x)}}\mathrm{e}^{-(h-\mathrm{i} x)/\epsilon} + \mathrm{c. c.},\qquad \epsilon \rightarrow 0.
\end{equation}
%We find that the total wave behaviour on $y = 0$ is given by
%\begin{equation}
%\xi_{\mathrm{exp}} \sim \frac{ \mathrm{e}^{-h/\epsilon}}{\sqrt{2\pi
%\epsilon}}\left[\frac{\mathrm{i}}{\sqrt{4h-3\mathrm{i} x}}\mathrm{e}^{\mathrm{i} x/\epsilon} -
%\frac{\mathrm{i}}{\sqrt{4h+3\mathrm{i} x}}\mathrm{e}^{-\mathrm{i} x/\epsilon}\right],\qquad \eps
%\rightarrow 0. 
%\end{equation}
%In the limit that $x$ becomes large in the negative direction, we obtain
%\begin{equation}
%R^{(N)}_{\mathrm{TOT}} \sim \frac{\mathrm{i} h^{5/12}}{8\times 3^{1/6}\sqrt{2 \pi  \epsilon}}\left(\frac{\mathrm{i}}{3x}\right)^{1/6}\mathrm{e}^{-(h-\mathrm{i} x)/\epsilon},\qquad \epsilon \rightarrow 0,\quad x\rightarrow -\infty.
%\end{equation}
In the limit that $x$ becomes large and negative, we find that the
amplitude of the capillary waves along $y = 0$ is given by 
\begin{equation}
\mathrm{Amplitude} \sim \frac{2 \mathrm{e}^{-h/\epsilon}}{\sqrt{6\pi\epsilon|x|}},\qquad x\rightarrow -\infty,\, \epsilon\rightarrow 0.\label{1:AsympPred}
\end{equation}
This provides us with the means to check 	the accuracy of our
approximation. We can compare the amplitude of the asymptotic results
with those of numerically-calculated free surface profiles, calculated
using an adaption of the method described \cite{Lustri2}.
%In order to
%obtain a solution that is constant in the far field, we compare the
%\textit{modified} amplitude, given by $\sqrt{|x|} \times
%\mathrm{Amplitude}$. This quantity tends to a constant as $x
%\rightarrow -\infty$, rather than decaying to zero. 

In figure \ref{F:SteadyAsympNum}, we illustrate the scaled numerical
amplitude (circles) against the asymptotic prediction from
\eqref{1:AsympPred}, computed for $h = 1$ over a range of $\epsilon$
values. It is apparent that there is strong agreement between the
numerical and the asymptotic results.
For values of $\epsilon$ smaller than those
depicted, it become numerically challenging to compute the wave
behaviour, due to the very small amplitude of the resulting waves. 

%Nonetheless, we see that the asymptotic prediction agrees with the
%numerically computed results, confirming the asymptotic behaviour
%obtained in section \ref{CH5_STEADYSMOOTHING}. Finally, we illustrate
%a comparison between the equal phase lines obtained using exponential
%asymptotics, and the numerically computed free surface for $\epsilon =
%0.2$. We see that the equal phase lines correspond to the wave
%behaviour throughout the solution domain
%%%%%%%%%%%%%%%%%%%%%%%%%%%%%%%%%%%%%%%%%%%%%%%%%%%%%%%%%%%%
%%       WHERE DO YOU DO THIS??   ^^^            %%%%%%%%%%%
%%%%%%%%%%%%%%%%%%%%%%%%%%%%%%%%%%%%%%%%%%%%%%%%%%%%%%%%%%%%

\section{Unsteady Flow}

\subsection{Formulation}
In this section, we consider the same flow configuration described in
section \ref{STEADY}; however, we permit the system to vary in
time. We  prescribe the initial state of the flow and
investigate the resultant unsteady behaviour.  

\subsubsection{Full problem}
We again consider three-dimensional potential flow with infinite depth
and a submerged point source at depth $H$ and upstream flow velocity
$U$. We normalise the fluid velocity with $U$ and distance by with an
unspecified length $L$, giving nondimensionalised source depth $h =
H/L$.
%, shown schematically in figure \ref{1:3dsteadyconfig2}.  

Denoting the (nondimensional) position of the free surface by $z =
\xi(x,y,t)$, the (nondimensional) velocity potential again satisfied
Laplace's equation (\ref{1:nlgeq}), however the kinematic and dynamic
boundary conditions  respectively become 
\begin{alignat}{2}
\label{2:nlbc1}\xi_x \phi_x + \xi_y \phi_y + \xi_t &= \phi_z,\qquad &z& = \xi(x,y,t),\\
\label{2:nlbc2}\phi_t + \frac{1}{2}(\phi_t + |\nabla\phi|^2- 1) + \epsilon\kappa &= 0,\qquad &z& = \xi(x,y,t),
\end{alignat}
where $\epsilon$ again denotes the inverse Weber number, and $\kappa$ the
curvature of the surface. The far field conditions are identical to
those in section \ref{STEADY}. The source condition is given by
(\ref{1:nlsc1}). As the problem is unsteady, we do not require a
radiation condition, but rather specify that the free surface must be
waveless in the far field. Finally, we require an initial condition,
as in \cite{Lustri4}, we specify that the initial state is given by
the leading order solution to the steady problem, given in
(\ref{1:phi0})--(\ref{1:xi0}). Hence the initial behaviour takes the
form 
\begin{align}
 \label{2:nlic1} \phi(x,y,z,0) &= \frac{\delta}{4\pi\sqrt{x^2 + y^2 + (z+h)^2}} - \frac{\delta}{4\pi\sqrt{x^2 + y^2 + (z-h)^2}},\\
 \label{2:nlic2} \xi(x,y,0) &= -\frac{\delta x h}{2 \pi (y^2 + h^2)\sqrt{x^2+y^2+h^2}}- \frac{\delta}{2\pi(y^2+h^2)}.
\end{align}
The reason for this choice of initial condition is that it enables us
to focus on wave generation rather than the bulk flow adjusting to the
presence of the source; in particular it guarantees that the
leading-order solution is steady, and hence that the leading-order
behaviour $\phi^{(0)}(x,y,z,t) = \phi(x,y,z,0)$ and $\xi^{(0)}(x,y,t)
= \xi(x,y,0)$. Note that it does not imply that any
subsequent order is steady.

\subsubsection{Linearisation}

We again linearise about uniform flow, and find that the governing equation (\ref{1:lgeq}) is valid in the unsteady problem. However, the boundary conditions become
%\begin{equation*}
% \phi = x + \delta\tilde{\phi},\qquad \xi = \delta\tilde{\xi},
%\end{equation*}
%to give, at leading order in $\delta$
\begin{alignat}{2}
%\label{1:lgeq} \nabla^2\tilde{\phi} &= 0 ,\qquad &-\infty < z& < 0,\\
\label{2:lbc1}\tilde{\phi}_z - \tilde{\xi}_x - \tilde{\xi}_t &= 0,\qquad & z&= 0,\\
\label{2:lbc2}\tilde{\phi}_x +\tilde{\phi}_t -\epsilon \left(\tilde{\xi}_{xx} + \tilde{\xi}_{yy}\right)&=0  ,\qquad & z &= 0,
\end{alignat}
where the boundary conditions are again applied on the fixed surface $z = 0$. The far-field conditions imply that $\tilde{\phi} \rightarrow 0$ as $x^2 + y^2 + z^2 \rightarrow \infty$, while near the source, (\ref{1:lsc1}) still holds. The initial condition is still given by \eqref{2:nlic1}--\eqref{2:nlic2}.

%\begin{align}
 %\label{2:lic1} \phi_0 &= \frac{1}{4\pi\sqrt{x^2 + y^2 + (z+h)^2}} - \frac{1}{4\pi\sqrt{x^2 + y^2 + (z-h)^2}},\\
 %\label{2:lic2} \xi_0 &= -\frac{x h}{2 \pi (y^2 + h^2)\sqrt{x^2+y^2+h^2}}.
%\end{align}
%\begin{equation}
%\label{1:lsc1}\tilde{\phi} \sim \frac{1}{4\pi\sqrt{x^2 + y^2 + (z+h)^2}} \qquad \mathrm{as} \quad (x,y,z) \rightarrow (0,0,-h).
%\end{equation}
%With the addition of a radiation condition, specifying that gravity waves may only propagate upstream, the system described in (\ref{1:lgeq})--(\ref{1:lsc1}) completely specifies the linearized version of the three-dimensional problem shown in figure \ref{1:3dsteadyconfig}.

\begin{figure}
\centering

\includegraphics[width=0.99\textwidth]{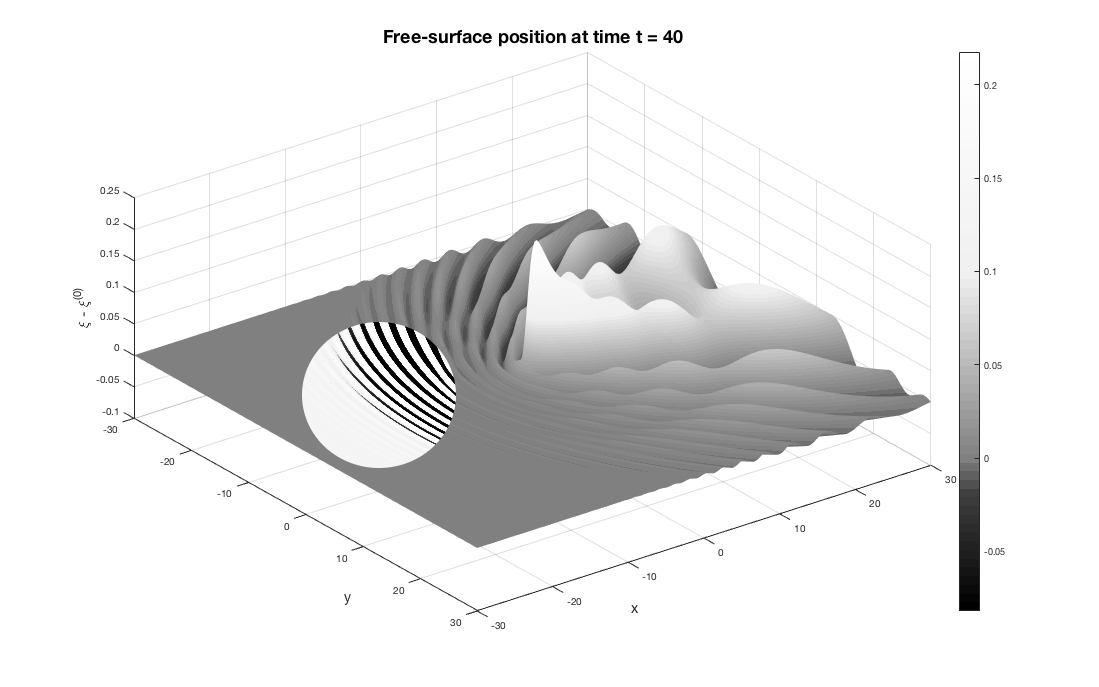}

\caption{Surface plot of the modified free-surface position $\xi -
  \xi^{(0)}$, where $\xi^{(0)}$ is the leading-order free-surface
  profile given in \eqref{1:xi0}, with $h = 1$ and $\epsilon = 0.2$. The
  initial condition of the flow is given by $\xi = \xi^{(0)}$, and
  therefore $\xi - \xi^{(0)} = 0$. This image corresponds to $t =
  40$. Flow is in the positive $x$-direction. We can see capillary
  waves extending upstream from the obstacle, however they have not
  propagated to the negative edge of the displayed region. In order to
  show the position of the wavefront clearly, the contrast has been
  increased in a circular region of the figure.} 
\label{F:3DNum2}
\end{figure}

We  again analytically continue the free surface such that $x, y \in
\mathbb{C}$, with the free surface still satisfying $z = 0$. We do
not, however, need to analytically continue $t$ in this
problem. Continuation does not change the form of
(\ref{1:lgeq})--(\ref{1:lsc1}), but it does mean that the
three-physical free surface (with two spatial and one time dimensions)
is now a subset of a five-dimensional complexified free surface. We
can again solve the linearised problem numerically using the method
from \cite{Lustri4}, to obtain numerical free-surface profiles such as
that illustrated in figure \ref{F:3DNum2}. 

\subsubsection{Series expression}\label{SERIES2}

Again, we expand the fluid potential and free-surface position as a power series in $\epsilon$. The governing equation is given by (\ref{1:sergeq}), while the boundary conditions become for $n\geq 0$
%\begin{equation}
% \label{1:series}\tilde{\phi} \sim \sum_{n=0}^{\infty}\epsilon^n\phi^{(n)},\qquad \tilde{\xi} \sim \sum_{n=0}^{\infty}\epsilon^n\xi^{(n)},
%\end{equation}
%to give for $n \geq 0$
\begin{alignat}{2}
%\label{1:sergeq} \nabla^2 \phi^{(n)} &= 0,\qquad &-\infty < z& < 0,\\
\label{2:serbc1}\phi^{(n)}_z - \xi^{(n)}_x- \xi^{(n)}_t &= 0,\qquad & z&= 0,\\
\label{2:serbc2}{\phi}_x^{(n)} + {\phi}_t^{(n)} -{\xi}_{xx}^{(n-1)} - {\xi}_{yy}^{(n-1)} &=0 ,\qquad & z &= 0,
\end{alignat}
again with the convention that $\xi^{(-1)} = 0$. The far-field behaviour tends to zero at all orders of $n$, and the singularity condition (\ref{1:lsc1}) is applied to the leading-order expression, giving the source condition in (\ref{1:sersc1}). The initial condition is obtained using \eqref{2:nlic1}--\eqref{2:nlic2}. As the leading-order behaviour is steady, we find that $\phi^{(0)} = \phi^{(0)}(x,y,z,0)$ and $\xi^{(0)} = \xi^{(0)}(x,y,0)$.
%\begin{align}
% \label{2:seric1} \phi^{(0)}(x,y,z,0) &= \frac{1}{4\pi\sqrt{x^2 + y^2 + (z+h)^2}} - \frac{1}{4\pi\sqrt{x^2 + y^2 + (z-h)^2}},\\
% \label{2:seric2} \xi^{(0)}(x,y,0) &= -\frac{x h}{2 \pi (y^2 + h^2)\sqrt{x^2+y^2+h^2}}- \frac{1}{2\pi(y^2+h^2)}.
%\end{align}
%As the leading-order behaviour is steady, we find that $\phi^{(0)} = \phi^{(0)}(x,y,z,0)$ and $\xi^{(0)} = \xi^{(0)}(x,y,0)$.

%\begin{equation}
%\label{1:sersc1}\phi^{(0)} \sim \frac{1}{4\pi\sqrt{x^2 + y^2 + (z+h)^2}} \qquad \mathrm{as} \quad (x,y,z) \rightarrow (0,0,-h).
%\end{equation}
%\begin{align}
% \label{1:phi0} \phi^{(0)} &= \frac{1}{4\pi\sqrt{x^2 + y^2 + (z+h)^2}} - \frac{1}{4\pi\sqrt{x^2 + y^2 + (z-h)^2}},\\
% \label{1:xi0} \xi^{(0)} &= -\frac{x h}{2 \pi (y^2 + h^2)\sqrt{x^2+y^2+h^2}}.
%\end{align}
%Through repeated iteration of (\ref{1:serbc1})--(\ref{1:serbc2}), we find that the position of singularities in subsequent terms of the series (\ref{1:series}) remains constant, while the power of the singularity increases at each order, as we expect for such singular perturbation problems \cite{Chapman3}.

\subsection{Late-Order Terms}

In order to optimally truncate the asymptotic series prescribed in (\ref{1:series}), we must determine the form of the late-order terms. To accomplish this, we make the new unsteady factorial-over-power ansatz \cite{Chapman1}
\begin{equation}
 \label{2:ansatz1} \phi^{(n)} \sim \frac{\Phi(x,y,z,t)\Gamma(n+\gamma)}{\chi(x,y,z,t)^{n+\gamma}},\qquad \xi^{(n)} \sim \frac{\Xi(x,y,t)\Gamma(n+\gamma)}{\chi(x,y,0,t)^{n+\gamma}}, \qquad \mathrm{as} \quad n \rightarrow \infty,
\end{equation}
which varies now in $t$, as well as the spatial dimensions. A nearly identical analysis to section \ref{CH5_SINGULANT1} gives the singulant equation on the free surface as
\begin{equation}
\label{2:eik}(\chi_x + \chi_t)^4 + (\chi_x^2 + \chi_y^2)^3 = 0.
\end{equation}
In considering the unsteady flow problem, we will restrict our attention to the singulants, ignoring the prefactor equation, and use the singulant behaviour to determine the position of Stokes lines and wave regions on the free surface.

%where $\gamma$ is a constant. In order that (\ref{1:ansatz1}) is the power series developed in section \ref{SERIES}, we require that the singulant, $\chi$, satisfies
%\begin{equation}
% \label{1:singularitycondition}  \chi = 0 \qquad \mathrm{on} \qquad x^2 + y^2 + (z\pm h)^2 = 0,
%\end{equation}
%where the sign chosen depends upon which of the two singularities is being considered. For complex values of  $x$, $y$ and $z$, this defines a four-dimensional hypersurface. Irrespective of which singularity is under consideration, this hypersurface intersects the four-dimensional complexified free surface on the two-dimensional hypersurface satisfying $x^2 + y^2 + h^2 = 0$.  

%It is important to note that the expression for $\xi^{(n)}$ is restricted to $z = 0$, as it describes the free-surface position. This does not pose a problem for the subsequent analysis, but does ensure that care must be taken at each stage to determine whether we are considering the full flow region, or just the free surface.

\subsubsection{Calculating the singulant}\label{CH5_SINGULANT2}

%Applying the ansatz expressions in (\ref{1:ansatz1}) to the governing equation (\ref{1:sergeq}) and taking the first two orders as $n \rightarrow \infty$ gives
%\begin{align}
%\label{2:ansgov1} \chi_x^2 + \chi_y^2 + \chi_z^2 &= 0,\\
%\label{2:ansgov2} 2\Phi_x\chi_x + 2\Phi_y\chi_y + 2\Phi_z\chi_z &= -(\chi_{xx}+\chi_{yy}+\chi_{zz}),
%\end{align}
%while the boundary conditions on $z = 0$ become, to leading order,
%\begin{align}
% \label{2:ansbc1} -\chi_z\Phi + \chi_x \Xi &= 0,\\
%\label{2:ansbc2} -\chi_x\Phi + (\chi_x^2 + \chi_y^2)\Xi  &= 0.
%\end{align}
%The system in (\ref{1:ansbc1})--(\ref{1:ansbc2}) has nonzero solutions when 
%\begin{equation}
%\label{2:eik0} \chi^2_x = \chi_z\left(\chi_x^2 + \chi_y^2\right).
%\end{equation}
%which gives the result
%\begin{equation}
%\label{2:chiz} \chi_z = \frac{\chi_x^2}{\chi_x^2+\chi_y^2}.
%\end{equation}
%Using (\ref{2:chiz}), we find a relationship between $\Phi$ and $\Xi$ by rearranging (\ref{2:ansbc1}) to obtain
%\begin{equation}\label{2:phixi}
%\Xi = \frac{\chi_x}{\chi_x^2+\chi_y^2} \Phi.
%\end{equation}

To determine the singulant behaviour on the free surface, we note that
the leading-order behaviour does have singularities on the
analytically-continued free surface located at $x^2 + y^2 + (z\pm h)^2
= 0$ for all time, and that these are identical to those described in
(\ref{CH5_SINGULANT2}). As a consequence, wave behaviour associated
with $\chi_{1}$ and $\overline{\chi}_{1}$ will be present on the free
surface. The presence of these waves is unsurprising, as the steady
wave behaviour satisfies (\ref{1:lbc1})--(\ref{1:lbc2}), as well as
the governing equation. 

However, we do see from figure \ref{1:realrays} that these singulants
lead to wave behaviour far upstream of the obstacle, which violates
the prescribed (waveless) far field condition. Consequently, we infer that there
must be another wave contribution, which introduces more complicated
Stokes line behaviour into the unsteady problem. 

\begin{figure}
\centering

\centering
\begin{tikzpicture}
[xscale=0.4,>=stealth,yscale= 0.4]

\begin{scope}
\clip (-5,5) -- (-5,-5) -- (5,-5) -- (5,5) -- cycle;

\fill[opacity=0.05] (-5,5) -- (-5,-5) -- (5,-5) -- (5,5) -- cycle;
\fill[opacity=0.05] (-5,5) -- (-5,-5) -- (5,-5) -- (5,5) -- cycle;
\fill[opacity=0.05] (-5,5) -- (-5,-5) -- (5,-5) -- (5,5) -- cycle;
\fill[opacity=0.05] (-5,5) -- (-5,-5) -- (5,-5) -- (5,5) -- cycle;
\fill[opacity=0.05] (-5,5) -- (-5,-5) -- (5,-5) -- (5,5) -- cycle;
\fill[opacity=0.05] (-5,5) -- (-5,-5) -- (5,-5) -- (5,5) -- cycle;
\fill[opacity=0.05] (-5,5) -- (-5,-5) -- (5,-5) -- (5,5) -- cycle;
\fill[opacity=0.05] (-5,5) -- (-5,-5) -- (5,-5) -- (5,5) -- cycle;
\fill[opacity=0.05] (-5,5) -- (-5,-5) -- (5,-5) -- (5,5) -- cycle;
\fill[opacity=0.05] (-5,5) -- (-5,-5) -- (5,-5) -- (5,5) -- cycle;
\fill[opacity=0.05] (-5,5) -- (-5,-5) -- (5,-5) -- (5,5) -- cycle;

\fill[white] (0,0) circle (4.537/3.5*5);
\fill[opacity=0.05] (0,0) circle (4.537/3.5*5);
\fill[opacity=0.05] (0,0) circle (4.537/3.5*5);
\fill[opacity=0.05] (0,0) circle (4.537/3.5*5);
\fill[opacity=0.05] (0,0) circle (4.537/3.5*5);
\fill[opacity=0.05] (0,0) circle (4.537/3.5*5);
\fill[opacity=0.05] (0,0) circle (4.537/3.5*5);
\fill[opacity=0.05] (0,0) circle (4.537/3.5*5);
\fill[opacity=0.05] (0,0) circle (4.537/3.5*5);
\fill[opacity=0.05] (0,0) circle (4.537/3.5*5);
\fill[opacity=0.05] (0,0) circle (4.537/3.5*5);

\fill[white] (0,0) circle (4.282/3.5*5);
\fill[opacity=0.05] (0,0) circle (4.282/3.5*5);
\fill[opacity=0.05] (0,0) circle (4.282/3.5*5);
\fill[opacity=0.05] (0,0) circle (4.282/3.5*5);
\fill[opacity=0.05] (0,0) circle (4.282/3.5*5);
\fill[opacity=0.05] (0,0) circle (4.282/3.5*5);
\fill[opacity=0.05] (0,0) circle (4.282/3.5*5);
\fill[opacity=0.05] (0,0) circle (4.282/3.5*5);
\fill[opacity=0.05] (0,0) circle (4.282/3.5*5);
\fill[opacity=0.05] (0,0) circle (4.282/3.5*5);

\fill[white] (0,0) circle (4.010/3.5*5);
\fill[opacity=0.05] (0,0) circle (4.010/3.5*5);
\fill[opacity=0.05] (0,0) circle (4.010/3.5*5);
\fill[opacity=0.05] (0,0) circle (4.010/3.5*5);
\fill[opacity=0.05] (0,0) circle (4.010/3.5*5);
\fill[opacity=0.05] (0,0) circle (4.010/3.5*5);
\fill[opacity=0.05] (0,0) circle (4.010/3.5*5);
\fill[opacity=0.05] (0,0) circle (4.010/3.5*5);
\fill[opacity=0.05] (0,0) circle (4.010/3.5*5);

\fill[white] (0,0) circle (3.719/3.5*5);
\fill[opacity=0.05] (0,0) circle (3.719/3.5*5);
\fill[opacity=0.05] (0,0) circle (3.719/3.5*5);
\fill[opacity=0.05] (0,0) circle (3.719/3.5*5);
\fill[opacity=0.05] (0,0) circle (3.719/3.5*5);
\fill[opacity=0.05] (0,0) circle (3.719/3.5*5);
\fill[opacity=0.05] (0,0) circle (3.719/3.5*5);
\fill[opacity=0.05] (0,0) circle (3.719/3.5*5);

\fill[white] (0,0) circle (3.403/3.5*5);
\fill[opacity=0.05] (0,0) circle (3.403/3.5*5);
\fill[opacity=0.05] (0,0) circle (3.403/3.5*5);
\fill[opacity=0.05] (0,0) circle (3.403/3.5*5);
\fill[opacity=0.05] (0,0) circle (3.403/3.5*5);
\fill[opacity=0.05] (0,0) circle (3.403/3.5*5);
\fill[opacity=0.05] (0,0) circle (3.403/3.5*5);

\fill[white] (0,0) circle (3.055/3.5*5);
\fill[opacity=0.05] (0,0) circle (3.055/3.5*5);
\fill[opacity=0.05] (0,0) circle (3.055/3.5*5);
\fill[opacity=0.05] (0,0) circle (3.055/3.5*5);
\fill[opacity=0.05] (0,0) circle (3.055/3.5*5);
\fill[opacity=0.05] (0,0) circle (3.055/3.5*5);

\fill[white] (0,0) circle (2.661/3.5*5);
\fill[opacity=0.05] (0,0) circle (2.661/3.5*5);
\fill[opacity=0.05] (0,0) circle (2.661/3.5*5);
\fill[opacity=0.05] (0,0) circle (2.661/3.5*5);
\fill[opacity=0.05] (0,0) circle (2.661/3.5*5);

\fill[white] (0,0) circle (2.199/3.5*5);
\fill[opacity=0.05] (0,0) circle (2.199/3.5*5);
\fill[opacity=0.05] (0,0) circle (2.199/3.5*5);
\fill[opacity=0.05] (0,0) circle (2.199/3.5*5);

\fill[white] (0,0) circle (1.607/3.5*5);
\fill[opacity=0.05] (0,0) circle (1.607/3.5*5);
\fill[opacity=0.05] (0,0) circle (1.607/3.5*5);

\fill[white] (0,0) circle (0.577/3.5*5);
\fill[opacity=0.05] (0,0) circle (0.577/3.5*5);

\end{scope}

\draw[line width=0.5mm] (0,0) circle (0.577/3.5*5);

\draw (-5,5) -- (-5,-5) -- (5,-5) -- (5,5) -- cycle;
\draw (-5,4.3) -- (-5.2,4.3) node[left] {\scriptsize{$3$}};
\draw (-5,2.85) -- (-5.2,2.85) node[left] {\scriptsize{$2$}};
\draw (-5,1.43) -- (-5.2,1.43) node[left] {\scriptsize{$1$}};
\draw (-5,0) -- (-5.2,0) node[left] {\scriptsize{$0$}};
\draw (-5,-4.3) -- (-5.2,-4.3) node[left] {\scriptsize{$3$}};
\draw (-5,-2.85) -- (-5.2,-2.85) node[left] {\scriptsize{$2$}};
\draw (-5,-1.43) -- (-5.2,-1.43) node[left] {\scriptsize{$1$}};
\draw (4.3,-5) -- (4.3,-5.2) node[below] {\scriptsize{$3$}};
\draw (2.85,-5) -- (2.85,-5.2) node[below] {\scriptsize{$2$}};
\draw (1.43 ,-5) -- (1.43 ,-5.2) node[below] {\scriptsize{$1$}};
\draw (0,-5) -- (0,-5.2) node[below] {\scriptsize{$0$}};
\draw (-4.3,-5) -- (-4.3,-5.2) node[below] {\scriptsize{$-3$}};
\draw (-2.85,-5) -- (-2.85,-5.2) node[below] {\scriptsize{$-2$}};
\draw (-1.43 ,-5) -- (-1.43 ,-5.2) node[below] {\scriptsize{$-1$}};

\node at (0,7) [above] {\small{$\mathrm{Re}(t^2 \chi)$}};

\draw[->] (-5,0) -- (5.8,0) node [right] {\scriptsize{$x-t$}};
\draw[->] (0,-5) -- (0,5.8) node [above] {\scriptsize{$y$}};

\begin{scope}
\clip (-5+15,5) -- (-5+15,-5) -- (5+15,-5) -- (5+15,5) -- cycle;

\fill[opacity=0.05] (-5+15,5) -- (-5+15,-5) -- (5+15,-5) -- (5+15,5) -- cycle;
\fill[opacity=0.05] (-5+15,5) -- (-5+15,-5) -- (5+15,-5) -- (5+15,5) -- cycle;
\fill[opacity=0.05] (-5+15,5) -- (-5+15,-5) -- (5+15,-5) -- (5+15,5) -- cycle;
\fill[opacity=0.05] (-5+15,5) -- (-5+15,-5) -- (5+15,-5) -- (5+15,5) -- cycle;
\fill[opacity=0.05] (-5+15,5) -- (-5+15,-5) -- (5+15,-5) -- (5+15,5) -- cycle;
\fill[opacity=0.05] (-5+15,5) -- (-5+15,-5) -- (5+15,-5) -- (5+15,5) -- cycle;
\fill[opacity=0.05] (-5+15,5) -- (-5+15,-5) -- (5+15,-5) -- (5+15,5) -- cycle;
\fill[opacity=0.05] (-5+15,5) -- (-5+15,-5) -- (5+15,-5) -- (5+15,5) -- cycle;
\fill[opacity=0.05] (-5+15,5) -- (-5+15,-5) -- (5+15,-5) -- (5+15,5) -- cycle;
\fill[opacity=0.05] (-5+15,5) -- (-5+15,-5) -- (5+15,-5) -- (5+15,5) -- cycle;
\fill[opacity=0.05] (-5+15,5) -- (-5+15,-5) -- (5+15,-5) -- (5+15,5) -- cycle;

\fill[white] (15,0) circle (4.185/3.5*5);
\fill[opacity=0.05] (15,0) circle (4.185/3.5*5);
\fill[opacity=0.05] (15,0) circle (4.185/3.5*5);
\fill[opacity=0.05] (15,0) circle (4.185/3.5*5);
\fill[opacity=0.05] (15,0) circle (4.185/3.5*5);
\fill[opacity=0.05] (15,0) circle (4.185/3.5*5);
\fill[opacity=0.05] (15,0) circle (4.185/3.5*5);
\fill[opacity=0.05] (15,0) circle (4.185/3.5*5);
\fill[opacity=0.05] (15,0) circle (4.185/3.5*5);
\fill[opacity=0.05] (15,0) circle (4.185/3.5*5);
\fill[opacity=0.05] (15,0) circle (4.185/3.5*5);

\fill[white] (15,0) circle (4.041/3.5*5);
\fill[opacity=0.05] (15,0) circle (4.041/3.5*5);
\fill[opacity=0.05] (15,0) circle (4.041/3.5*5);
\fill[opacity=0.05] (15,0) circle (4.041/3.5*5);
\fill[opacity=0.05] (15,0) circle (4.041/3.5*5);
\fill[opacity=0.05] (15,0) circle (4.041/3.5*5);
\fill[opacity=0.05] (15,0) circle (4.041/3.5*5);
\fill[opacity=0.05] (15,0) circle (4.041/3.5*5);
\fill[opacity=0.05] (15,0) circle (4.041/3.5*5);
\fill[opacity=0.05] (15,0) circle (4.041/3.5*5);

\fill[white] (15,0) circle (3.891/3.5*5);
\fill[opacity=0.05] (15,0) circle (3.891/3.5*5);
\fill[opacity=0.05] (15,0) circle (3.891/3.5*5);
\fill[opacity=0.05] (15,0) circle (3.891/3.5*5);
\fill[opacity=0.05] (15,0) circle (3.891/3.5*5);
\fill[opacity=0.05] (15,0) circle (3.891/3.5*5);
\fill[opacity=0.05] (15,0) circle (3.891/3.5*5);
\fill[opacity=0.05] (15,0) circle (3.891/3.5*5);
\fill[opacity=0.05] (15,0) circle (3.891/3.5*5);

\fill[white] (15,0) circle (3.725/3.5*5);
\fill[opacity=0.05] (15,0) circle (3.725/3.5*5);
\fill[opacity=0.05] (15,0) circle (3.725/3.5*5);
\fill[opacity=0.05] (15,0) circle (3.725/3.5*5);
\fill[opacity=0.05] (15,0) circle (3.725/3.5*5);
\fill[opacity=0.05] (15,0) circle (3.725/3.5*5);
\fill[opacity=0.05] (15,0) circle (3.725/3.5*5);
\fill[opacity=0.05] (15,0) circle (3.725/3.5*5);

\fill[white] (15,0) circle (3.540/3.5*5);
\fill[opacity=0.05] (15,0) circle (3.540/3.5*5);
\fill[opacity=0.05] (15,0) circle (3.540/3.5*5);
\fill[opacity=0.05] (15,0) circle (3.540/3.5*5);
\fill[opacity=0.05] (15,0) circle (3.540/3.5*5);
\fill[opacity=0.05] (15,0) circle (3.540/3.5*5);
\fill[opacity=0.05] (15,0) circle (3.540/3.5*5);

\fill[white] (15,0) circle (3.332/3.5*5);
\fill[opacity=0.05] (15,0) circle (3.332/3.5*5);
\fill[opacity=0.05] (15,0) circle (3.332/3.5*5);
\fill[opacity=0.05] (15,0) circle (3.332/3.5*5);
\fill[opacity=0.05] (15,0) circle (3.332/3.5*5);
\fill[opacity=0.05] (15,0) circle (3.332/3.5*5);

\fill[white] (15,0) circle (3.091/3.5*5);
\fill[opacity=0.05] (15,0) circle (3.091/3.5*5);
\fill[opacity=0.05] (15,0) circle (3.091/3.5*5);
\fill[opacity=0.05] (15,0) circle (3.091/3.5*5);
\fill[opacity=0.05] (15,0) circle (3.091/3.5*5);

\fill[white] (15,0) circle (2.797/3.5*5);
\fill[opacity=0.05] (15,0) circle (2.797/3.5*5);
\fill[opacity=0.05] (15,0) circle (2.797/3.5*5);
\fill[opacity=0.05] (15,0) circle (2.797/3.5*5);

\fill[white] (15,0) circle (2.409/3.5*5);
\fill[opacity=0.05] (15,0) circle (2.409/3.5*5);
\fill[opacity=0.05] (15,0) circle (2.409/3.5*5);

\fill[white] (15,0) circle (1.732/3.5*5);
\fill[opacity=0.05] (15,0) circle (1.732/3.5*5);

\end{scope}

\draw (-5+15,5) -- (-5+15,-5) -- (5+15,-5) -- (5+15,5) -- cycle;
\draw (-5+15,4.3) -- (-5.2+15,4.3) node[left] {\scriptsize{$3$}};
\draw (-5+15,2.85) -- (-5.2+15,2.85) node[left] {\scriptsize{$2$}};
\draw (-5+15,1.43) -- (-5.2+15,1.43) node[left] {\scriptsize{$1$}};
\draw (-5+15,0) -- (-5.2+15,0) node[left] {\scriptsize{$0$}};
\draw (-5+15,-4.3) -- (-5.2+15,-4.3) node[left] {\scriptsize{$3$}};
\draw (-5+15,-2.85) -- (-5.2+15,-2.85) node[left] {\scriptsize{$2$}};
\draw (-5+15,-1.43) -- (-5.2+15,-1.43) node[left] {\scriptsize{$1$}};
\draw (4.3+15,-5) -- (4.3+15,-5.2) node[below] {\scriptsize{$3$}};
\draw (2.85+15,-5) -- (2.85+15,-5.2) node[below] {\scriptsize{$2$}};
\draw (1.43 +15,-5) -- (1.43 +15,-5.2) node[below] {\scriptsize{$1$}};
\draw (0+15,-5) -- (0+15,-5.2) node[below] {\scriptsize{$0$}};
\draw (-4.3+15,-5) -- (-4.3+15,-5.2) node[below] {\scriptsize{$-3$}};
\draw (-2.85+15,-5) -- (-2.85+15,-5.2) node[below] {\scriptsize{$-2$}};
\draw (-1.43 +15,-5) -- (-1.43 +15,-5.2) node[below] {\scriptsize{$-1$}};

\node at (0+15,7) [above] {\small{$\mathrm{Im}(t^2 \chi)$}};

\draw[->] (-5+15,0) -- (5.8+15,0) node [right] {\scriptsize{$x-t$}};
\draw[->] (0+15,-5) -- (0+15,5.8) node [above] {\scriptsize{$y$}};

%%%%%

\draw (7+17,5) -- (8+17,5) -- (8+17,-5) -- (7+17,-5) -- cycle;

\draw[line width=0.5mm] (15,0) circle (1.732/3.5*5);

\fill[opacity = 0.05]  (7+17,5) -- (8+17,5) -- (8+17,-5+1*0.8333) -- (7+17,-5+1*0.8333) -- cycle;
\fill[opacity = 0.05]  (7+17,5) -- (8+17,5) -- (8+17,-5+2*0.8333) -- (7+17,-5+2*0.8333) -- cycle;
\fill[opacity = 0.05]  (7+17,5) -- (8+17,5) -- (8+17,-5+3*0.8333) -- (7+17,-5+3*0.8333) -- cycle;
\fill[opacity = 0.05]  (7+17,5) -- (8+17,5) -- (8+17,-5+4*0.8333) -- (7+17,-5+4*0.8333) -- cycle;
\fill[opacity = 0.05]  (7+17,5) -- (8+17,5) -- (8+17,-5+5*0.8333) -- (7+17,-5+5*0.8333) -- cycle;
\fill[opacity = 0.05]  (7+17,5) -- (8+17,5) -- (8+17,-5+6*0.8333) -- (7+17,-5+6*0.8333) -- cycle;
\fill[opacity = 0.05]  (7+17,5) -- (8+17,5) -- (8+17,-5+7*0.8333) -- (7+17,-5+7*0.8333) -- cycle;
\fill[opacity = 0.05]  (7+17,5) -- (8+17,5) -- (8+17,-5+8*0.8333) -- (7+17,-5+8*0.8333) -- cycle;
\fill[opacity = 0.05]  (7+17,5) -- (8+17,5) -- (8+17,-5+9*0.8333) -- (7+17,-5+9*0.8333) -- cycle;
\fill[opacity = 0.05]  (7+17,5) -- (8+17,5) -- (8+17,-5+10*0.8333) -- (7+17,-5+10*0.8333) -- cycle;
\fill[opacity = 0.05]  (7+17,5) -- (8+17,5) -- (8+17,-5+11*0.8333) -- (7+17,-5+11*0.8333) -- cycle;

\draw[line width=0.5mm] (6.8+17,-5+2*0.8333) -- (8+17,-5+2*0.8333);
\draw (7+17,5) -- (6.8+17,5) node[left] {\scriptsize{$10$}};
\draw (7+17,-5+2*0.8333) -- (6.8+17,-5+2*0.8333) node[left] {\scriptsize{$0$}};
\draw (7+17,-5) -- (6.8+17,-5) node[left] {\scriptsize{$-2$}};

%\draw (-9,5) -- (-10,5) -- (-10,-5) -- (-9,-5) -- cycle;
%\fill[opacity = 0.07]  (-9,-5) -- (-10,-5) -- (-10,-5+1*0.8333) -- (-9,-5+1*0.8333) -- cycle;
%\fill[opacity = 0.07]  (-9,-5) -- (-10,-5) -- (-10,-5+2*0.8333) -- (-9,-5+2*0.8333) -- cycle;
%\fill[opacity = 0.07]  (-9,-5) -- (-10,-5) -- (-10,-5+3*0.8333) -- (-9,-5+3*0.8333) -- cycle;
%\fill[opacity = 0.07]  (-9,-5) -- (-10,-5) -- (-10,-5+4*0.8333) -- (-9,-5+4*0.8333) -- cycle;
%\fill[opacity = 0.07]  (-9,-5) -- (-10,-5) -- (-10,-5+5*0.8333) -- (-9,-5+5*0.8333) -- cycle;
%\fill[opacity = 0.07]  (-9,-5) -- (-10,-5) -- (-10,-5+6*0.8333) -- (-9,-5+6*0.8333) -- cycle;
%\fill[opacity = 0.07]  (-9,-5) -- (-10,-5) -- (-10,-5+7*0.8333) -- (-9,-5+7*0.8333) -- cycle;
%\fill[opacity = 0.07]  (-9,-5) -- (-10,-5) -- (-10,-5+8*0.8333) -- (-9,-5+8*0.8333) -- cycle;
%\fill[opacity = 0.07]  (-9,-5) -- (-10,-5) -- (-10,-5+9*0.8333) -- (-9,-5+9*0.8333) -- cycle;
%\fill[opacity = 0.15]  (-9,-5) -- (-10,-5) -- (-10,-5+10*0.8333) -- (-9,-5+10*0.8333) -- cycle;
%\fill[opacity = 0.07]  (-9,-5) -- (-10,-5) -- (-10,-5+11*0.8333) -- (-9,-5+11*0.8333) -- cycle;

%\fill[opacity = 0.03]  (-9,5) -- (-10,5) -- (-10,0) -- (-9,0) -- cycle;

%\draw[line width=0.5mm] (-8.8,-5+10*0.8333) -- (-10,-5+10*0.8333);
%\draw (-9,5) -- (-8.8,5) node[right] {\scriptsize{$10$}};
%\draw (-9,-5+10*0.8333) -- (-8.8,-5+10*0.8333) node[right] {\scriptsize{$0$}};
%\draw (-9,-5) -- (-8.8,-5) node[right] {\scriptsize{$-2$}};

\end{tikzpicture}

\caption{Singulant behaviour $\chi_U$, obtained by solving (\ref{2:eik1}) with boundary data (\ref{2:cauchy}). The expression is scaled by $t^2$ and presented in a frame moving with the flow, so that the resultant behaviour is constant in time.The remaining singulant expressions associated with unsteady behaviour are given by $-\chi_U$, as well as $\pm\overline{\chi}_U$.  We see that Stokes switching is possible for $\chi_U$ as $\mathrm{Re}(\chi)>0$ across the Stokes line satisfying $\mathrm{Re}(\chi) = 0$. There is an anti-Stokes line inside the Stokes line, meaning that any surface wave behaviour would become exponentially large as this line is crossed. Consequently, we conclude that the unsteady exponentially small contribution cannot be present inside the Stokes line, and must instead be switched on outside. The same is true of the contribution associated with $\overline{\chi}_U$. However, $\mathrm{Re}(-\chi_U)$ and $\mathrm{Re}(-\overline{\chi}_U)$ are both negative as the Stokes line is crossed, and hence these singulants cannot produce any Stokes switching on the free surface.
}
\label{1:realrays2}
\end{figure}
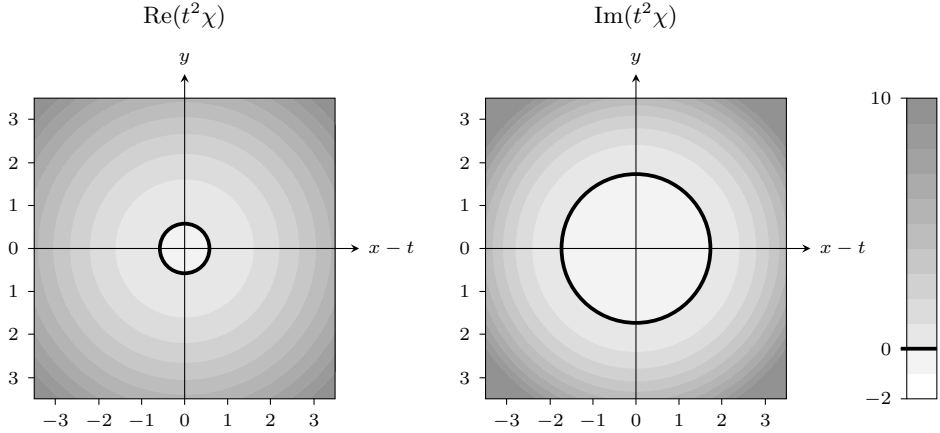

Specifically, we note that a second singularity is present in the unsteady problem, introduced in the unsteady second-order terms. As in \cite{Chapman5} and \cite{Lustri4}, we require that all characteristics pass through the disturbance located at $x^2 + y^2 + (z\pm h)^2 =0$ when $t=0$. As in \cite{Lustri4}, we observe that this singularity corresponds to the instantaneous initial change introduced into the flow at $t=0$.   Hence, we apply the boundary conditions
\begin{equation}
 \label{2:sscauchy}x_0 = s, \qquad y_0 = \pm\mathrm{i}\sqrt{s^2+h^2},\qquad \chi_0 = 0,\qquad t = 0.
\end{equation}
The singulant equation (\ref{2:eik}) may again be solved using Charpit's method, however the analysis is simpler if we note that the solution may be expressed in a reduced set of coordinates
\begin{equation}
\tau = t,\qquad \rho = \sqrt{(x-t)^2 + y^2},
\end{equation}
implying that the solution is radially symmetric about the propagating point $x = t$. This is consistent with the boundary data and reduces the singulant equation (\ref{2:eik}) to
\begin{equation}
\label{2:eik1}\chi_{\tau}^4 + \chi_p^6 = 0,
\end{equation}
with the boundary conditions becoming
\begin{equation}
\label{2:cauchy}\tau_0 = 0,\qquad \rho_0 = \pm\mathrm{i} h,\qquad \chi_0 = 0.
\end{equation}
Solving this much simpler equation using Charpit's method gives four nonzero solutions, which take the form
\begin{equation}
\label{2:movingchi}\chi = \pm \frac{4\mathrm{i}(\rho\pm\mathrm{i} h)^3}{27\tau^2},
\end{equation}
where the signs may be chosen independently. We will refer to the solution with the first sign being positive and the second being negative as $\chi_U$, and hence the remaining possible singulant expressions are given by $-\chi_U$ and $\pm\overline{\chi}_U$. We illustrate this singulant behaviour in figure \ref{1:realrays2}. Importantly, we see from (\ref{2:movingchi}) that 
\begin{equation}
\mathrm{Re}(\chi_U) = 0 \quad \mathrm{on} \quad \rho = \frac{h}{\sqrt{3}},\qquad \qquad\mathrm{Im}(\chi_U) = 0 \quad \mathrm{on}\quad \rho = \sqrt{3}h.
\end{equation}
The first of these conditions describes the location of anti-Stokes lines, while the second describes the location of Stokes lines. These may be seen clearly in figure \ref{1:realrays2}, where the anti-Stokes and Stokes lines are described by concentric circles about $x = t$. Importantly, the anti-Stokes lines are always contained within the Stokes lines, meaning that any waves contained within the Stokes line circle will produce exponentially large behaviour on the free surface. Consequently, we conclude the free surface can only contain wave behaviour outside the Stokes lines.

Furthermore, we see that only $\chi_U$ and $\overline{\chi}_U$ have $\mathrm{Re}(\chi) > 0$ as the Stokes line is crossed. Therefore, it is only these contributions that will be switched across the Stokes lines. Hence, inside the Stokes line, there are no exponentially small free-surface waves associated with the unsteady contribution, but as the Stokes line is crossed, waves associated with $\chi_U$ and $\overline{\chi}_U$ will be switched on.

\subsection{Stokes line interactions}

We have shown that there are two sets of Stokes line behaviours on the free surface, associated with $\chi_{S}$, $\chi_{U}$ and their complex conjugate expressions, across which the leading-order behaviour switched on exponentially small contributions to the free surface behaviour. However, to fully describe the free surface behaviour, we must consider Stokes lines caused by the interaction between $\chi_{S}$ and $\chi_U$, as well as the interaction between $\overline{\chi}_S$ and $\overline{\chi}_U$. In this section, we will restrict our attention to $\chi_{S}$ and $\chi_U$, noting that the same switching behaviour will be demonstrated by the complex conjugate expressions.

In previous analyses of the Stokes structure of partial differential equations \cite{Chapman4, Howls1}, it was found that Stokes switching may also occur when one exponentially subdominant contribution switches on a further subdominant contribution. Hence, we find that Stokes switching also occurs on curves satisfying $\mathrm{Im}(\chi_{S}) = \mathrm{Im}(\chi_{U})$ and $\mathrm{Re}(\chi_{U}) < \mathrm{Re}(\chi_{S})$, across which the capillary wave wave behaviour associated with $\chi_{U}$ is switched on. 

\begin{figure}
\centering

\begin{tikzpicture}
[xscale=0.65,>=stealth,yscale= 0.65]

%\draw[white] (-9,-8) -- (10,-8) -- (10,7) -- (-9,7) -- cycle; 

\draw (-6.5,7) -- (-6.5,-7) -- (7.5,-7) -- (7.5,7) -- cycle;
\draw[gray,->] (-6.5,0) -- (8,0) node[right,black] {\scriptsize{$x$}};
\draw[gray,->] (0,-7) -- (0,7.5) node[above,black] {\scriptsize{$y$}};

%\draw[line width = 0.4mm, red, dashed] plot[smooth] file {SL_Unsteady5A};

\draw[line width=0.75mm,gray,opacity=0.7] plot[smooth] file {SL_Unsteady1.txt};

\draw[line width=0.4mm,dashed] plot[smooth] file {SL_Unsteady3.txt};
\draw[line width=0.4mm,dashed] plot[smooth] file {SL_Unsteady6.txt};
\draw[line width=0.4mm] plot[smooth] file {SL_Unsteady4.txt};

\draw[line width=0.75mm,black]  plot[smooth] file {SL_Unsteady2.txt};

\draw[line width=0.75mm,gray,opacity=0.7]  (8,6) -- (8.75,6) node[black,right,opacity=1] {\scriptsize{$\mathrm{Im}(\chi_S) = 0$}};
%\node at (8.75,5.5) [black,right] {\scriptsize{Line}};
\draw[ line width=0.75mm]  (8,4.5) -- (8.75,4.5) node[black,right] {\scriptsize{$\mathrm{Im}(\chi_U)=0$}};
%\node at (8.75,4) [right] {\scriptsize{Line}};
\draw[line width=0.4mm]  (8,3) -- (8.75,3) node[black,right] {\scriptsize{$\mathrm{Im}(\chi_U)=\mathrm{Im}(\chi_S)$}};
\node at (8.75,2.25) [right] {\scriptsize{$\mathrm{Re}(\chi_U)>\mathrm{Re}(\chi_S)$}};
%\node at (8.75,2) [right] {\scriptsize{(Active)}};
\draw[dashed,line width=0.4mm]  (8,-2.5) -- (8.75,-2.5) node[black,right]  {\scriptsize{$\mathrm{Im}(\chi_U)=\mathrm{Im}(\chi_S)$}};
\node at (8.75,-3.25) [right]  {\scriptsize{$\mathrm{Re}(\chi_S)>\mathrm{Re}(\chi_U)$}};

\filldraw (8.375,-4.5) circle (0.15);
\node at (8.75,-4.5) [right] {\scriptsize{Stokes Crossing}};
\node at (8.75,-5) [right] {\scriptsize{Point (Active)}};
\draw[line width=0.4mm] (8.375,-6) circle (0.15);
\node at (8.75,-6) [right] {\scriptsize{Stokes Crossing}};
\node at (8.75,-6.5) [right] {\scriptsize{Point (Inactive)}};

\filldraw (4.5,1.65) circle (0.15);
\filldraw (4.5,-1.65) circle (0.15);
\filldraw[white] (5.4,-1.675) circle (0.15);
\draw[line width=0.4mm] (5.4,-1.675) circle (0.15);
\filldraw[white] (5.4,1.675) circle (0.15);
\draw[line width=0.4mm] (5.4,1.675) circle (0.15);

\draw (5,-7) node[below] {\scriptsize{$5$}} -- (5,-6.8);
\draw (-5,-7) node[below] {\scriptsize{$-5$}} -- (-5,-6.8);
\draw (-6.5,5) node [left] {\scriptsize{$5$}} -- (-6.3,5);
\draw (-6.5,-5) node [left] {\scriptsize{$-5$}} -- (-6.3,-5);
\node at (-6.5,0) [left] {\scriptsize{$0$}};
\node at (0,-7) [below] {\scriptsize{$0$}};

\end{tikzpicture}

\caption{Relevant Stokes and higher-order Stokes lines present on the free surface for $t=5$ and $h=1$. The wide gray and black curves are the Stokes lines associated with the steady and unsteady contributions (satisfying $\mathrm{Im}(\chi_{S}) = 0$ and $\mathrm{Im}(\chi_U) = 0$) respectively. The solid narrow curve is a Stokes line across which the unsteady contribution switches the steady contribution (satisfying $\mathrm{Im}(\chi_{S}) = \mathrm{Im}(\chi_U)$) and $\mathrm{Re}(\chi_U) > \mathrm{Re}(\chi_{S})$). The dashed narrow curves are Stokes lines across which the steady contribution would switch the unsteady contribution, however these contributions are inactive. The filled circles are Stokes crossing points, at which Stokes lines terminate. The empty circles are potential Stokes crossing points where at least one of the contributions is inactive, and therefore nothing occurs. }
\label{2:movingsl}
\end{figure}
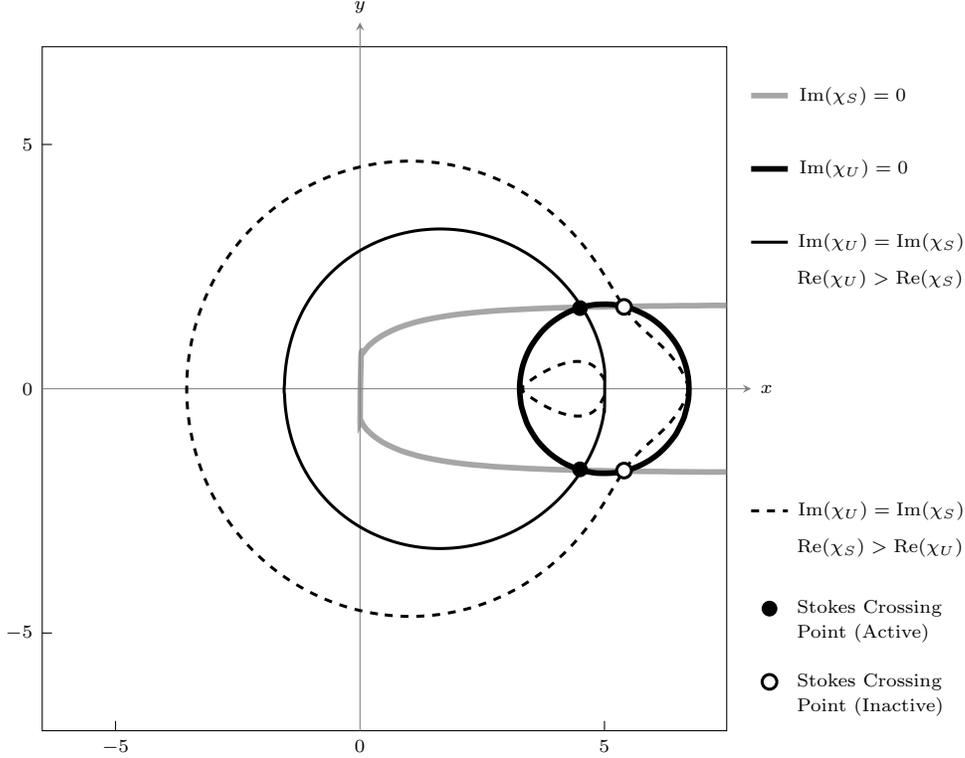

Consequently, the complete Stokes structure contains three sets of equal phase lines, which are illustrated in figure \ref{2:movingsl} for $t=5$ and $h=1$, although the equal phase line following $y=0$ has been omitted, as it was established in section \ref{STEADY} to be inactive. We have also illustrated the anti-Stokes line along which $\mathrm{Re}(\chi_{S}) = \mathrm{Re}(\chi_U)$. 

\begin{figure}
\centering

\subfloat[$t=5$]{
\centering
\begin{tikzpicture}
[xscale=0.35,>=stealth,yscale= 0.35]

\fill[opacity=0.05]  (-6.5+1.5,7) -- (-6.5+1.5,-7) -- (7.5+1.5,-7) -- (7.5+1.5,7) -- cycle;
\fill[white]  plot[smooth] file {SL_Unsteady2.txt};

\begin{scope}
\clip plot[smooth] (-6.5+1.5,7) -- (4.535,7) -- (4.535,-7) -- (-6.5+1.5,-7) -- cycle;
\clip plot[smooth] file {SL_Unsteady1.txt} -- (7.5+1.5,-7) -- (-6+1.5,-7) -- (-6+1.5,7) -- (7.5+1.5,7) -- cycle;
\fill[opacity=0.2]  plot[smooth] file {SL_Unsteady4.txt};
\end{scope}

\draw (-6.5+1.5,7) -- (-6.5+1.5,-7) -- (7.5+1.5,-7) -- (7.5+1.5,7) -- cycle;
\draw[gray] (-6.5+1.5,0) -- (7.5+1.5,0);
\draw[gray] (0,-7) -- (0,7);

\begin{scope}
\clip plot[smooth] file {SL_Unsteady1.txt} -- (7.5+1.5,-7) -- (-6+1.5,-7) -- (-6+1.5,7) -- (7.5+1.5,7) -- cycle;
\draw plot[smooth] file {SL_Unsteady4.txt};
\end{scope}

\begin{scope}
\clip plot[smooth] (-6.5+1.5,7) -- (4.535,7) -- (4.535,-7) -- (-6.5+1.5,-7) -- cycle;
\draw[line width=0.4mm] plot[smooth] file {SL_Unsteady1.txt};
\end{scope}

\draw[line width=0.4mm,dashed]  plot[smooth] file {SL_Unsteady2.txt};

\draw(0,-7) -- (0,-7.2) node[below] {\scriptsize{0}};
\draw(2,-7) -- (2,-7.2) node[below] {\scriptsize{2}};
\draw(4,-7) -- (4,-7.2) node[below] {\scriptsize{4}};
\draw(6,-7) -- (6,-7.2) node[below] {\scriptsize{6}};
\draw(0,-7) -- (0,-7.2) node[below] {\scriptsize{0}};
\draw(-2,-7) -- (-2,-7.2) node[below] {\scriptsize{-2}};
\draw(-4,-7) -- (-4,-7.2) node[below] {\scriptsize{-4}};
\draw(8,-7) -- (8,-7.2) node[below] {\scriptsize{8}};

\draw(-6.5+1.5,0) -- (-6.7+1.5,0) node[left] {\scriptsize{0}};
\draw(-6.5+1.5,2) -- (-6.7+1.5,2) node[left] {\scriptsize{2}};
\draw(-6.5+1.5,4) -- (-6.7+1.5,4) node[left] {\scriptsize{4}};
\draw(-6.5+1.5,6) -- (-6.7+1.5,6) node[left] {\scriptsize{6}};
\draw(-6.5+1.5,-2) -- (-6.7+1.5,-2) node[left] {\scriptsize{-2}};
\draw(-6.5+1.5,-4) -- (-6.7+1.5,-4) node[left] {\scriptsize{-4}};
\draw(-6.5+1.5,-6) -- (-6.7+1.5,-6) node[left] {\scriptsize{-6}};

\draw[->] (7.5+1.5,0) -- (8+1.5,0) node[right]{\scriptsize{$x$}};
\draw[->] (0,7) -- (0,7.5) node[above]{\scriptsize{$y$}};

\draw[line width=0.4mm,->] (6,1) -- (7,1);
\draw[line width=0.4mm,->] (6,-1) -- (7,-1);
\draw[line width=0.4mm,->] (2.2,2.9) -- (2.5,3.7);
\draw[line width=0.4mm,->] (2.2,-2.9) -- (2.5,-3.7);
\draw[line width=0.4mm,->] (-1+0.1,1.5-0.2) -- (-1.8+0.1,2-0.2);
\draw[line width=0.4mm,->] (-1+0.1,-1.5+0.2) -- (-1.8+0.1,-2+0.2);

\filldraw (4.535,1.67) circle (0.1);
%\filldraw (5.395,1.686) circle (0.1);
\filldraw (4.535,-1.67) circle (0.1);
%\filldraw (5.395,-1.686) circle (0.1);

\draw[line width=0.4mm,dashed] (-5.25+0.5,-8.5) -- (-6+0.5,-8.5) -- (-6+0.5,-9.25) -- (-5.25+0.5,-9.25) -- cycle;
\node at (-5+0.5,-8.825) [right] {\scriptsize{No waves}};
\draw (-5.25+0.5,-8.5-1-0.5) -- (-6+0.5,-8.5-1-0.5) -- (-6+0.5,-9.25-1-0.5) -- (-5.25+0.5,-9.25-1-0.5) -- cycle;
\fill[opacity=0.05] (-5.25+0.5,-8.5-1-0.5) -- (-6+0.5,-8.5-1-0.5) -- (-6+0.5,-9.25-1-0.5) -- (-5.2+0.5,-9.25-1-0.5) -- cycle;
\node at (-5+0.5,-8.825-1-0.5) [right] {\scriptsize{Unsteady ripple}};

\fill[opacity=0.2] (-0.25+1+1.5,-8.5) -- (0.5+1+1.5,-8.5) -- (0.5+1+1.5,-9.25) -- (-0.25+1+1.5,-9.25) -- cycle;
\draw[line width=0.4mm] (-0.25+1+1.5,-8.5) -- (0.5+1+1.5,-8.5) -- (0.5+1+1.5,-9.25) -- (-0.25+1+1.5,-9.25) -- cycle;
\node at (0.525+1+1.5,-8.825) [right] {\scriptsize{Steady waves and}};
\node at (1+1+1.5,-8.825-0.85) [right] {\scriptsize{unsteady ripple}};

\end{tikzpicture}
}

\subfloat[$t=3$]{
\centering
\begin{tikzpicture}
[xscale=0.3,>=stealth,yscale= 0.3]

\draw (-3,4) -- (-3,-4) -- (5,-4) -- (5,4) -- cycle;

\fill[opacity=0.05]  (-3,4) -- (-3,-4) -- (5,-4) -- (5,4)-- cycle;
\fill[white]  plot[smooth] file {SL_Unsteady2_t3.txt};

\begin{scope}
\clip plot[smooth] (-3,4) -- (2.2,4) -- (2.2,-4) -- (-3,-4) -- cycle;
\clip plot[smooth] file {SL_Unsteady1_t3.txt} -- (5,-4) -- (-3,-4) -- (-3,4) -- (5,4) -- cycle;
\fill[opacity=0.2]  plot[smooth] file {SL_Unsteady3_t3.txt};
\end{scope}
\draw[gray] (0,-4) -- (0,4);
\draw[gray] (-3,0) -- (5,0);
\begin{scope}
\clip plot[smooth] file {SL_Unsteady1_t3.txt} -- (5,-4) -- (-3,-4) -- (-3,4) -- (5,4) -- cycle;
\draw plot[smooth] file {SL_Unsteady3_t3.txt};
\end{scope}

\begin{scope}
\clip plot[smooth] (-3,4) -- (2.2,4) -- (2.2,-4) -- (-3,-4) -- cycle;
\draw[line width=0.4mm] plot[smooth] file {SL_Unsteady1_t3.txt};
\end{scope}

\draw[line width=0.4mm,dotted]  plot[smooth] file {SL_Unsteady2_t3.txt};

\draw[->] (5,0) -- (5.6,0) node[right] {\scriptsize{$x$}};
\draw[->] (0,4) -- (0,4.6) node[above] {\scriptsize{$y$}};
\draw (0,-4) -- (0,-4.2) node[below] {\scriptsize{$0$}};
\draw (2,-4) -- (2,-4.2) node[below] {\scriptsize{$2$}};
\draw (4,-4) -- (4,-4.2) node[below] {\scriptsize{$4$}};
\draw (-2,-4) -- (-2,-4.2) node[below] {\scriptsize{$-2$}};
\draw (-3,0) -- (-3.2,0) node[left] {\scriptsize{$0$}};
\draw (-3,2) -- (-3.2,2) node[left] {\scriptsize{$2$}};
\draw (-3,-2) -- (-3.2,-2) node[left] {\scriptsize{$-2$}};
\draw (-3,4) -- (-3.2,4) node[left] {\scriptsize{$4$}};
\draw (-3,-4) -- (-3.2,-4) node[left] {\scriptsize{$-4$}};

\end{tikzpicture}
}
\subfloat[$t=10$]{
\centering
\begin{tikzpicture}
[xscale=0.12,>=stealth,yscale= 0.12]

\draw (-7.5,10) -- (-7.5,-10) -- (12.5,-10) -- (12.5,10) -- cycle;

\fill[opacity=0.05]  (-7.5,10) -- (-7.5,-10) -- (12.5,-10) -- (12.5,10) -- cycle;
\fill[white]  plot[smooth] file {SL_Unsteady2_t10.txt};

\begin{scope}
\clip plot[smooth] (-7.5,10) -- (9.8,10) -- (9.8,-10) -- (-7.5,-10) -- cycle;
\clip plot[smooth] file {SL_Unsteady1_t10.txt} -- (12.5,10) -- (-7.5,10) -- (-7.5,-10) -- (12.5,-10) -- cycle;
\fill[opacity=0.2]  plot[smooth] file {SL_Unsteady3_t10.txt};
\end{scope}
\draw[gray] (0,-10) -- (0,10);
\draw[gray] (-7.5,0) -- (12.5,0);
\begin{scope}
\clip plot[smooth] (-7.5,10) -- (9.8,10) -- (9.8,-10) -- (-7.5,-10) -- cycle;
\clip plot[smooth] file {SL_Unsteady1_t10.txt} -- (12.5,10) -- (-7.5,10) -- (-7.5,-10) -- (12.5,-10) -- cycle;
\draw plot[smooth] file {SL_Unsteady3_t10.txt};
\end{scope}

\begin{scope}
\clip plot[smooth] (-7.5,10) -- (9.5,10) -- (9.5,-10) -- (-7.5,-10)-- cycle;
\draw[line width=0.4mm] plot[smooth] file {SL_Unsteady1_t10.txt};
\end{scope}

\draw[line width=0.4mm,dotted]  plot[smooth] file {SL_Unsteady2_t10.txt};

\draw[->] (12.5,0) -- (14,0) node[right] {\scriptsize{$x$}};
\draw[->] (0,10) -- (0,11.5) node[above] {\scriptsize{$y$}};
\draw (0,-10) -- (0,-10.5) node[below] {\scriptsize{$0$}};
\draw (5,-10) -- (5,-10.5) node[below] {\scriptsize{$5$}};
\draw (10,-10) -- (10,-10.5) node[below] {\scriptsize{$10$}};
\draw (-5,-10) -- (-5,-10.5) node[below] {\scriptsize{$-5$}};
\draw (-7.5,0) -- (-8,0) node[left] {\scriptsize{$0$}};
\draw (-7.5,5) -- (-8,5) node[left] {\scriptsize{$5$}};
\draw (-7.5,-5) -- (-8,-5) node[left] {\scriptsize{$-5$}};
\draw (-7.5,10) -- (-8,10) node[left] {\scriptsize{$10$}};
\draw (-7.5,-10) -- (-8,-10) node[left] {\scriptsize{$-10$}};

\end{tikzpicture}
}
\subfloat[$t=20$]{
\centering
\begin{tikzpicture}
[xscale=0.06,>=stealth,yscale= 0.06]

\draw (-15,20) -- (-15,-20) -- (25,-20) -- (25,20) -- cycle;

\fill[opacity=0.05]  (-15,20) -- (-15,-20) -- (25,-20) -- (25,20) -- cycle;
\fill[white]  plot[smooth] file {SL_Unsteady2_t20.txt};

\draw[gray] (0,-20) -- (0,20);
\draw[gray] (-15,0) -- (25,0);

\begin{scope}
\clip plot[smooth] (-15,20) -- (20,20) -- (20,-20) -- (-15,-20) -- cycle;
\clip plot[smooth] file {SL_Unsteady1_t20.txt} -- (25,20) -- (-15,20) -- (-15,-20) -- (25,-20) -- cycle;
\fill[opacity=0.2]  plot[smooth] file {SL_Unsteady3_t20.txt};
\end{scope}
\draw[gray] (0,-20) -- (0,20);
\draw[gray] (-15,0) -- (25,0);

\begin{scope}
\clip plot[smooth] (-15,20) -- (20,20) -- (20,-20) -- (-15,-20) -- cycle;
\clip plot[smooth] file {SL_Unsteady1_t20.txt} -- (25,20) -- (-15,20) -- (-15,-20) -- (25,-20) -- cycle;
\draw plot[smooth] file {SL_Unsteady3_t20.txt};
\end{scope}

\begin{scope}
\clip plot[smooth] (-15,20) -- (20,20) -- (20,-20) -- (-15,-20)-- cycle;
\draw[line width=0.4mm] plot[smooth] file {SL_Unsteady1_t20.txt};
\end{scope}

\draw[line width=0.4mm,dotted]  plot[smooth] file {SL_Unsteady2_t20.txt};

\draw[->] (25,0) -- (28,0) node[right] {\scriptsize{$x$}};
\draw[->] (0,20) -- (0,23) node[above] {\scriptsize{$y$}};
\draw (0,-20) -- (0,-21) node[below] {\scriptsize{$0$}};
\draw (10,-20) -- (10,-21) node[below] {\scriptsize{$10$}};
\draw (20,-20) -- (20,-21) node[below] {\scriptsize{$20$}};
\draw (-10,-20) -- (-10,-21) node[below] {\scriptsize{$-10$}};
\draw (-15,0) -- (-16,0) node[left] {\scriptsize{$0$}};
\draw (-15,10) -- (-16,10) node[left] {\scriptsize{$10$}};
\draw (-15,-10) -- (-16,-10) node[left] {\scriptsize{$-10$}};
\draw (-15,20) -- (-16,20) node[left] {\scriptsize{$20$}};
\draw (-15,-20) -- (-16,-20) node[left] {\scriptsize{$-20$}};

\end{tikzpicture}
}

\caption{Regions of the free surface containing exponentially small contributions to the surface behaviour for $h=1$ and $t = 3$, $5$, $10$ and $20$. In the white region, no contributions are present. In the light grey region, the expanding ripple centred about $x=t$ is present, while in the dark grey region, both the unsteady ripple and the steady waves are present on the free surface. The arrows in (a) indicate the direction in which the Stokes lines move over time. The waveless region has constant radius of $\sqrt{3}h$ and is present downstream from the obstacle. The region containing steady waves tends to an expanding circular region with a narrow section removed.
}
\label{2:waveregions}
\end{figure}
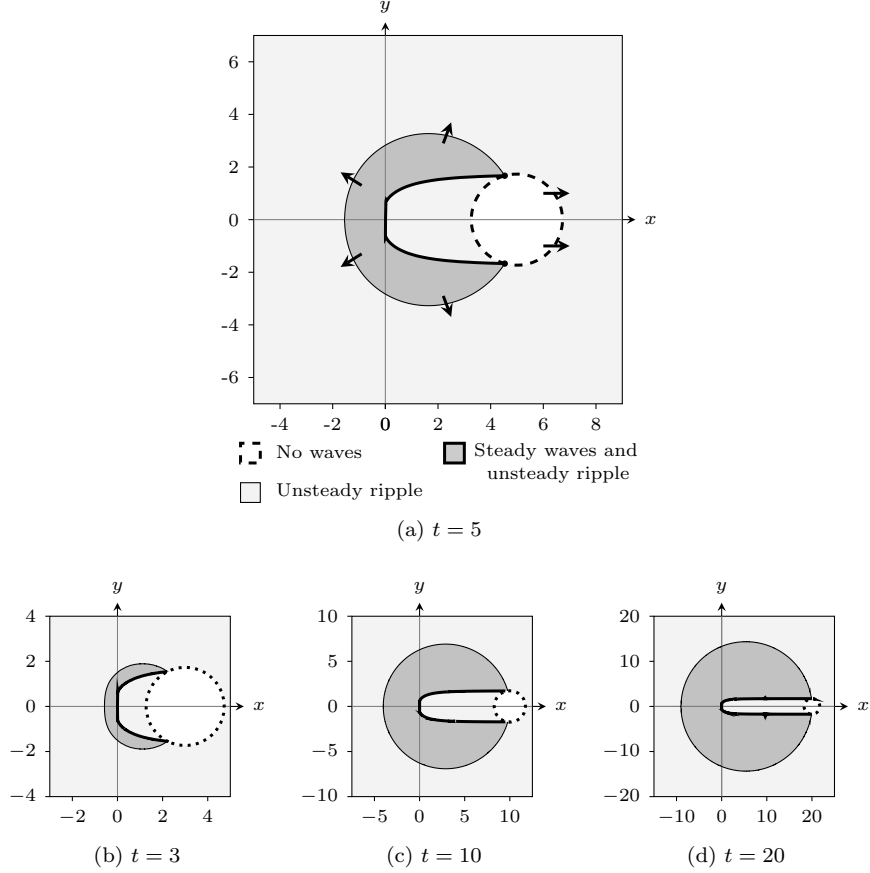

It is not possible, however, for all three sets of potential Stokes
lines to be active throughout the domain. We recall from
\cite{Howls1}, and \cite{Chapman4} that Stokes lines may become
inactive as they cross higher order Stokes lines, which originate at
Stokes crossing points (SCP). A Stokes crossing point is found when
three different Stokes lines intersect at a single point, and is
represented in figure \ref{2:movingsl} as a black circle. Noting that
Stokes lines can terminate only at Stokes crossing points, we
determine that the region in which the unsteady ripple and steady
waves are present are those indicated in figure \ref{2:waveregions},
again for  $h=1$, over a range of times.  

In this figure, we see that dashed curve satisfying
$\mathrm{Im}(\chi_{S}) = \mathrm{Im}(\chi_U)$ does not contribute to
the free surface behaviour. This is because the steady wave
contribution would exponentially dominate (and therefore switch) the
unsteady ripple across this curve. However, the capillary wave
contribution is switched off along the inner curve, and therefore is
not present in this region, and therefore no Stokes switching
occurs. We therefore find that the free surface wave behaviour
consists of an unsteady ripple present outside a circular region of
growing radius, and an expanding region containing steady waves that
spreads outwards from $(x,y) = (0,0)$. As $t \rightarrow \infty$, the
radius of this expanding region will become infinite, and the steady
wave behaviour will be equivalent to that obtained for the steady
problem in section \ref{STEADY}. 

By solving 
\begin{equation}
\mathrm{Im}(\chi_S) = \mathrm{Im}(\chi_U),
\end{equation}
we can determine the position of the expanding capillary wavefront. This becomes
\begin{equation}
x = \frac{4 ((t - x)^2-3h^2) \sqrt{(t - x)^2}}{27 t^2}.
\end{equation}
We recall that on $y = 0$, the singulant is given by $\chi_S = h \pm \mathrm{i} x$ for $x < 0$. If we define a moving frame $\eta = x + t/2$, and equate $\mathrm{Im}(\chi_S)$ with the imaginary part of the corresponding unsteady singulant from \eqref{2:movingchi}, we find that
\begin{equation}
 %- \frac{t}{2} +\eta = -\frac{t}{2} + \eta + \frac{2(h^2 - \eta^2)}{3t} + \mathcal{O}(t^{-2})\qquad \mathrm{as} \qquad t \rightarrow \infty,
0 = \frac{2(h^2 - \eta^2)}{3t} + \mathcal{O}(t^{-2})\qquad \mathrm{as} \qquad t \rightarrow \infty,
\end{equation}
Matching this expression at $\mathcal{O}(t^{-1})$ as $t \rightarrow \infty$ gives the boundary of the expanding capillary wave region on $y = 0$ as one of $\eta = h$ or $\eta = -h$ in this limit. We see from figure \ref{2:movingsl} that the active Stokes line is located at the interior of these two points, and therefore that the front position tends to $\eta \rightarrow h$ as $t \rightarrow \infty$, or
\begin{equation}
x \sim -t/2 + h\qquad as \qquad t \rightarrow \infty.
\end{equation}

\subsection{Results and Comparison}

\begin{figure}
\centering

\subfloat[$t = 10$]{
\begin{tikzpicture}
[xscale=0.125,>=stealth,yscale= 0.25]
\draw[line width=0.5mm,gray,opacity=0.7] (-4.022,-5) -- (-4.022,5);
\draw[gray](-25,5) -- (-25,-5) -- (0,-5) -- (0,5) -- cycle;
\draw[gray,dotted] (-20,-5) node[below,black] {\scriptsize{$-20$}} -- (-20,5);
\draw[gray,dotted] (-15,-5) node[below,black] {\scriptsize{$-15$}} -- (-15,5);
\draw[gray,dotted] (-10,-5) node[below,black] {\scriptsize{$-10$}} -- (-10,5);
\draw[gray,dotted] (-5,-5) node[below,black] {\scriptsize{$-5$}} -- (-5,5);
\node at (0,-5) [below ] {\scriptsize{$0$}}; 
\node at (-25,-5) [below ] {\scriptsize{$-25$}}; 
\draw[gray] (-25,0) node[left,black] {\scriptsize{$0$}}-- (0,0);
\node at (-25.5,5) [ left] {\scriptsize{$5\times10^{-4}$}};
\node at (-25.5,-5) [ left] {\scriptsize{$-5\times10^{-4}$}};
\draw[gray] (-25,5) -- (-26,5);
\draw [gray](-25,-5) -- (-26,-5);
\begin{scope}
\clip[] (-25,5) -- (-25,-5) -- (0,-5) -- (0,5) -- cycle;
\draw[-] plot[smooth] file {unsteady_t_10_eps_0p15_m1000.txt};
\end{scope}
\draw[->] (0,0) -- (1.5,0) node[right] {\scriptsize{$x$}};
\draw[->] (-25,5) -- (-25,5.75) node[above] {\scriptsize{$\xi-\xi_0$}};
\node at (-4.022,5) [above] {\scriptsize{$-4.022$}};
\end{tikzpicture}
}
\subfloat[$t = 20$]{
\begin{tikzpicture}
[xscale=0.125,>=stealth,yscale= 0.25]
\draw[line width=0.5mm,gray,opacity=0.7] (-9.011,-5) -- (-9.011,5);
\draw[gray](-25,5) -- (-25,-5) -- (0,-5) -- (0,5) -- cycle;
\draw[gray,dotted] (-20,-5) node[below,black] {\scriptsize{$-20$}} -- (-20,5);
\draw[gray,dotted] (-15,-5) node[below,black] {\scriptsize{$-15$}} -- (-15,5);
\draw[gray,dotted] (-10,-5) node[below,black] {\scriptsize{$-10$}} -- (-10,5);
\draw[gray,dotted] (-5,-5) node[below,black] {\scriptsize{$-5$}} -- (-5,5);
\node at (0,-5) [below ] {\scriptsize{$0$}}; 
\node at (-25,-5) [below ] {\scriptsize{$-25$}}; 
\draw[gray] (-25,0) -- (0,0);
\begin{scope}
\clip[] (-25,5) -- (-25,-5) -- (0,-5) -- (0,5) -- cycle;
\draw[-] plot[smooth] file {unsteady_t_20_eps_0p15_m1000.txt};
\end{scope}
\node at (-9.011,5) [above] {\scriptsize{$-9.011$}};
\end{tikzpicture}
}
\subfloat[$t = 30$]{
\begin{tikzpicture}
[xscale=0.125,>=stealth,yscale= 0.25]
\draw[line width=0.5mm,gray,opacity=0.7] (-14.001,-5) -- (-14.001,5);
\draw[gray](-25,5) -- (-25,-5) -- (0,-5) -- (0,5) -- cycle;
\draw[gray,dotted] (-20,-5) node[below,black] {\scriptsize{$-20$}} -- (-20,5);
\draw[gray,dotted] (-15,-5) node[below,black] {\scriptsize{$-15$}} -- (-15,5);
\draw[gray,dotted] (-10,-5) node[below,black] {\scriptsize{$-10$}} -- (-10,5);
\draw[gray,dotted] (-5,-5) node[below,black] {\scriptsize{$-5$}} -- (-5,5);
\node at (0,-5) [below ] {\scriptsize{$0$}}; 
\node at (-25,-5) [below ] {\scriptsize{$-25$}}; 
\draw[gray] (-25,0)-- (0,0);
\begin{scope}
\clip[] (-25,5) -- (-25,-5) -- (0,-5) -- (0,5) -- cycle;
\draw[-] plot[smooth] file {unsteady_t_30_eps_0p15_m1000.txt};
\end{scope}
\draw[gray](-25,5) -- (-25,-5) -- (0,-5) -- (0,5) -- cycle;
\node at (-14.001,5) [above] {\scriptsize{$-14.001$}};
\end{tikzpicture}
}

\subfloat[$t = 40$]{
\begin{tikzpicture}
[xscale=0.125,>=stealth,yscale= 0.25]
\draw[line width=0.5mm,gray,opacity=0.7] (-19,-5) -- (-19,5);
\draw[gray](-25,5) -- (-25,-5) -- (0,-5) -- (0,5) -- cycle;
\draw[gray,dotted] (-20,-5) node[below,black] {\scriptsize{$-20$}} -- (-20,5);
\draw[gray,dotted] (-15,-5) node[below,black] {\scriptsize{$-15$}} -- (-15,5);
\draw[gray,dotted] (-10,-5) node[below,black] {\scriptsize{$-10$}} -- (-10,5);
\draw[gray,dotted] (-5,-5) node[below,black] {\scriptsize{$-5$}} -- (-5,5);
\node at (0,-5) [below ] {\scriptsize{$0$}}; 
\node at (-25,-5) [below ] {\scriptsize{$-25$}}; 
\draw[gray] (-25,0) node[left,black] {\scriptsize{$0$}}-- (0,0);
\node at (-25.5,5) [ left] {\scriptsize{$5\times10^{-4}$}};
\node at (-25.5,-5) [ left] {\scriptsize{$-5\times10^{-4}$}};
\draw[gray] (-25,5) -- (-26,5);
\draw [gray](-25,-5) -- (-26,-5);
\begin{scope}
\clip[] (-25,5) -- (-25,-5) -- (0,-5) -- (0,5) -- cycle;
\draw[-] plot[smooth] file {unsteady_t_40_eps_0p15_m1000.txt};
\end{scope}
\draw[->] (0,0) -- (1.5,0) node[right] {\scriptsize{$x$}};
\draw[->] (-25,5) -- (-25,5.75) node[above] {\scriptsize{$\xi-\xi_0$}};
\node at (-17.000,5) [above] {\scriptsize{$-19.000$}};
\end{tikzpicture}
}
\subfloat[$t = 50$]{
\begin{tikzpicture}
[xscale=0.125,>=stealth,yscale= 0.25]
\draw[line width=0.5mm,gray,opacity=0.7] (-24,-5) -- (-24,5);
\draw[gray](-25,5) -- (-25,-5) -- (0,-5) -- (0,5) -- cycle;
\draw[gray,dotted] (-20,-5) node[below,black] {\scriptsize{$-20$}} -- (-20,5);
\draw[gray,dotted] (-15,-5) node[below,black] {\scriptsize{$-15$}} -- (-15,5);
\draw[gray,dotted] (-10,-5) node[below,black] {\scriptsize{$-10$}} -- (-10,5);
\draw[gray,dotted] (-5,-5) node[below,black] {\scriptsize{$-5$}} -- (-5,5);
\node at (0,-5) [below ] {\scriptsize{$0$}}; 
\node at (-25,-5) [below ] {\scriptsize{$-25$}}; 
\draw[gray] (-25,0) -- (0,0);
\begin{scope}
\clip[] (-25,5) -- (-25,-5) -- (0,-5) -- (0,5) -- cycle;
\draw[-] plot[smooth] file {unsteady_t_50_eps_0p15_m1000.txt};
\end{scope}
\node at (-22,5) [above] {\scriptsize{$-24.000$}};
\end{tikzpicture}
}
\subfloat[$t = 60$]{
\begin{tikzpicture}
[xscale=0.125,>=stealth,yscale= 0.25]
\draw[gray](-25,5) -- (-25,-5) -- (0,-5) -- (0,5) -- cycle;
\draw[gray,dotted] (-20,-5) node[below,black] {\scriptsize{$-20$}} -- (-20,5);
\draw[gray,dotted] (-15,-5) node[below,black] {\scriptsize{$-15$}} -- (-15,5);
\draw[gray,dotted] (-10,-5) node[below,black] {\scriptsize{$-10$}} -- (-10,5);
\draw[gray,dotted] (-5,-5) node[below,black] {\scriptsize{$-5$}} -- (-5,5);
\node at (0,-5) [below ] {\scriptsize{$0$}}; 
\node at (-25,-5) [below ] {\scriptsize{$-25$}}; 
\draw[gray] (-25,0) -- (0,0);
\begin{scope}
\clip[] (-25,5) -- (-25,-5) -- (0,-5) -- (0,5) -- cycle;
\draw[-] plot[smooth] file {unsteady_t_60_eps_0p15_m1000.txt};
\end{scope}
\end{tikzpicture}
}

\caption{$\epsilon = 0.15$. The $x$ position of the Stokes line is marked by a vertical gray stripe. This is where waves should be half amplitude, decaying exponentially as this line is crossed. The axes are only shown on the first figure of each row, but are identical for each figure.}
\label{2:unsteady_num}
\end{figure}
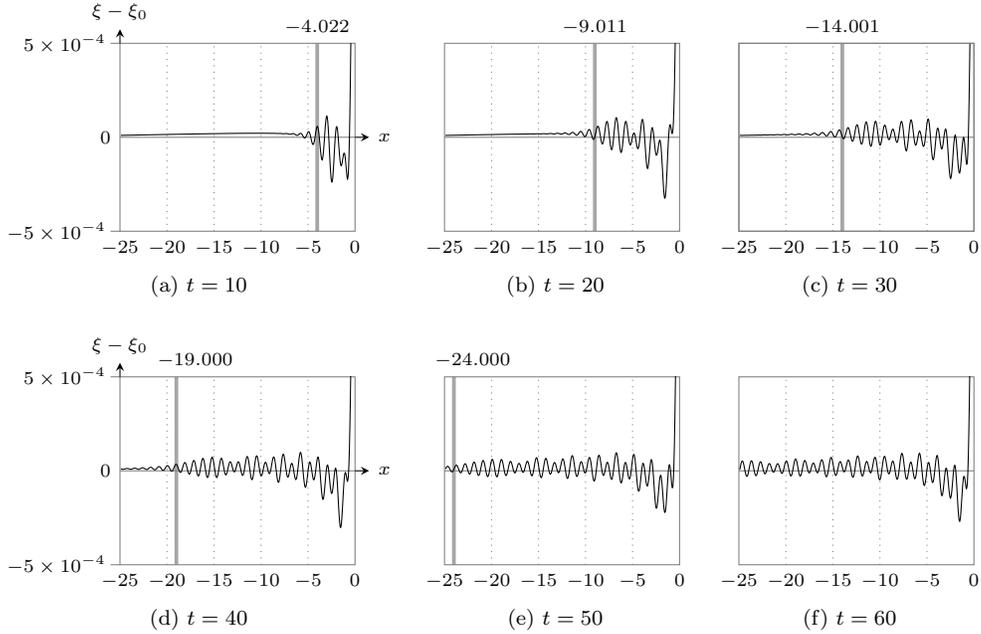

In figure \eqref{2:unsteady_num}, we compare these results to numerical computations, obtained using the numerical scheme adapted from the algorithm detailed in \cite{Lustri4}. In this figure we show that the expanding front matches the position obtained by solving $\mathrm{Im}(\chi_S) = \mathrm{Im}(\chi_U)$ exactly. In each case, we expect that the waves will switch on as the Stokes line is crossed, and consequently that the waves have half amplitude at this point, and rapidly decay as it is crossed. This is consistent with the computed free-surface behaviour. We see that the position of the Stokes line accurately describes the boundary of the capillary wave region, and therefore the propagation of these capillary waves.

\begin{figure}
\centering

\includegraphics[width=0.99\textwidth]{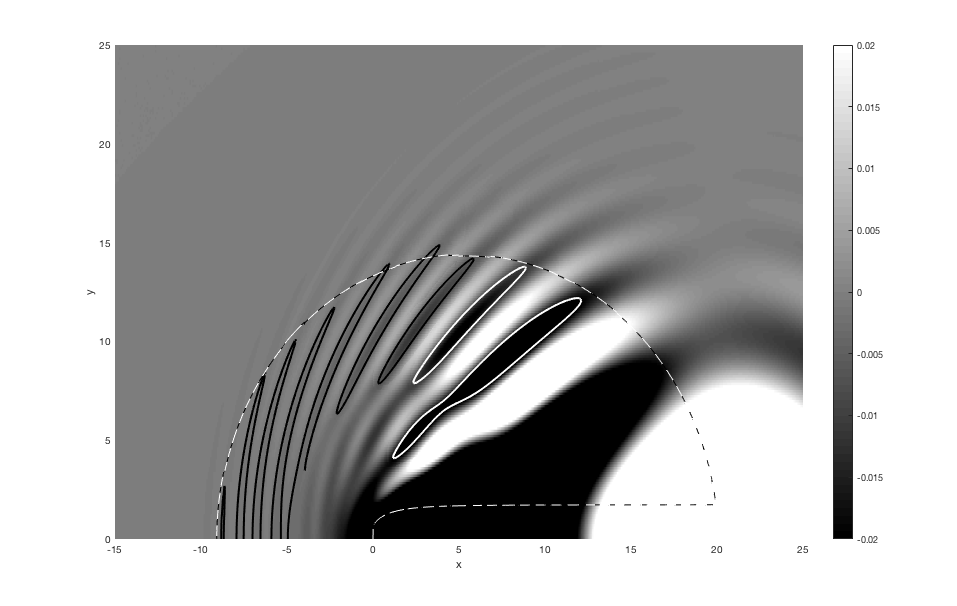}

\caption{Modified free surface at $t = 20$, with $\epsilon = 0.15$,
  obtained numerically. The dashed curve illustrates the Stokes
  curve. The black and white solid curves denote the point at which
  each wave takes half of the maximum wave amplitude for that
  particular wave trough. It can be seen that each of the indicated
  troughs reduces to half amplitude approximately as the Stokes line
  is crossed. There is some visible wave behaviour outside this
  region, due to the fact that Stokes switching behaviour is smooth
  for finite $\epsilon$.}
\label{F:3DNumUnstead1}
\end{figure}

Finally, in figure \ref{F:3DNumUnstead1}, we show the full computed
two-dimensional system for $t = 20$ and $\epsilon = 0.15$, with the
position of the Stokes line overlaid. We see that the capillary waves
are clearly switched on in the interior of the predicted Stokes line,
decaying to half of the maximum wave amplitude at the Stokes line, and
rapidly decaying away as the Stokes line is crossed into the exterior
region. 
%
%\begin{figure}
%\centering
%
%\includegraphics[width=1.0\textwidth]{Unsteady_Cap_t20_temp.eps};
%
%\caption{Surface plot of the modified free-surface position $\xi - \xi^{(0)}$, where $\xi^{(0)}$ is the leading-order free-surface profile given in \eqref{1:xi0}, with $h = 1$ and $\epsilon = 0.2$. Flow is in the positive $x$-direction. We can clearly see capillary waves extending backwards in the direction of the flow origin.}
%\label{F:3DNumUnstead}
%\end{figure}

%

%
%
%
%
%
\section{Discussion and Conclusions}

\subsection{Conclusions}

In this investigation, we calculated the behaviour of steady and unsteady capillary waves on the free surface of flow over a point source in three dimensions in the low surface tension limit. We considered the source to be weak, and therefore linearised the problem about the undisturbed solution. In the analysis of the unsteady capillary wave problem, the flow was initially set to be waveless. 

We subsequently applied exponential asymptotic techniques in order to determine the behaviour of the resultant capillary waves. By analysing the Stokes switching behaviour present in the solution to the problem, we were able to determine the form of the waves on the free surface. Examples of this behaviour are illustrated in figure \ref{F:3DNum} for the steady flow problem, and figure \ref{F:3DNum2} for the unsteady flow problem. We note that the steady wave behaviour seen in figure \ref{F:3DNum}, with the behaviour illustrated in figure \ref{1:realrays} (a) is qualitatively similar to the flow induced by a whirligig beetle in figure 2 of \cite{Tucker1}.

In the steady case, the far-field amplitude of these capillary waves was compared to numerical solutions to the linearised equations in figure \ref{F:SteadyAsympNum}. The numerical results were obtained by formulating the solution to the linearised system as an integral equation using methods similar to those given in \cite{Lustri2}, and evaluating the integral numerically. The comparison showed agreement between the asymptotic and numerical wave amplitudes.

We then considered the behaviour of unsteady capillary waves, in order to determine how these waves propagate over time. We found that there is a transient component of the surface behaviour generated by the initial disturbance. This transient surface behaviour switched on capillary waves across a second-generation Stokes line, which moves in space as $t$ increases. We also determined the location of higher-order Stokes phenomenon, which determined locations at which Stokes curves terminate; this was required in order to complete the Stokes structure of the unsteady problem, illustrated in figure \ref{2:movingsl}. This analysis showed that the steady capillary waves are restricted to a circular region of increasing size, with the downstream region removed.

Finally, we compared the position of the spreading wavefront predicted by the asymptotics with numerical solutions to the unsteady problem. The numerical solutions were obtained by formulating an integral expression for the surface behaviour in a similar fashion to \cite{Lustri4} and computing the solution to the integrals. The results are seen in figure \ref{2:unsteady_num} and \ref{F:3DNumUnstead1}, and show agreement between the asymptotic and numerical results.

The natural next step in this investigation is to study the Stokes
structure that appears in systems in which both gravity and capillary
waves play an important role. Following the work of \cite{Trinh3,
  Trinh4}, we expect that the behaviour of surface waves in these
systems requires determining not only the individual gravity and
capillary wave contributions, but also the higher-order and
second-generation Stokes interactions within the system. A brief
analysis of the combined gravity-capillary wave problem is included in
Appendix \ref{APP_GravCap}, in which the singulant equation is
obtained; however, solving this singulant equation is a challenging
numerical problem that is beyond the scope of this study.

\appendix

\section{Finding the prefactor}\label{PREFACTOR_APP}
\subsection{Prefactor equation}
In order to solve the prefactor equation (\ref{1:ansgov2}), we will express the equation on the free surface entirely in terms of $x$ and $y$ derivatives. This will result in an equation that has the exact same ray structure as the singulant equation (\ref{1:eik}), and hence the solution may be obtained in terms of the same characteristic variables. To accomplish this, we must eliminate the $z$ derivatives from all relevant quantities. Equations (\ref{1:chiz}) and (\ref{1:phiz}) give appropriate expressions for $\chi_z$ and $\Phi_z$ respectively, however we must still consider the second derivative terms that will appear in the equation.

Taking derivatives of (\ref{1:chiz}) and rearranging gives
\begin{align}
\label{1:chixz}\chi_{xz} &= -\frac{2\chi_x\chi_{xx}}{\chi_x^2+\chi_y^2} + \frac{2\chi_x^2(\chi_x\chi_{xx}+\chi_y\chi_{xy})}{(\chi_x^2+\chi_y^2)^2},\\
\label{1:chiyz}\chi_{yz} &= -\frac{2\chi_x\chi_{xy}}{\chi_x^2+\chi_y^2} + \frac{2\chi_x^2(\chi_x\chi_{xy}+\chi_y\chi_{yy})}{(\chi_x^2+\chi_y^2)^2},\\
\label{1:chizz}\chi_{zz}&= \frac{4\chi_x^2\chi_y^2(\chi_{xx}\chi_y^2 - 2\chi_x\chi_y\chi_{xy} + \chi_x^2\chi_{yy})}{(\chi_x^2+\chi_y^2)^4}.
\end{align}

Using (\ref{1:chixz})--(\ref{1:chizz}), as well as (\ref{1:chiz}), (\ref{1:phixi}), and (\ref{1:phiz}) we are finally able to write the prefactor equation (\ref{1:ansgov2}) in terms of $x$ and $y$ derivatives on $z=0$ as
\begin{equation}
 \label{1:sspfeq}\left[4\chi_x^3 + 6\chi_x(\chi_x^2+\chi_y^2)\right]\Phi_x + \left[6\chi_y(\chi_x^2 + \chi_y^2)\right]\Phi_y = G(x,y)\Phi, 
\end{equation}
where
\begin{align*}
G(x,y) = \Bigg[\frac{6\chi_x^2\chi_y^2(\chi_x^2-\chi_y^2)}{(\chi_x^2+\chi_y^2)^2}-(\chi_x^2+\chi_y^2)^2  \Bigg]&\chi_{xx} +\\
\Bigg[\frac{8\chi_x^3 \chi_y (\chi_x^2 - 2\chi_y^2)}{(\chi_x^2+\chi_y^2)^2}\Bigg]\chi_{xy}&+\Bigg[\frac{2\chi_x^4(\chi_x^2-5\chi_y^2)}{(\chi_x^2+\chi_y^2)^2}-(\chi_x^2+\chi_y^2)^2  \Bigg]\chi_{yy}.
\end{align*}
This equation may be solved using the method of characteristics, giving the ray equations (with characteristic variable $u$) as
\begin{equation}
\label{1:ssrays2}\frac{\mathrm{d}x}{\mathrm{d}u} = 4\chi_x^3 + 6\chi_x(\chi_x^2+\chi_y^2)^2,\qquad \frac{\mathrm{d}y}{\mathrm{d}u} = 6\chi_y(\chi_x^2 + \chi_y^2)^2,\qquad \frac{\mathrm{d}\Phi}{\mathrm{d}u} = G(x,y)\Phi.
\end{equation}
The first two of these equations govern the ray paths, and importantly, are identical to the ray equations associated with (\ref{1:eik}). This allows (\ref{1:ssrays2}) to be written in terms of the associated Charpit variables, and solved to give 
\begin{equation}
 \label{1:Phiss1} \Phi(s,u) = \Phi(s,0)\left[1+\frac{6s^6 u(4s^4+6s^2h^2-3h^4)}{h^7(2s^4+3h^4)}\right]^{\tfrac{s^2(24s^8+34s^6h^2+36s^4h^4+35h^6s^2+14h^8)}{6h^6(3h^4-6s^2h^2-4s^4)}},\end{equation}
where the characteristic variable $u$ is given by
\begin{equation}
u = -\frac{h^{11}(s-x)}{2s^9(3s^2+2)}.
\end{equation}
Selecting the corresponding expression for $s$ in terms of $x$ and $y$ from (\ref{1:sspoly}) gives the solution in terms of the physical coordinates $x$ and $y$. To find an expression for $\Phi(s,0)$, the behaviour of the system in the neighbourhood of $u = 0$ must be computed and matched to this outer solution. 

\subsection{Inner problem}\label{CH5_INNER1}

%In actuality, we are only getting switching from one-quarter of the full contribution. There is $\sigma_j$ and $-\sigma_j$ for $j = 1, 2, 3, 4$. We have been ignoring the later ones, but they matter in the inner region at late orders. This is where the mystery one quarter is coming from.

To solve the inner problem, we first consider the behaviour of $\chi_{L1}$ near the singularity at $x^2 + y^2 + (z+h)^2 = 0$, which takes the form
\begin{equation}
\label{1:innerchi1} \chi_{L1} \sim \frac{x^2}{2h^3}\left(x^2 + y^2 + (z+h)^2\right).
\end{equation}
In the prefactor equation (\ref{1:Phiss}), we see that the unknown coefficient is a function of $s$. From (\ref{1:sscauchy}), it follows that $s \sim x$ near the singularity at $t = 0$. Hence, we define a system of inner coordinates given by
\begin{equation}
\label{1:innercoords} \epsilon \sigma_1 = \frac{x^2}{2h^3}\left(x^2 + y^2 + (z+h)^2\right),\qquad\epsilon \sigma_2 = \frac{x^2}{2h^3}\left(x^2 + y^2 + (z-h)^2\right),\qquad\lambda = x.
\end{equation}
% The partial derivatives subsequently become
% \begin{align*}
%  \pdiff{}{x} &= \left[\frac{hx}{\lambda^2 \epsilon} - 2 \lambda \sigma_1 \right]\pdiff{}{\sigma_1} + \left[\frac{hx}{\lambda^2 \epsilon} - 2 \lambda \sigma_2 \right]\pdiff{}{\sigma_2}+ \pdiff{}{\lambda},\\         
%  \pdiff{}{y} &= \left[\frac{hy}{\lambda^2 \epsilon}\right]\pdiff{}{\sigma_1} + \left[\frac{hy}{\lambda^2 \epsilon}\right]\pdiff{}{\sigma_2},\\
%  \pdiff{}{z} &= \left[\frac{h(z+h)}{\lambda^2\epsilon}\right]\pdiff{}{\sigma_1} + \left[\frac{h(z-h)}{\lambda^2\epsilon}\right]\pdiff{}{\sigma_2}   .  
% \end{align*}
To leading order in $\epsilon$, the linearised governing equation (\ref{1:lgeq}) becomes (omitting the bars) 
% \begin{equation*}
%  2\lambda^2\sigma_1\phi_{\sigma_1\sigma_1} + 2\lambda^2\sigma_2\phi_{\sigma_2\sigma_2} - (z+h)\phi_{\lambda\sigma_2} - (z-h)\phi_{\lambda\sigma_1} = 0.
% \end{equation*}
\begin{equation*}
 3\sigma_1\phi_{\sigma_1\sigma_1} + 3\sigma_2\phi_{\sigma_2\sigma_2} + \lambda\phi_{\lambda\sigma_2} + \lambda\phi_{\lambda\sigma_1} = 0,
\end{equation*}
where terms containing derivatives with respect to both $\sigma_1$ and $\sigma_2$ were disregarded due to the form of the inner expansion, (\ref{1:localseries1}).
% where terms that are small compared to the leading-order behaviour as $\epsilon \rightarrow 0$ have been discarded. Furthermore, we find that
% \begin{equation*}
%  z \sim \frac{\eps(\sigma_1 - \sigma_2)}{4h},
% \end{equation*}
% and hence disregarding terms that are small compared to the leading-order behaviour leaves 
% \begin{equation*}
%  2\lambda^2\sigma_1\phi_{\sigma_1\sigma_1} + 2\lambda^2\sigma_2\phi_{\sigma_2\sigma_2} - h\phi_{\lambda\sigma_2} + h\phi_{\lambda\sigma_1} = 0.
% \end{equation*}
Similarly, the boundary conditions (\ref{1:lbc1})-(\ref{1:lbc2}) become 
\begin{alignat}{2}
\label{1:localbc1} h\phi_{\sigma_1} - h\phi_{\sigma_2} - \lambda\xi_{\sigma_1} -\lambda\xi_{\sigma_2}&= 0 \qquad & \mathrm{on} \quad \sigma_1 &= \sigma_2,\\
\label{1:localbc2} h\phi_{\sigma_1} +h\phi_{\sigma_2} -\lambda\xi_{\sigma_1\sigma_1} - \lambda\xi_{\sigma_2\sigma_2} &= 0  \qquad & \mathrm{on} \quad \sigma_1 &= \sigma_2.
\end{alignat}
Finally, by expressing the leading-order behaviour (\ref{1:phi0}) in terms of the local variables, we find that
\begin{equation}
\label{1:localphi0} \phi^{(0)} \sim \frac{\lambda\sqrt{2}}{8\pi h^{3/2}\epsilon^{1/2}\sigma_1^{1/2}}-\frac{\lambda\sqrt{2}}{8\pi h^{3/2}\epsilon^{1/2}\sigma_2^{1/2}}.
\end{equation}
We now define the series expansion near the singularity on the complexified free surface as
\begin{align}
\label{1:localseries1} \phi \sim \sum_{n=0}^{\infty} \left[\frac{a_n(\lambda)\Gamma(n+1/2)}{\sigma_1^{n+1/2}} + \frac{b_n(\lambda)\Gamma(n+1/2)}{\sigma_2^{n+1/2}}\right], \qquad \xi \sim \sum_{n=0}^{\infty} \left[\frac{2c_n(\lambda)\Gamma(n+1/2)}{\sigma_1^{n+1/2}}\right],
\end{align}
where the latter expression is only valid on the free-surface itself, on which $\sigma_1 = \sigma_2$. The factor of two is included for subsequent algebraic convenience, and has no effect on the solution to the problem as $c_n$ is  unknown at this stage. From (\ref{1:localphi0}), we have
\begin{equation}
\label{1:locala0b0} a_0(\lambda) = \frac{\lambda\sqrt{2}}{8\pi h^{3/2}},\qquad b_0(\lambda) = -\frac{\lambda\sqrt{2}}{8\pi h^{3/2}}.
\end{equation}
We are interested in the behaviour of the terms on the complexified free surface in the neighbourhood of the singularity at $x^2 + y^2 + h^2 = 0$. Consequently, we apply the series expression to (\ref{1:localbc1}) on the surface (defined by $\sigma_1 = \sigma_2$) and match in the limit that $\sigma_1$ (and therefore $\sigma_2$) tend to zero, giving
\begin{equation}
\label{1:localsurf1} -h(a_n - b_n) - 2\lambda c_n = 0, \qquad n \geq 0.
\end{equation}
Applying the series expansion to (\ref{1:localbc2}) and matching in the same limit gives
\begin{equation}
\label{1:localsurf2} -h(n+3/2)(a_n + b_n) + 2 c_{n+1} = 0, \qquad n \geq 0.
\end{equation}
We are interested in the behaviour on the complexified free-surface; however, restricting the domain in this fashion means that it is impossible to distinguish between the contributions from the series in $\sigma_1$ and the series in $\sigma_2$. However, we see that the two contributions have equal magnitude in (\ref{1:localphi0}). As the singular behaviour of the problem is preserved in all higher orders \cite{Dingle1}, we conclude that this must be true of the contributions at all subsequent orders. We therefore specify that $|a_n| = |b_n|$ in order to maintain consistency with the leading-order behaviour. This may only be accomplished if we divide the two equations given in (\ref{1:localsurf1})--(\ref{1:localsurf2}) into four equations such that
\begin{alignat*}{2}
 -h a_n- \lambda c_n &= 0, \qquad\qquad &-h (n+1/2) a_n+ \lambda c_{n+1} &= 0,\\
 h b_n- \lambda c_n  &= 0, \qquad\qquad &-h (n+1/2) b_n+ \lambda c_{n+1}&= 0.
\end{alignat*}
We will consider only the first two of these equations, noting that the remaining equations imply that $b_n = (-1)^n a_n$. Eliminating $c_n$ from this system gives
\begin{equation*}
 a_{n+1} =  (n+1/2)a_n = \frac{a_0 \Gamma(n+1/2)}{\Gamma(1/2)}.
\end{equation*}
Hence, using the expression for $a_0$ given in (\ref{1:locala0b0}), we may match the local series expression given in (\ref{1:localseries1}) with the prefactor given in (\ref{1:Phiss}). Noting that $\lambda$ is the local expression for $s$ in the outer solution, and that $\Phi(s,0)$ in the outer coordinates matches with $a_n(\lambda) + b_n(\lambda)$ in the inner coordinates, we find that
%\begin{equation}
%\label{1:arbitraryprefactor} \Phi(s,0) = \frac{\sqrt{2}s}{8\pi^{3/2} h^{3/2}}.
%\end{equation}
\begin{equation}
\label{1:arbitraryprefactor} \Phi(s,0) = \frac{s\sqrt{2}}{4\pi^{3/2} h^{3/2}}.
\end{equation}
Hence, we are able to completely describe the late-order behaviour of terms in (\ref{1:series}), with the complete expression given in (\ref{1:Phiss}).

\section{Stokes Smoothing}\label{CH5_STEADYSMOOTHING}

The asymptotic series given in (\ref{1:series}) may be truncated to give
\begin{equation*}
 \label{1:series2}\overline{\phi} = \sum_{n=0}^{N-1}\epsilon^n\phi^{(n)} + R^{(N)},\qquad \overline{\xi} = \sum_{n=0}^{N-1}\epsilon^n\xi^{(n)} + S^{(N)},
\end{equation*}
where $N$ will be chosen in order to minimise the remainders $R^{(N)}$ and $S^{(N)}$. Applying this series expression to (\ref{1:lgeq}) gives
\begin{equation}
 \label{1:ssge}\nabla^2 R^{(N)} = 0,
\end{equation}
while the boundary conditions (\ref{1:lbc1})--(\ref{1:lbc2}) become on $z = 0$
\begin{align}
 \label{1:ssit0}  R^{(N)}_z - S^{(N)}_x &= 0,\\
 \label{1:ssit} R^{(N)}_x + \ \epsilon \left( S^{(N)}_{xx} + S^{(N)}_{yy} \right)&= -\epsilon^N(\xi^{(N-1)}_{xx} - \xi^{(N-1)}_{yy}),
\end{align}
having made use of the relationship in (\ref{1:serbc2}) and the fact that $\phi_x^{(0)} = 0$. The homogeneous form of (\ref{1:ssge})--(\ref{1:ssit}) is satisfied as $\epsilon \rightarrow 0$ by
\begin{equation*}
  \label{1:ssrt2}R^{(N)} \sim \Phi\mathrm{e}^{-\chi/\epsilon},\qquad S^{(N)} \sim \Xi\mathrm{e}^{-\chi/\epsilon},
\end{equation*}
where $\chi$ is one of the singulants determined from (\ref{1:sschiray2})--(\ref{1:sspoly}). 

 We therefore set the remainder terms for the inhomogeneous problem to take the form
\begin{equation}
  \label{1:ssrt}R^{(N)} = A(x,y,z)\Phi\mathrm{e}^{-\chi/\epsilon},\qquad S^{(N)} = B(x,y)\Xi\mathrm{e}^{-\chi/\epsilon},
\end{equation}
where $A$ and $B$ are Stokes switching parameters. From (\ref{1:ssit0}), we see that $A = B$ on $z=0$. 

To determine the late order term behaviour, we will require the first correction term for the prefactors, and we therefore set
\begin{equation*}
\Phi = \Phi_0 + \epsilon \Phi_1 + \ldots, \qquad \Xi = \Xi_0 + \epsilon \Xi_1 + \ldots.
\end{equation*}
Applying the remainder forms given in (\ref{1:ssrt}) to the boundary conditions , (\ref{1:ssit0}) and (\ref{1:ssit}), gives after some rearrangement
\begin{align*}
-A\chi_x\Xi_1 + A\chi_z\Phi_1 =& \,A\Xi_{0,x}  + A_x\Xi_0- A\Phi_{0,z} -A_z\Phi_0 ,\\
-A\chi_x\Phi_1 +A (\chi_x^2 + \chi_y^2)\Xi_1  =&\,  2A\chi_x\Xi_{0,x} + 2A\chi_y\Xi_{0,y}  + 2A_x\chi_x\Xi_{0}\\& + 2A_y\chi_y\Xi_{0} + A(\chi_{xx}+\chi_{yy})\Xi_0 \\&- A\Phi_{0,x} - A_x\Phi_{0} + \epsilon^N \mathrm{e}^{\chi/\epsilon} (\xi^{(N-1)}_{xx} +  \xi^{(N-1)}_{yy}).
\end{align*}
Combining these expressions, and making use of (\ref{1:phiz}) to eliminate terms and (\ref{1:serbc2}) to simplify the right-hand side gives
\begin{equation}
A_z\Phi_0 - A_x \Xi_0 + 2A_y\chi_y\Xi_0 + 2A_x\chi_x\Xi_0 - A_x\Phi_0 \sim - \epsilon^N \phi_x^{(N)} \mathrm{e}^{\chi/\epsilon}.
\end{equation}
As only the leading order prefactor behaviour appears in the final expression, we will no longer retain the subscripts. Applying the late-order ansatz gives
\begin{equation*}
A_z\Phi - A_x \Xi + 2A_y\chi_y\Xi + 2A_x\chi_x\Xi - A_x\Phi \sim  \epsilon^N \frac{\chi_x\Phi\Gamma(N+3/2)}{\chi^{N+3/2}} \mathrm{e}^{\chi/\epsilon}.
\end{equation*}
Motivated by the homogeneous solution, we express the equation in terms of $\chi$ and $y$, and apply (\ref{1:phixi}) to obtain 
\begin{equation*}
 A_{\chi}  = \epsilon^{N}\mathrm{e}^{{\chi}/\epsilon}\frac{\Gamma(N+3/2)}{{\chi}^{N+3/2}}.
\end{equation*}
The optimal truncation point is given by $N \sim |\chi|/\epsilon$ in the limit that $\epsilon \rightarrow 0$. We write ${\chi} = r\mathrm{e}^{\mathrm{i}\theta}$, with $r$ and $\theta$ real so that $N = r/\epsilon + \alpha$, where $\alpha$ is necessary to make $N$ an integer. Since $N$ depends on $r$ but not $\theta$, we write
\begin{equation*}\label{CH2S1_TruncatedSeriesRescaling1}
 \frac{\partial}{{\partial\chi}} = -\frac{\mathrm{i}\mathrm{e}^{-\mathrm{i}\theta}}{r}\frac{\partial}{\partial\theta}.
\end{equation*}
Using Stirling's formula on the resultant expression gives 
\begin{equation*}
 A_{\theta}\sim\frac{\mathrm{i} \sqrt{2\pi r}}{\epsilon}\exp\left(\frac{r}{\epsilon}\left(\mathrm{e}^{\mathrm{i}\theta}-1\right) - \mathrm{i}\theta\left(\frac{r}{\epsilon}+\alpha-\frac{1}{2}\right)\right).
\end{equation*}
This variation is exponentially small, except in the neighbourhood of the Stokes line, given by $\theta = 0$, where it is algebraically large To investigate the rapid change in $A$ in the vicinity of the Stokes line, we set $\theta = \epsilon^{1/2}\hat{\theta}$, giving
\begin{equation*}
 A_{\hat{\theta}} \sim {\mathrm{i} \sqrt{\frac{2\pi r }{\epsilon}}}\mathrm{e}^{-r\hat{\theta}^2/2},
\end{equation*}
so that
\begin{equation*}
 A\sim \mathrm{i}\sqrt{\frac{2\pi}{\epsilon}}\int_{-\infty}^{\theta \sqrt{r/\epsilon}} \mathrm{e}^{-t^2/2} \mathrm{d} t + C,
\end{equation*}
where $C$ is constant. Thus, as the Stokes line is crossed, $A$ rapidly increases from 0 to $2\pi\mathrm{i}\epsilon^{-1/2}$. Using \eqref{1:ssrt}, we find the variation in the fluid potential, and we subsequently use \eqref{1:ssit0} to relate $B$ to $A$, and therefore find the variation in the free surface behaviour as the Stokes line is crossed. The Stokes line variation for the potential and free surface position are respectively given by
\begin{equation}
 \label{1:ssvariation} \left[R^{(N)}\right]_-^+ =\frac{2\pi\mathrm{i}\Phi}{\sqrt{\epsilon}}\mathrm{e}^{-\chi_S/\epsilon},\qquad \left[S^{(N)}\right]_-^+ =\frac{2\pi\mathrm{i}\Xi}{\sqrt{\epsilon}}\mathrm{e}^{-\chi_S/\epsilon},
\end{equation}
where we have reintroduced the specific singulant form, $\chi_S$. Hence, if we determine the prefactor and singulant behaviour associated with each contribution, (\ref{1:ssvariation}) gives an expression for the behaviour switched on across the appropriate Stokes line. The combined expression for the exponentially small terms in regions where they are active is therefore given by \eqref{1:RnSn}.

\section{Gravity-Capillary Waves}\label{APP_GravCap}
The natural sequel to this work is to combine capillary and gravity waves, in order to determine how the two wave contributions interact. It is likely that the Stokes structure will be significantly more complicated than the Stokes structure for either capillary or gravity waves alone. Previous work on the two-dimensional problem by \cite{Trinh3, Trinh4} shows that the interaction between Stokes lines associated with gravity and capillary waves plays an important role in the behaviour of waves on the free surface. 

Including both gravity and capillary effects in the analysis 
requires scaling both the Weber number $\mathrm{We}= \rho L U^2/\sigma$, and
the Froude number 
$F= U/\sqrt{gL}$, where as before $\sigma$ is the
surface tension, $\rho$ is the fluid density, $U$ is the background
fluid velocity,  $L$ is a representative length scale, and $g$ is the
acceleration due to gravity. 

As determined by \cite{Trinh3, Trinh4}, we see that the scaling in
which both gravity and capillary waves play an important role is given
by setting $F^2 =  \beta \epsilon$ and $\mathrm{We}^{-1} = \beta\tau\epsilon^2$ as $\epsilon
\rightarrow 0$. The variables $\beta$ and $\tau$ determine the
relationship between the Froude and Weber number. 

This gives a system that is nearly identical to
\eqref{1:nlgeq}--\eqref{1:nlsc1}, with the dynamic condition
\eqref{1:nlbc2} now given by 
\begin{equation}
\frac{\beta\epsilon}{2}(|\nabla\phi^2| - 1) + \xi = \beta\tau\epsilon^2\kappa \qquad \mathrm{on} \qquad z = \xi(x,y).
\end{equation}
After linearisation, this boundary condition becomes
\begin{equation}
\beta \epsilon\phi_x + \xi = \beta \tau \epsilon^2 (\xi_{xx} + \xi_{yy})\qquad \mathrm{on} \qquad z = 0.
\end{equation}

Applying late-order techniques to the linearised system in a similar fashion to section \ref{LOT_steady} yields the singulant equation
\begin{equation}\label{d:sing}
\beta^2\chi_x^4 + (\chi_x^2 + \chi_y^2)[\beta\tau(\chi_x^2+\chi_y^2) - 1]^2 = 0,
\end{equation}
with the boundary condition
\begin{equation}\label{d:singbc}
\chi = 0 \qquad \mathrm{on} \qquad x^2 + y^2 + h^2 = 0.
\end{equation}
We see that when $\tau = 0$, this system gives the gravity wave singulant from \cite{Lustri2}, while for $\tau = 1$, the capillary wave singulant \eqref{1:eik} is obtained in the limit $\beta \rightarrow \infty$.

The Stokes surfaces can be obtained by obtaining the full set of solutions to this system, and determining the Stokes surfaces. This is a challenging problem involving complex ray tracking, as seen in \cite{Stone1}, and is beyond the scope of the present study.

\bibliographystyle{plain}
\bibliography{thesisrefs}

%\bibliography{jfm-instructions}

\end{document}